\documentclass[aps,prd,twocolumn,nofootinbib,preprintnumbers,superscriptaddress,showpacs,amsmath,amssymb]{revtex4-1}

\usepackage{comment}
\usepackage{booktabs,makecell}
\usepackage{graphicx}
\usepackage{amsmath,amssymb}
\usepackage{amsfonts}
\usepackage{xspace} 
\usepackage[usenames]{color}
\usepackage{dcolumn}
\usepackage{bm}
\usepackage{mathrsfs}
\usepackage[colorlinks=true]{hyperref}
\usepackage[all]{hypcap} 
\usepackage[utf8]{inputenc} 
\usepackage{slashed}
\usepackage{multirow}
\usepackage{rotating}
\usepackage{nicematrix}
\definecolor{orange}{rgb}{1,0.5,0}
\usepackage{tabularx}

\graphicspath{{./}{Figures/}}
\begin{document}
%
%

\title{
Fastest spinning millisecond pulsars: indicators for quark matter in neutron stars? 
}

\author{Christoph Gärtlein} \email{christoph.gartlein@tecnico.ulisboa.pt} 
\affiliation{Centro de Astrof\'{\i}sica e Gravita\c c\~ao  - CENTRA, Departamento de F\'{\i}sica, Instituto Superior T\'ecnico - IST, Universidade de Lisboa - UL, Av. Rovisco Pais 1, 1049-001 Lisboa, Portugal}
\affiliation{CFisUC, Department of Physics, University of Coimbra, Rua Larga 3004-516, Coimbra, Portugal}
\affiliation{Institute of Theoretical Physics, University of Wroclaw, 50-204 Wroclaw, Poland}

\author{Violetta Sagun} 
\email{v.sagun@soton.ac.uk}
\affiliation{Mathematical Sciences and STAG Research Centre, University of Southampton, Southampton SO17 1BJ, United Kingdom}
\affiliation{CFisUC, Department of Physics, University of Coimbra, Rua Larga 3004-516, Coimbra, Portugal}

\author{Oleksii Ivanytskyi} \email{oleksii.ivanytskyi@uwr.edu.pl}  
\affiliation{Incubator of Scientific Excellence---Centre for Simulations of Superdense Fluids, University of Wrocław, 50-204, Wroclaw, Poland}

\author{David Blaschke}
\email{david.blaschke@uwr.edu.pl} 
\affiliation{Institute of Theoretical Physics, University of Wroclaw, 50-204 Wroclaw, Poland}
\affiliation{Helmholtz-Zentrum Dresden-Rossendorf (HZDR), Bautzner Landstrasse 400, 01328 Dresden, Germany}
\affiliation{Center for Advanced Systems Understanding (CASUS), Untermarkt 20, 02826 G\"orlitz, Germany}

\author{Ilidio Lopes}
\email{ilidio.lopes@tecnico.ulisboa.pt} 
\affiliation{Centro de Astrof\'{\i}sica e Gravita\c c\~ao  - CENTRA, Departamento de F\'{\i}sica, Instituto Superior T\'ecnico - IST, Universidade de Lisboa - UL, Av. Rovisco Pais 1, 1049-001 Lisboa, Portugal}

\date{\today}
\begin{abstract}

We study rotating hybrid stars, with a particular emphasis on the effect of a deconfinement phase transition on their properties at high spin. Our analysis is based on a hybrid equation of state with a phase transition from hypernuclear matter to color-superconducting quark matter, where both phases are described within a relativistic density functional approach. By varying the vector meson and diquark couplings in the quark matter phase, we obtain different hybrid star sequences with varying extension of the quark matter core, ensuring consistency with astrophysical constraints from mass, radius and tidal deformability measurements. As a result, we demonstrate the impact of an increasing rotational frequency on the maximum gravitational mass, the central energy density of compact stars, the appearance of the quasi-radial oscillations and non-axisymmetric instabilities. We demonstrate that for the most favorable parameter sets with a strong vector coupling, hybrid star configurations with color superconducting quark matter core can describe the fastest spinning and heaviest galactic neutron star J0952-0607, while it is out of reach for the purely hadronic hypernuclear star configuration.
We also revise the previously proposed empirical relation between the Kepler frequency, gravitational mass, and radius of non-rotating neutron stars, obtained based on the assumption that all neutron stars, up to the heaviest, are hadronic. We show how the phase transition to quark matter alters this relation and, consequently, the constraints on the dense matter equation of state. Our findings reveal that incorporating the hybrid equation of state has significant implications for the constraints on the properties of strongly interacting matter and neutron stars, placing the upper limit on $R_{1.4} \leq 14.90$ km and $R_{0.7}<11.49$ km (considering 716 Hz frequency limit from J1748+2446ad) and $R_{1.4}\leq$11.90~km (for 1000 Hz).

\end{abstract}
\maketitle

\section{Introduction}
\label{intro}

The most rapidly rotating neutron stars (NSs) are known as millisecond pulsars (MSPs) due to their millisecond-range rotation periods. In comparison, most pulsars are observed with a spin period in the 0.1 to 10-second range. MSPs are formed by the accretion-induced spin-up of the old NS in a close binary system. In the low-mass X-ray binary, the angular momentum transfer from the companion star onto the NS with infalling matter causes a ``recycling" spin-up~\cite{1991PhR...203....1B}. This process leads to the formation of such unique objects as MSPs, characterized by the extreme rotation frequency, remarkable rotational stability, high age ($\sim10^9$ yr), low magnetic fields ($\sim10^{8-9}$ G)~\cite{1998Natur.394..344W, Guillot:2019vqp} and possibly also high masses \citet{Romani:2022jhd}.

Therefore, MSPs attract much attention as the unprecedented laboratory to test fundamental physics such as the question for quark matter inside neutron stars \cite{Alford:2006vz,Koberlein:2024}. 
Recently this stimulated a splash of interest to the physics of rotating hybrid stars \cite{Rather:2021yxo,Espino:2021adh,Sen:2022qol,Tsaloukidis:2022rus,Moreno:2023xez}.
Thus, high-precision pulse measurements in compact binaries allow unique tests of the theories of gravity with the double-pulsar system J0737-3039~\cite{Kramer:2006nb} and obtaining the spin-orbit coupling, which allows extracting the moment of inertia~\cite{Bejger:2005jy,Kramer:2009zza}. The latter, together with the simultaneous measurements of pulsar mass and radius with the NICER telescope, aim to constrain the dense matter equation of state (EoS) and probe the interior composition of NSs~\cite{Raaijmakers:2019qny}. MSPs in binary systems, e.g. ``spider'' pulsar binaries, are also targets for mass determination~\cite{Romani:2021xmb,Romani:2022jhd}.

Although theoretical models of NSs allow much higher spin rotation rates above 1 kHz, the fastest known pulsar, PSR J1748-2446ad, has a spin period of 1.4 ms (716 Hz)~\cite{Hessels:2006ze} only. This fact raised a lot of discussions regarding whether the observed limit of $\sim$700 Hz corresponds to the spin frequency cutoff related to the Kepler (mass-shedding) limit above which matter becomes unbound~\cite{Haskell:2018nlh}, it is related to the binary system evolution~\cite{Patruno:2017oum} or the result of the deconfinement phase transition in the NS core causing either the change of the frequency distribution~\cite{Glendenning:2000zz} or the rotational frequency cutoff~\cite{Bejger:2016emu}. 

\begin{figure}[ht]
\includegraphics[width=\columnwidth]{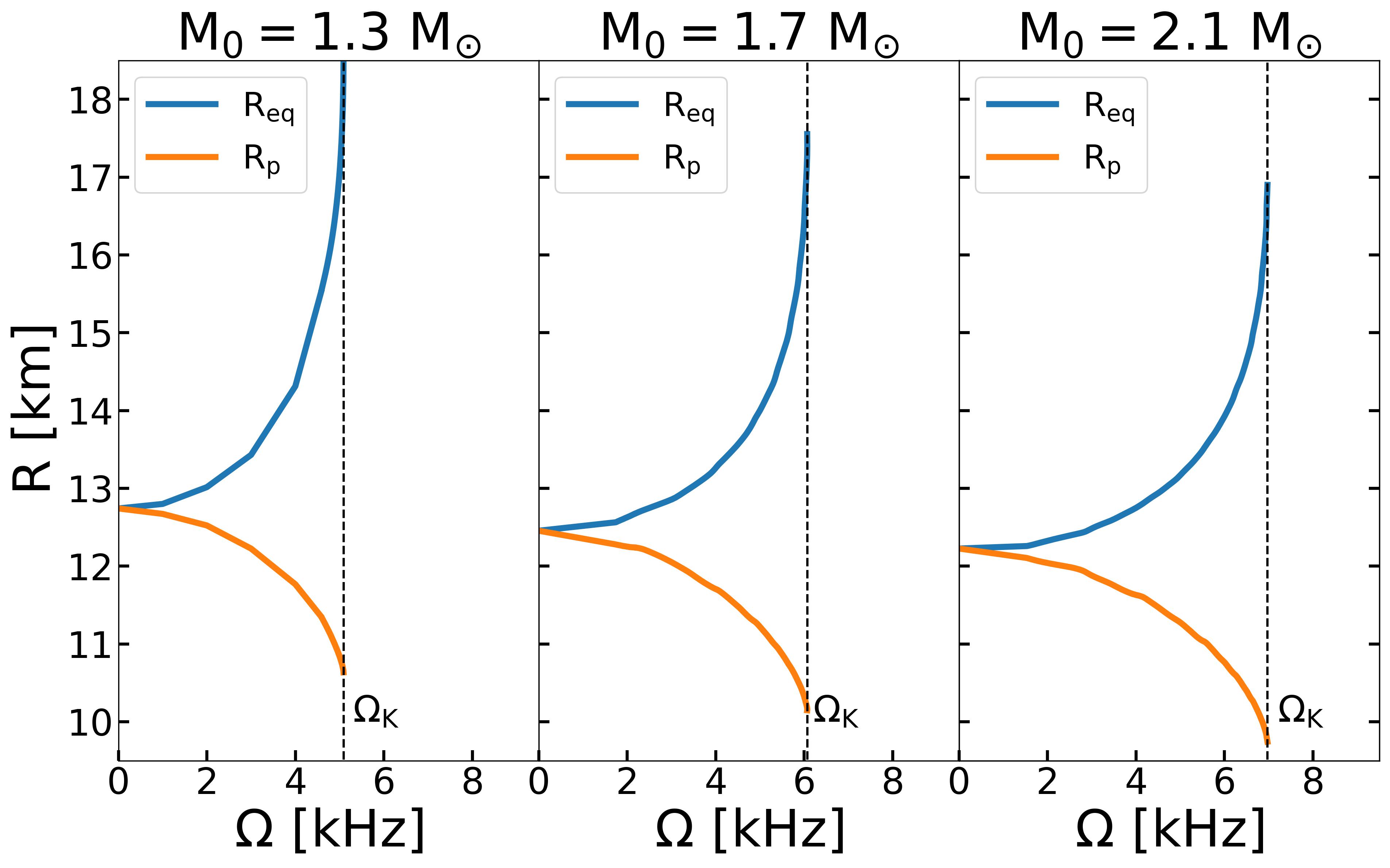}
\caption{The increase (decrease) of the equatorial (polar) radius as a function of their rotation frequency $\Omega$ for neutron stars with three different masses: $M_0/M_\odot=1.3,1.7,2.1$.}
\label{fig1}
\end{figure}

\begin{figure}[ht]
\includegraphics[width=\columnwidth]{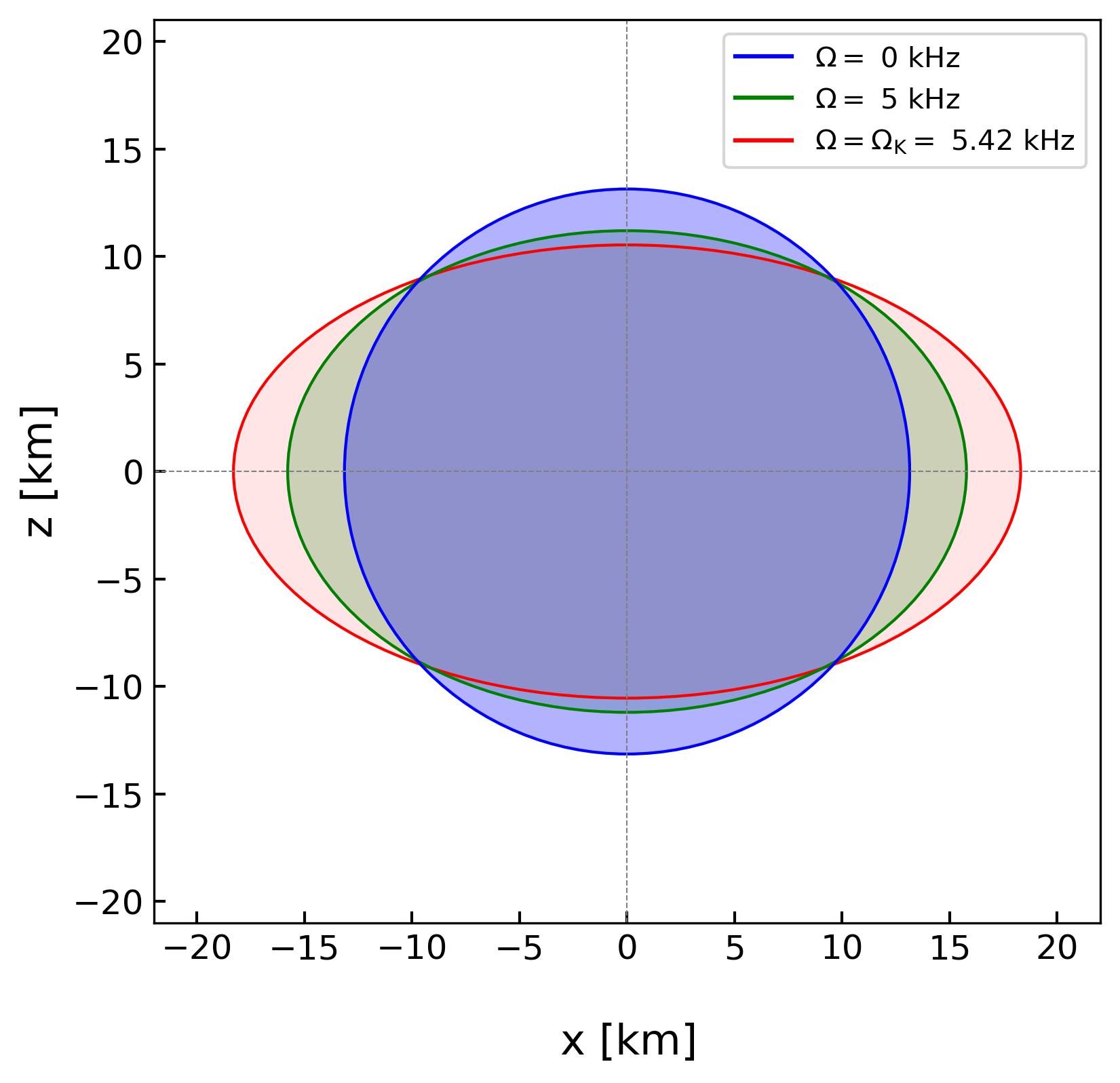}
\caption{The change of the shape for a neutron star with rest mass of  $\approx 1.5M_{\odot}$  upon rotation
up to its Kepler frequency $\Omega_K=5.42$ kHz depicted by a projection onto the $x-z$ plane including the rotation axis $z$. 
}
\label{fig1a}
\end{figure}

On the other hand, modelling the rapidly rotating MSPs involves advanced numerical methods compared to the slow-rotating NSs, which require solving additional coupled differential equations. Rotation causes an NS to deform into an oblate spheroid, resulting in a larger equatorial radius and an increased gravitational mass compared to a non-rotating NS, which is related to an increase of the centrifugal force (see the change of the equatorial/polar radii in Fig.~\ref{fig1} and illustration of star's deformation in Fig.~\ref{fig1a} as a function of their rotation frequency). As rotation affects the star's compactness, it alters the matter composition and, consequently, its dynamical properties, e.g., moment of inertia~\cite{Friedman:1986tx,Komatsu:1989zz,Cook:1993qj,Cook:1993qr,Stergioulas:1994ea,Kruger:2021zta}. Another example of such rotation-driven changes is the thermal evolution of NSs~\cite{Weber:2013uja, Beznogov:2022wae}. An increase in the central density can trigger new processes, such as the direct Urca process, which is forbidden at lower central densities but may become active as the star spins down, leading to its rapid cooling~\cite{Krastev:2007wh,Weber:2013uja}. The universal relations derived for rapidly rotating NSs, which relate various essential quantities, enable constraints on the EoS and provide the NS properties at arbitrary rotation rates~\cite{Largani:2021hjo,Kruger:2023olj}.

As the maximum rotational frequency and star's properties are determined by the EoS of cold dense strongly interacting matter, it is necessary to reproduce the observational data together with realistic models that satisfy existing constraints and account for the possibility of the existence of an exotic phase in the NS core, e.g. hyperons, phase transition to quark matter. Thus, in this work, we explore the effect of rotation on the hybrid star properties characteri\-zed by the different quark matter properties and onset density of the phase transition. This study is especially important due to the discussed in literature rotationally induced phase transition to the quark matter in the NS core~\cite{Zdunik:2005kh,Dimmelmeier:2009vw}. As a star gradually spins down over time, the increasing compression of matter could trigger a phase transition to deconfined quarks at the corresponding density.

While the rapidly rotating hybrid stars have been studied in several papers in the past, e.g.~\citet{Ippolito:2007hn,Dhiman:2010imr, Ayvazyan:2013cva,Bhattacharyya:2017tos,Largani:2021hjo}, this topic got new insights due to the recent observations of NICER, heavy pulsars, and detection of many new MSPs with the ongoing radio surveys. Thus, the study of the fast-spinning NSs and their maximum spin frequency (i.e., the Kepler frequency)~\cite{Shapiro:1983du} is especially important after the three recent NICER measurements, PSR J0030+0451~\cite{Miller:2019cac,Riley:2019yda}, PSR J0740+6620~\cite{Miller:2021qha,Riley:2021pdl} and PSR J0437-4715~\cite{Choudhury:2024xbk}, together with the stringent maximum mass limit~\cite{Antoniadis:2013pzd,Romani:2021xmb,Romani:2022jhd}. 

In the present work, the quark phase is described by the effective model that captures the aspects of confinement and color superconductivity, as well as perturbative QCD corrections~\cite{Ivanytskyi:2022oxv}. In~\citet{Gartlein:2023vif} the fit of the utilized model by the Alford-Braby-Paris-Reddy (ABPR) parameterization~\cite{Alford:2004pf} is presented, providing a simple functional relation between the key parameters of the EoS and the microscopic parameters of the initial Lagrangian.

This article presents a thorough analysis of the impact of rotation on hybrid star properties and stability against quasi-radial and non-axisymmetric oscillations utilizing astrophysical and gravitational wave (GW) constraints on the cold, dense matter EoS, together with the fastest MSP measurements. It allows us to revise the previously found empirical formulas relating the mass and radius of the star with the Kepler frequency. We demonstrate how the first-order phase transition in the hybrid star interior alters the previously found relations. 

The article is organized as follows. In Sec.~\ref{eos}, we describe the hadronic and quark EoSs and how we construct the hybrid stars. Sec.~\ref{rot} presents the theoretical framework to model uniformly rotating stars and the impact of rotation on the star's properties. Sec.~\ref {kepler} discusses the Kepler frequency and empirical relations. In Sec.~\ref{res}, we show our findings for the effect of rotation on the stability of hybrid stars and constraints on their interior composition. Finally, Sec.~\ref{concl} summarizes the results. Throughout the article, we use the natural unit system in which $\hbar=c=G=1$.

\section{Hybrid stars}
\label{eos}

\subsection{EoS of hadronic matter}
\label{eos_hadr}

The hadronic matter is described within the relativistic density functional DD2npY-T EoS~\cite{Shahrbaf:2022upc} that includes nucleonic and hyperonic degrees of freedom. The DD2npY-T EoS satisfies the maximum mass constraint~\cite{Antoniadis:2013pzd,Romani:2021xmb,Romani:2022jhd}, the LIGO-Virgo tidal deformability measurements of GW170817~\cite{LIGOScientific:2018cki} binary NS merger and NICER results~\cite{Miller:2019cac,Riley:2019yda,Miller:2021qha,Riley:2021pdl,Choudhury:2024xbk}.

Below the saturation density, the DD2npY-T model is supplemented by the generalized relativistic density functional (RDF) EoS for the crust~\cite{Typel:2018wmm}. It incorporates nuclei in a body-centred cubic lattice with a uniform background of electrons supplemented with a neutron gas above the neutron drip line. The transition between the generalized RDF and DD2npY-T models is consistently described within the unified approach.

\subsection{EoS of quark matter}
\label{eos_qgp}

Quark matter is modeled based on the confining RDF approach with color superconductivity~\cite{Ivanytskyi:2022oxv,Ivanytskyi:2022bjc}.
The confining aspect is connected to the dynamical breaking of chiral symmetry, which, at small temperatures and densities, leads to large effective quark masses above 0.7 GeV and efficiently suppresses quark excitations. The approach is equivalent to a chiral quark model with medium-dependent couplings of the scalar and pseudo\-scalar channels, which in the general case are different due to the breaking of chiral symmetry but coincide when the chiral symmetry gets restored \cite{Ivanytskyi:2022oxv}. The medium dependence of the vector and diquark couplings is adjusted in order to reach the conformal limit of quark matter \cite{Ivanytskyi:2022bjc}. Fitting the pion mass and decay constant, sigma meson mass and chiral condensate in vacuum fixes all the model parameters except the vector and diquark couplings (see Ref. \cite{Ivanytskyi:2022oxv} for details). The latter are parameterized in terms of their dimensionless ratios to the scalar coupling evaluated in the vacuum, i.e. $\eta_V$ and $\eta_D$ being the only free parameters of the model.

In this work, we use the ABPR parameterization of the RDF EoS, which was proposed in Ref. \cite{Gartlein:2023vif}
\begin{eqnarray}
\label{eqV}
p=\frac{A_4\mu^4}{2\pi^2}+\frac{\Delta^2\mu^2}{\pi^2}-B.
\end{eqnarray}
The parameters $A_4, \Delta$, and $B$ in Eq. (\ref{eqV}) are the effective number of degrees of freedom, pairing gap, and bag pressure.
They are unambiguously defined by the dimensionless couplings $\eta_V$ and $\eta_D$.
The corresponding functional dependencies can be found in Appendix~\ref{app:A}. 

\begin{figure}[!ht]
\includegraphics[width=\columnwidth]{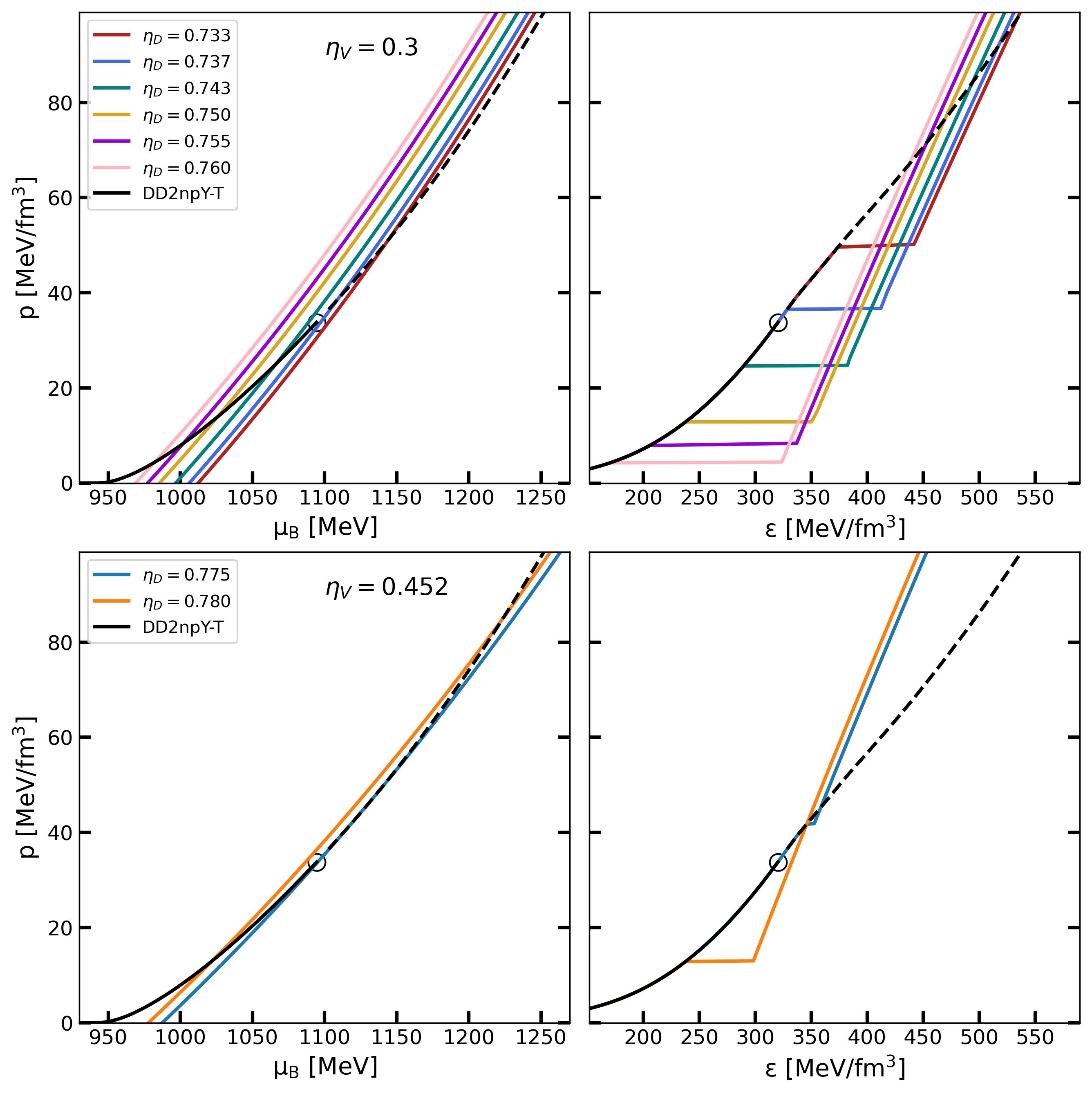}
\caption{
{Left panels: Pressure versus baryon chemical potential for the hadronic DD2npY-T EOS (solid line) and differnt parametrizations of the quark matter EOS. The crossing points determine the critical chemical potential and the transition pressure for the Maxwell construction of a first-order phase transition.
Right panels: Pressure versus energy density for the hybrid EOS obtained by the Maxwell construction.
For details, see text. 
}}
\label{fig:EOS}
\end{figure}

{
\subsection{Hybrid EOS construction}
}
The hybrid star EoS is constructed within the framework of the so-called  two-phase approach by matching an EoS for the hadronic phase, $p_h(\mu)$, in the outer core with that of the quark matter phase, $p_q(\mu)$, of the inner core by the Maxwell construction. 
The resulting EoS is characterized by a critical pressure 
\begin{equation}
    p_{\rm onset}=p_h(\mu_c)=p_q(\mu_c)
\end{equation}
for the onset of the transition from the hadronic phase $p_h(\mu)$ for $\mu\le \mu_c$ to the 2SC quark matter phase $p_q(\mu)$ for $\mu\ge \mu_c$ with a jump in energy density
\begin{equation}
    \Delta \varepsilon=\varepsilon_q(\mu_c)-\varepsilon_h(\mu_c)
    = \mu_c\left(\frac{dp_q}{d\mu} - \frac{dp_q}{d\mu} \right)\bigg|_{\mu=\mu_c}
\end{equation}
at the intercept of both pressure curves at $\mu=\mu_c$.
In the left panels of Fig. \ref{fig:EOS}, we show the hadronic EoS overlaid to the quark matter EoS
for several values of the diquark coupling $\eta_D$ at fixed vector coupling $\eta_V$, while in the right panels the corresponding hybrid neutron star EoS after the Maxwell construction are shown. 
In the upper panels of that figure,  we consider $\eta_V=0.30$, while in the lower panels  the results for $\eta_V=0.452$ are shown.

The quark-hadron phase transition modelled by the Maxwell construction is of first order because the jump in the energy density across the phase boundary is related to a jump in the baryon density as an order parameter of the transition, $\Delta n = \Delta \varepsilon/\mu_c$.
As has been pointed out by \citet{Glendenning:1992vb}, in the presence of several conserved charges like in the present case the baryon number and the electric charge, the pressure may not remain constant throughout the transition, namely when the electric charge is not conserved locally but rather globally. In this case, even under the influence of strong gravity, an extended mixed phase of quark and hadron matter can occur in the phase transition region.
While in this Glendenning construction the effects of a surface tension between quark and hadron phases are neglected, their inclusion together with electric charge screening leads to the formation of structures (so-called "pasta phases") in the mixed phase as the most general case, see \citet{Maslov:2018ghi} for a recent application.
In the works of \citet{Voskresensky:2001jq,Voskresensky:2002hu}, it has been demonstrated that for physically relevant values of surface tension and screening, the Glendenning construction effectively is very close to the Maxwell construction. Therefore, we apply the latter throughout this work. 

It is also worth mentioning that construction of the quark-hadron phase transition weakly affects the results of modeling NSs with quark cores. 
As it was reported earlier, changing the Maxwell construction to the replacement interpolation method \cite{Abgaryan:2018gqp, Cierniak:2021knt} or to the two-zone interpolation scheme \cite{Ivanytskyi:2022wln}, which effectively mimic the Glendenning construction and pasta phases, just slightly modifies the mass-radius relation of NSs in the vicinity of the onset mass of quark matter.
This allows us to conclude that the results presented below are not sensitive to the details of phase conversion in the NS interiors.

Previous work~\cite{Gartlein:2023vif} has already explored the roles of these parameters. 
As can be seen from Fig.~\ref{fig:EOS} and from Tab.~\ref{tablemasses}, the diquark coupling $\eta_D$ governs the onset of the deconfinement phase transition, with only a minor impact on the maximum mass of hybrid stars. In contrast, variation of the vector coupling $\eta_V$ affects the stiffness of the EoS and, therefore, the maximum mass of hybrid stars. 
The lower panels are for a strong vector coupling of $\eta_V=0.452$ resulting in relatively small jumps in energy density while the upper panels are for a medium vector coupling $\eta_V=0.3$ resulting in early onset of deconfinement and large energy density jumps, for earliest onsets fulfilling the Seidov criterion \cite{1971SvA15347S} 
\begin{equation}
    X = 2\Delta \varepsilon - \varepsilon_h(\mu_c) - 3 p_{\rm onset} > 0
\end{equation}
of gravitational instability, see Tab.~\ref{tablemasses}. 
This instability can appear in the case of a significant softening of the NS matter caused by a strong first order phase transition with large $\Delta\varepsilon$ signaled by positive $X$.
In this case the mass-radius relation of NSs is characterized by two disconnected branches including the NS configurations of the same mass but different radii, twin stars \cite{Blaschke:2019tbh,Goncalves:2022phg,Chanlaridis:2024rov}.
As is seen from Tab. \ref{tablemasses} and Fig. \ref{fig2}, the mass-radius relations of NSs with EoS characterized by small onset densities, indeed, have twin star configurations.

Another phenomenon, which is related to phase transformations in NSs is the so-called back-bending \cite{Glendenning:1997fy,Glendenning:2000zz,Zdunik:2005kh}.
It is manifested by an S-shaped dependence of the moment of inertia on the rotational frequency, analogous to the back-bending phenomenon known from the rotational properties of atomic nuclei.  It has been conjectured in \citet{Glendenning:2000zz} that a phase transition in the NS interior can manifest itself by a  spin clustering in the population of accreting NS in low-mass X-ray binaries.
A detailed investigation of the interconnection between the gravitational instability 
($dM/d\varepsilon_c<0$) that is a necessary condition for the ocurrence of a third family of compact stars and the back-bending in the moment of inertia vs. frequency has been given in \citet{Zdunik:2005kh}.
The authors have demonstrated that this instability entails an instability in the angular momentum ($dJ/d\varepsilon_c|_{M_b=const} >0$) which corresponds to backbending in the moment of inertia. However, as the authors demonstrate, the converse is not true. Backbending in the moment of inertia can occur together with a stable angular momentum $dJ/d\varepsilon_c|_{M_b=const} <0$.

One notices that the onset densities in Tab. \ref{tablemasses} are below twice the saturation density in several cases and may reach even close to the saturation density itself. This may not be a problem because of the strong asymmetry dependence of the onset density which entails that an onset density in symmetric matter, like in heavy-ion collisions where 3-5 times nuclear saturation densities are reached, may be a factor 2-3 larger than in the case of neutron star matter with a proton fraction of only 10\%, see Fig.~1 of \citet{Sagert:2008ka}. 
In this work, we vary $\eta_V$ and $\eta_D$ in a range where all the available astrophysical constraints are respected~\cite{Gartlein:2023vif}.
\\[5mm$dJ/d\varepsilon_c|_{M_b=const} >0$]

\begin{table*}[!ht]
\begin{tabular}{|c|c|c|c|c|c|c|c|c|c|c|c|}
\hline
$\eta_V$ & $\eta_D$ & $n_{\rm onset}$ & $p_{\rm onset}$ 
         & $\varepsilon_{\rm onset}$  & $\Delta\varepsilon$ & X
         & $M^{\rm TOV}_{\rm onset}$&$ M^{\rm TOV}_{\rm max}$ &$ M^{K}_{\rm onset}$ & $ M^{K}_{\rm max}$\\ 
         &          & $[\rm fm^{-3}]$ & $\rm [MeV~fm^{-3}]$
         & $\rm [MeV~fm^{-3}]$ & $\rm [MeV~fm^{-3}]$ & $\rm [MeV~fm^{-3}]$
         & $[ M_{\odot}]$ & $[ M_{\odot}]$ & $[M_{\odot}]$ & $[M_{\odot}]$  \\ \hline
\multirow{6}{*}{0.3}  & 0.733 & 0.3715 & 51.164&379.564& 64.298&-404.460&1.488 & 2.114  & 1.927 & 2.539 \\ 
                      & 0.737 & 0.3311 & 38.695&335.518& 79.705&-292.193&1.292 & 2.112 & 1.662& 2.534 \\ 
                      & 0.743 & 0.2951 & 25.130&291.536& 90.928&-185.070&1.053 & 2.116 &1.319&2.540 \\
                      & 0.750 & 0.2455 & 13.916&243.618&109.700& -65.966&0.686 & 2.133 & 0.843& 2.578 \\ 
                      & 0.755 & 0.2138 & 8.352 &209.351&127.736&  21.065&0.488 & 2.148 & 0.573& 2.621 \\ 
                      & 0.760 & 0.1778 & 4.399 &172.352&151.438& 117.327&0.295 & 2.167 &0.322& 2.688 \\ \hline
\multirow{2}{*}{0.452}& 0.775 & 0.3467 & 41.837&346.421& 6.405 &-459.122&1.347 & 2.403 & 1.692 & 2.935 \\ 
                      & 0.780 & 0.2455 & 13.000&238.713& 59.795&-158.123&0.666 & 2.414 & 0.739 & 2.965 
                      \\ \hline
\end{tabular}
\caption{
Parameters the considered set of hybrid EoS including baryon density $n_{onset}$, pressure $p_{onset}$, energy density $\varepsilon_{onset}$, its jump $\Delta\varepsilon$, the Seidov criterion variable $X$ (see text for detail), NS mass $M_{onset}$ at the onset of quark matter as well as the corresponding maximum mass of NS $M_{max}$.
The superscripts $\rm TOV$ and $\rm K$ represent the static and Kepler limits.}
\label{tablemasses}
\end{table*}

\section{Rotating Neutron stars}
\label{rot}

\subsection{Theoretical framework}
\label{theor}

For the rapidly rotating relativistic compact stars, we have to apply an appropriate framework. The starting point is an axisymmetric spacetime metric~\cite{Cipolletta_2015}
\begin{eqnarray}
\label{eqVI}
ds^2=e^{2\nu}dt^2-e^{2\psi}(d\phi-\omega dt)^2-e^{2\lambda}(dr^2+r^2d\theta^2).
\end{eqnarray}
The metric functions $\nu,\psi,\omega$ and $\lambda$ only depend on $r$ and $\theta$. The choice, $e^{\psi}=r^2\sin ^2\theta B^2(r,\theta)e^{-\nu}$ shows the differe\-nce with respect to the static NS metric. For rotating stars, the centrifugal force causes its flattering, resulting in a larger equatorial radius and an increased gravitational mass compared to a non-rotating NS. Moreover, the so-called \textit{dragging} of the local inertial frame due to the gravitational field of the source is a purely relativistic effect~\cite{Chakraborty:2014qba,Paschalidis:2016vmz}. The spacetime fabric around the rotating gravitational object is distorted and the orbit of nearby test particles will start to precess. It is often called a ``gravitomagnetic" effect due to the analogue of this phenomenon in electromagnetism~\cite{Poisson:2016wtv}. In order to account for the dragging of the inertial frames, one introduces the Zero-Angular-Momentum-Observers (ZAMO) frame~\cite{Baarden:1973ula}. From the view of an observer at rest at infi\-ni\-ty, observers in a local ZAMO frame would move with the angular velocity $\omega$ (see Eq. (\ref{eqVI})).

The considered axisymmetric metric appears under the following conditions: (i) there are two Killing vector fields, $t^{\alpha}$ as usual, and $\phi^{\alpha}$ due to axial symmetry; (ii) asymptotically flat spacetime.

In addition, we approximate the NS interior as a perfect fluid:
\begin{eqnarray}
\label{eqVII}
T^{\alpha \beta}=(\varepsilon+p)u^{\alpha}u^{\beta}+pg^{\alpha\beta} \ ,
\end{eqnarray}
where $\varepsilon$ and $p$ denote the energy density and pressure of the fluid, respectively, while $u^{\alpha}$ denotes the fluid 4-velocity vector.

Taking advantage of the symmetries of our system, the 4-velocity can be written in terms of the two Killing vectors:
\begin{eqnarray}
\label{eqVIII}
u^{\alpha}=\frac{e^{-\nu}}{\sqrt{1-v^2}}(t^{\alpha}+\Omega \phi^{\alpha}) \ ,
\end{eqnarray}
where the fluid 3-velocity vector $v$ with respect to the local ZAMO frame and $\Omega=\frac{d\phi}{dt}$, the angular velocity measured by an observer at rest at infinity were introduced~\cite{Cipolletta_2015}. The two quantities are related as 
\begin{eqnarray}
\label{eqIX}
v=(\Omega-\omega)e^{\psi-\nu} \ .
\end{eqnarray}
To derive the corresponding field equations, we need to solve the Einstein equations. In detail, the conservation of the energy-momentum tensor in Eq. (\ref{eqVII}) including the metric components of Eq. (\ref{eqVI}) and the Killing vectors of Eq. (\ref{eqVIII}) needs to be solved. The hydrostatic equilibrium equation will then read
\begin{eqnarray}
\label{eqX}
p_{,i}=-(\varepsilon+p)\bigg{[}\nu_{,i}+\frac{1}{1-v^2} \bigg{(}vv_{,i}+v^2\frac{\Omega_{,i}}{\Omega-\omega}\bigg{)}\bigg{]} .
\end{eqnarray}
In the limit of non-rotating NSs, the functions $\omega$ and $\Omega$ should both vanish as none of the observers will observe any rotation. For $v\rightarrow 0$, Eq. (\ref{eqX}) equals to
\begin{eqnarray}
\label{eqXI}
p_{,i}=-(\varepsilon+p)\nu_{,i} ,
\end{eqnarray}
as in the case of deriving the Tolman-Oppenheimer-Volkoff (TOV) equations~\cite{Tolman:1939jz,Oppenheimer:1939ne} for static gravitational objects. 

\begin{figure}[ht!]
\includegraphics[width=\columnwidth]{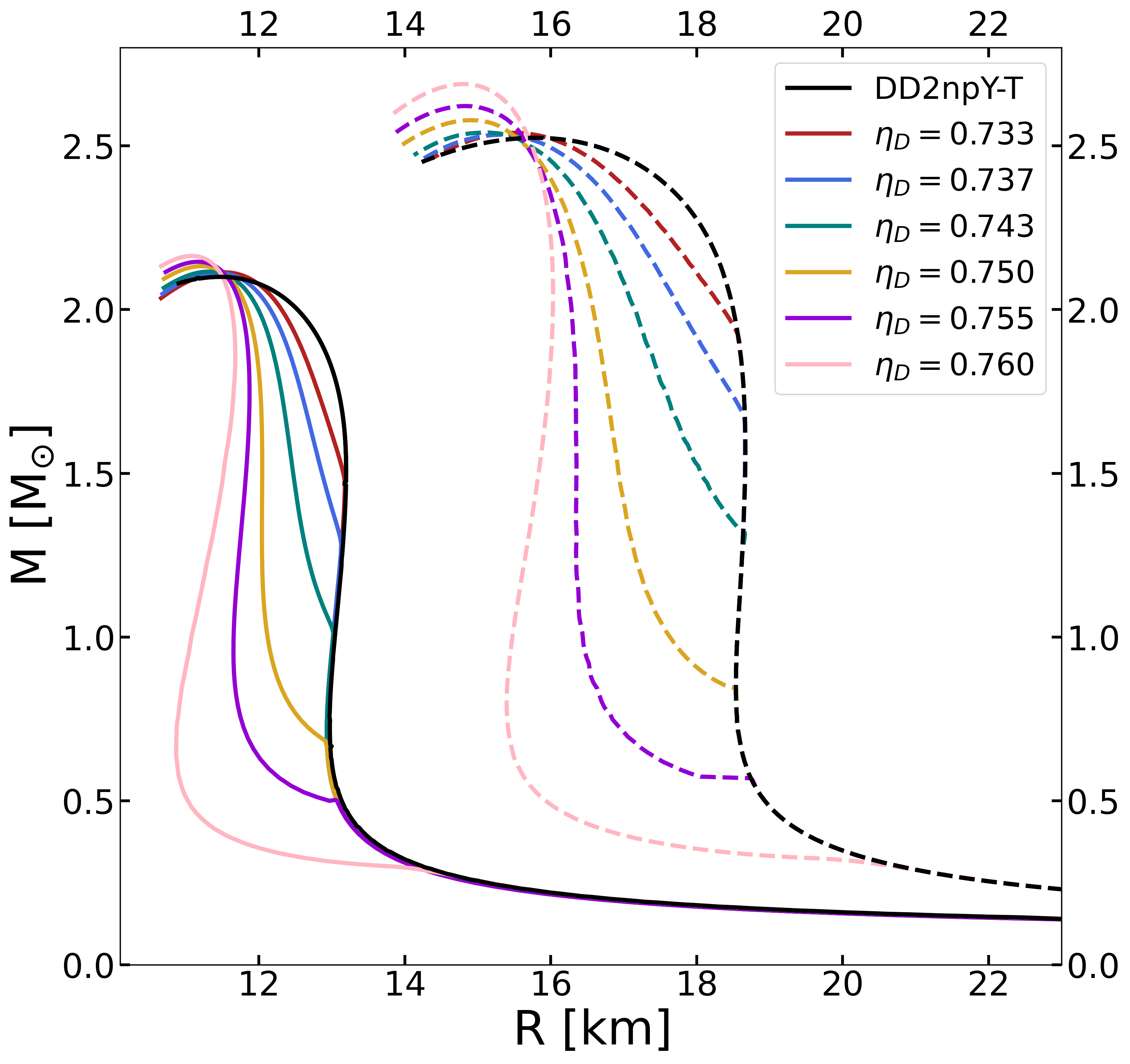}
\caption{Mass-radius diagram for a set of static (solid curves) and rotating stars with the Kepler frequency (dashed curves). The black and color curves depict the baryonic DD2npY-T EoS as well as hybrid stars for the fixed vector coupling $\eta_V=0.30$ and different values of the diquark coupling $\eta_D$, respectively. The radius corresponds to the equatorial radius. The allowed configurations for hybrid stars are located between the solid and dashed curves of the same color. }
\label{fig2}
\end{figure}

These equations for the static objects read~\cite{Tolman:1939jz,Oppenheimer:1939ne}
\begin{eqnarray}
\dfrac{dp}{dr}& = & - (\varepsilon + p) \frac{m+4 \pi p r^3}{r^2-2 r m},
\label{eqXII}
\\
\dfrac{d{ m}}{dr} & = & 4 \pi r^2 \varepsilon.
\label{eqXIII}
\end{eqnarray}
The differential equations in the rotating as well as in the static cases are solved by applying the same bounda\-ry conditions: (i) $m(0)=0$, $p(0)=p_c$, with $p_c$ being the pressure in the center of a star; (ii) pressure and energy density vanish at the surface of the star $r=R$. By fixing the value of the central pressure and varying it up to $p(R)=0$ we are able to numerically obtain the radius R and enclosed gravitational mass $m(R)=M$ of the NS. For each considered value of the angular velo\-ci\-ty, the $M-R$ curve is obtained by utilizing the publicly available \textit{RNS}\footnote{\url{https://github.com/cgca/rns}} code that calculates the properties of rapidly rotating relativistic compact stars~\cite{Stergioulas:1998hx}. We consider uniformly rotating stars described by the hybrid EoS with different onsets of the deconfinement phase transition and quark matter properties. Fig.~\ref{fig2} shows a set of $M-R$ curves for static (solid curves) and rotating hybrid stars with the Kepler frequency (dashed curves). Following~\citet{Gartlein:2023vif} to account for uncertainties in the dense matter EoS we utilize a set of six hybrid star configurations with different properties. The color represents the unique value of the diquark coupling defining the onset of the phase transition, while the value of the vector coupling $\eta_V$ is fixed at 0.30. The intriguing feature of the utilized RDF model is that it could describe hybrid stars with a wide range of properties, including the different onset and strength of the deconfinement phase transition. The set also includes hybrid stars with an early deconfinement phase transition. Table~\ref{tablemasses} lists the considered values of the vector and diquark couplings as well as the corresponding onset mass of the deconfinement phase transition and maximum gravitational mass for static configurations presented in Fig.~\ref{fig2}.

All these configurations are characterized by a different size of the quark core, the strength of the first-order phase transition (jump in the energy density at the phase transition), and the speed of sound values in the quark phase (for more details see~\cite{Gartlein:2023vif}). As can be seen in Fig.~\ref{fig2}, due to the centrifugal force, the rotating configurations (see the dashed curves of the corresponding color) are more massive in comparison to non-rotating counterparts, whereas the main features of static NSs, e.g. the onset density of the deconfinement phase transition, the special point (the point in which a set of $M-R$ curves intersect~\cite{Cierniak:2020eyh}), etc., are preserved.

As shown in Refs.~\cite{1965gtgc.book.....H,1995A&A...294..747G, Sagun:2020qvc}, non-rotating stars remain stable along the M-R curve up to the maximum gravitational mass, where the frequency of the fundamental eigenmode becomes zero. Beyond this critical value, the star becomes unstable, and the frequency of the radial oscillation mode turns imaginary. At such high densities, the fundamental mode of infinitesimal radial perturbations becomes unstable, leading to the collapse of the star into a black hole, which corresponds to the instability criterion ${\partial { M}}/{\partial \varepsilon_c} < 0$. 

Fig.~\ref{fig2} indicates the stability region for rotating NSs. Thus, hybrid stars with the early deconfinement phase transition are stable between the solid and dashed light pink curves, which correspond to the configurations of non-rotating and the ones with the Kepler frequency, respectively. The region between the solid and dashed dark red curves represents the sequence of stars with the late phase transition.

\subsection{Impact of rotation on compact star structure}

\begin{figure}[ht!]
\centering
\includegraphics[width=0.97\columnwidth]{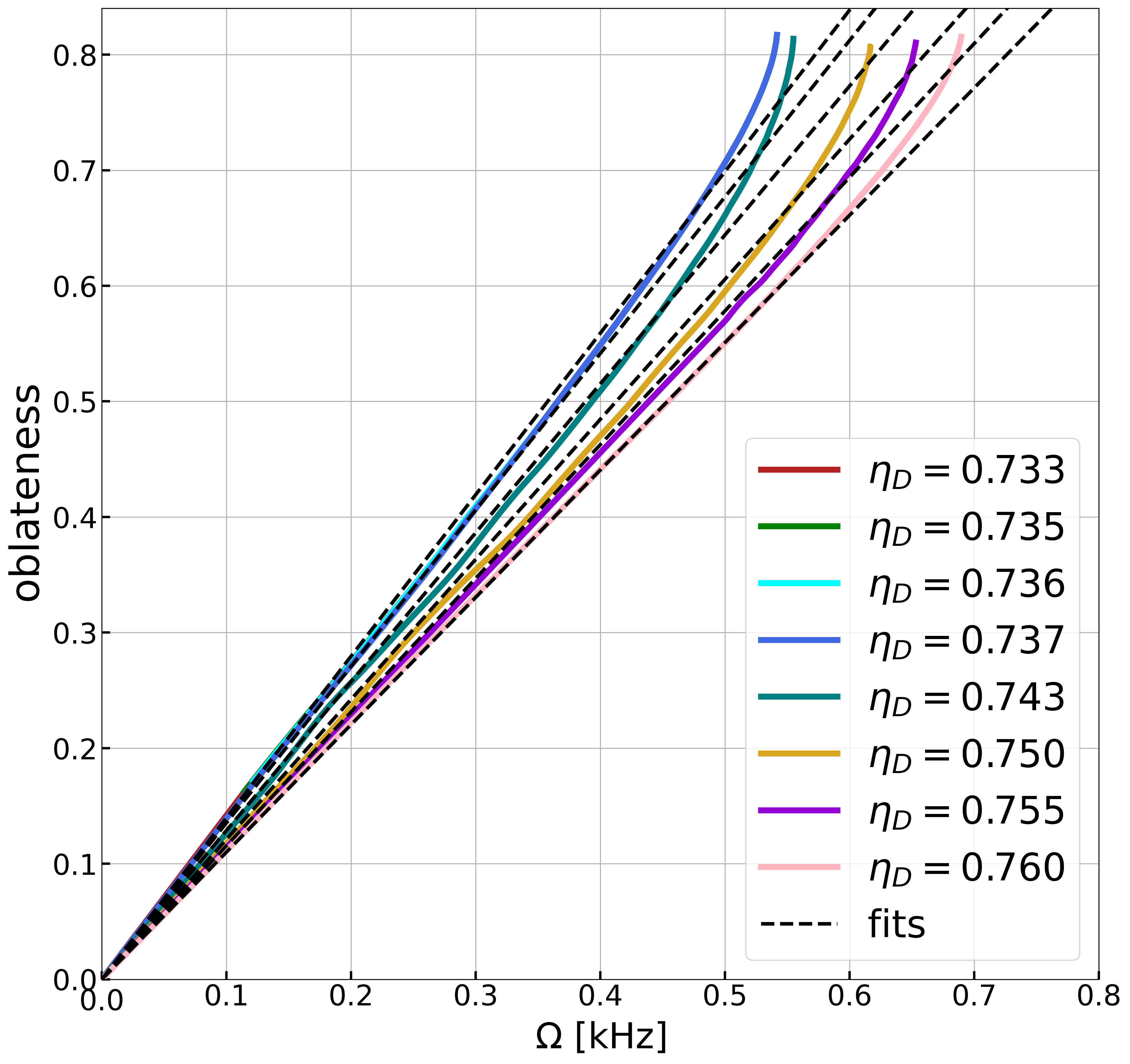}
\includegraphics[width=0.995\columnwidth]{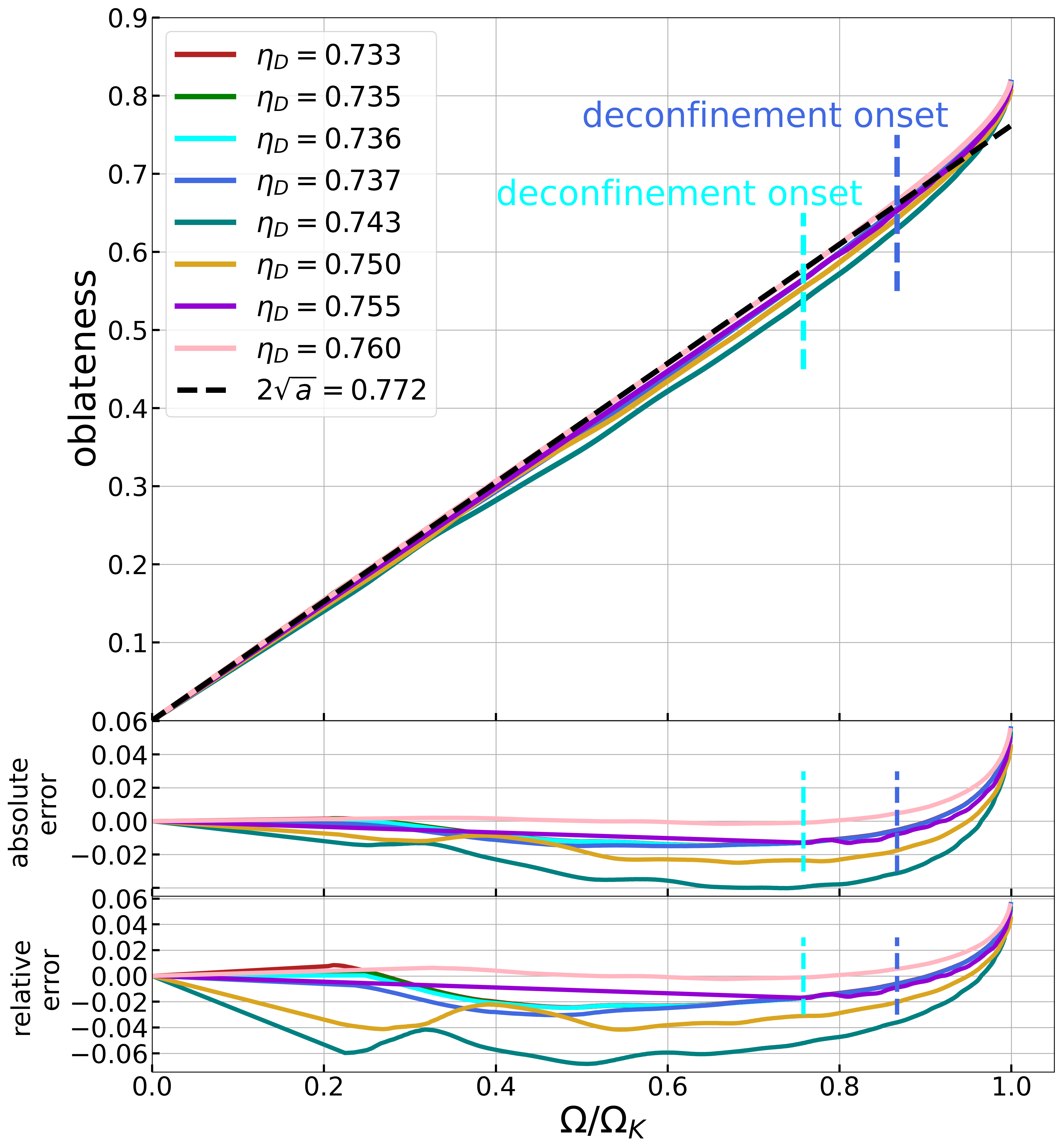}
\caption{{\bf Upper panel:} The oblateness/eccentricity $\rm e$ as a function of the angular velocity for the fixed value of the vector coupling $\eta_{V}=0.3$ and varying diquark coupling $\eta_{D}$. {\bf Lower panel:} The oblateness/eccentricity ${\rm e}$ as a function of the normalized angular velocity $\Omega$ to the Kepler (mass-shedding) velocity $\Omega_K$. The vertical dashed light and dark blue lines show the frequency of the phase transition onset at the corresponding solid curves. The black dashed line shows the optimal fit for $2\sqrt{a}$. The absolute error according to this fit is shown below. On both panels, the curves corresponding to the lowest $\eta_D$ values lay behind the blue curve as they coincide up to the high mass. The colors are the same as in Fig.~\ref{fig2}.}
\label{fig3}
\end{figure}

\textit{Oblateness.} The oblateness or ellipticity of a star is defined as 
\begin{equation}
\label{eqXIV}
{\rm e}=\sqrt{1-(R_{\rm p}/R_{\rm eq})^2},
\end{equation}
where $R_{\rm p}$ and $R_{\rm eq}$ are the polar and equatorial radii, respectively. Assuming that for not too large ellipticities holds $R_{\rm p,eq}=R\mp \Delta R$, then follows that
\begin{equation}
\label{eqXV}
{\rm e}=2\sqrt{R_{\rm eq}/R -1}.
\end{equation}
As it is schematically shown in Fig.~\ref{fig1} the increase of the spin frequency causes a deviation from the spherically symmetric configuration that reaches its maximum at the Kepler frequency. Let us assume that the deformation of the star, i.e. its oblateness, is proportional to the ratio of rotational, { T}, and gravitational, { W}, energy, which for a homogeneous Newtonian star is 
$T/W= (\Omega/\Omega_0)^2$, where $\Omega_0^2=4\pi \rho(0)$ and $\rho(0)$ is the mass density at the center of the star. This ratio is the small expansion parameter in the perturbative treatment of general relativistic rotating star configurations \cite{1968Ap......4...87S,1968Ap......4..227S,Chubarian:1999yn}.
Then according to Hooke's principle, the deformation should be proportional to the ratio $\Delta R/R = a (\Omega/\Omega_K)^2$, where we used the proportionality between $\Omega_0$ and the Kepler frequency $\Omega_K=2\pi f_K$ (for its definition, see below) and introduced the coefficient $a$ for the elasticity (deformability) of matter. The equatorial radius in the lowest order is then
$R_{\rm eq}=R(1+ a (\Omega/\Omega_K)^2) $ so that from Eq. (\ref{eqXV}) follows the linear response of the oblateness to the rotation frequency (angular velocity) 
\begin{equation}
\label{eqXVI}
{\rm e}=2\sqrt{a}~ \Omega/\Omega_K.
\end{equation}
Fig.~\ref{fig3} shows the relation between the oblateness and angular velocity (upper panel) as well as the normalized angular velocity for the Kepler velocity (lower panel) for a star of the fixed rest mass $M=1.5~M_\odot$ ($M_{\rm grav} \approx 1.4~M_\odot$). Note that the rest mass is equal to the baryon mass. The curves in Fig.~\ref{fig3} are obtained for $\eta_{V}=0.3$ and different values of the diquark coupling  $\eta_{D}=0.733 - 0.760$. On the upper panel, the behavior of all curves is strictly linear and universal with 2$\sqrt{a}\approx 0.772$ (see the dashed black lines), even for angular velocities as close as 95\% of the mass-shedding velocity. The deviation in oblateness is related to the stiffness of the quark matter defined by the value of the diquark coupling $\eta_D$. The lower panels of Fig.~\ref{fig3} show the absolute and relative deviations of the fitting function from the data. The latter does not exceed 6\%. The critical diquark coupling for which a star of this mass shall undergo a deconfinement transition is $\eta_D= 0.737$. Remarkably, we do not observe a significant change in the slope at the deconfinement phase transition (see the dashed vertical lines on the lower panel of Fig.~\ref{fig3} depicting the onset of the phase transition for $\eta_D= 0.736$ and $\eta_D= 0.737$) that results in a good agreement with 2$\sqrt{a}\approx 0.772$ value for all curves.

Our study confirms the results of~\cite{Konstantinou:2022vkr}, which showed that the changes in polar and equatorial radii are symmetric, with the polar radius shrinking at the same rate that the equatorial radius expands. 

In addition, the commonly used quantity as a star's compactness, $\mathcal{C}={M}/{R}$, which is defined for a static star, loses its meaning for rapidly-spinning NSs. This definition does not apply to rapidly rotating compact stars, especially those near the Kepler frequency. 
In principle, the compactness can be derived from the $g_{00}$ component of the corresponding metric. In the static case, this corresponds to $g_{00}=-\big{(}1-2\mathcal{C}\big{)}$. However, when considering rotation, the loss of symmetries significantly complicates the situation. As discussed by ~\citet{Hartle:1967he,Hartle:1968si}, even the metric for a slowly rotating (approximately spherical) star is non-trivial. In the lowest order of the multipole expansion, the $g_{00}$ component of the metric is given by
\begin{eqnarray}
\label{eqXVII}
g_{00}=-\big{(}1-2C\big{)}\bigg{(}1-\frac{2\delta M}{r-2M}+\frac{2J^2}{r^3(r-2M)}+\text{...}\bigg{)},
\end{eqnarray}
where (...) include even higher orders, $\delta M$ denotes the change of mass due to rotation and $J$ is the corresponding angular momentum. Consequently, the compactness $\mathcal{C}$ for rotating stars needs a redefinition.

\begin{figure}[t]
\includegraphics[width=\columnwidth]{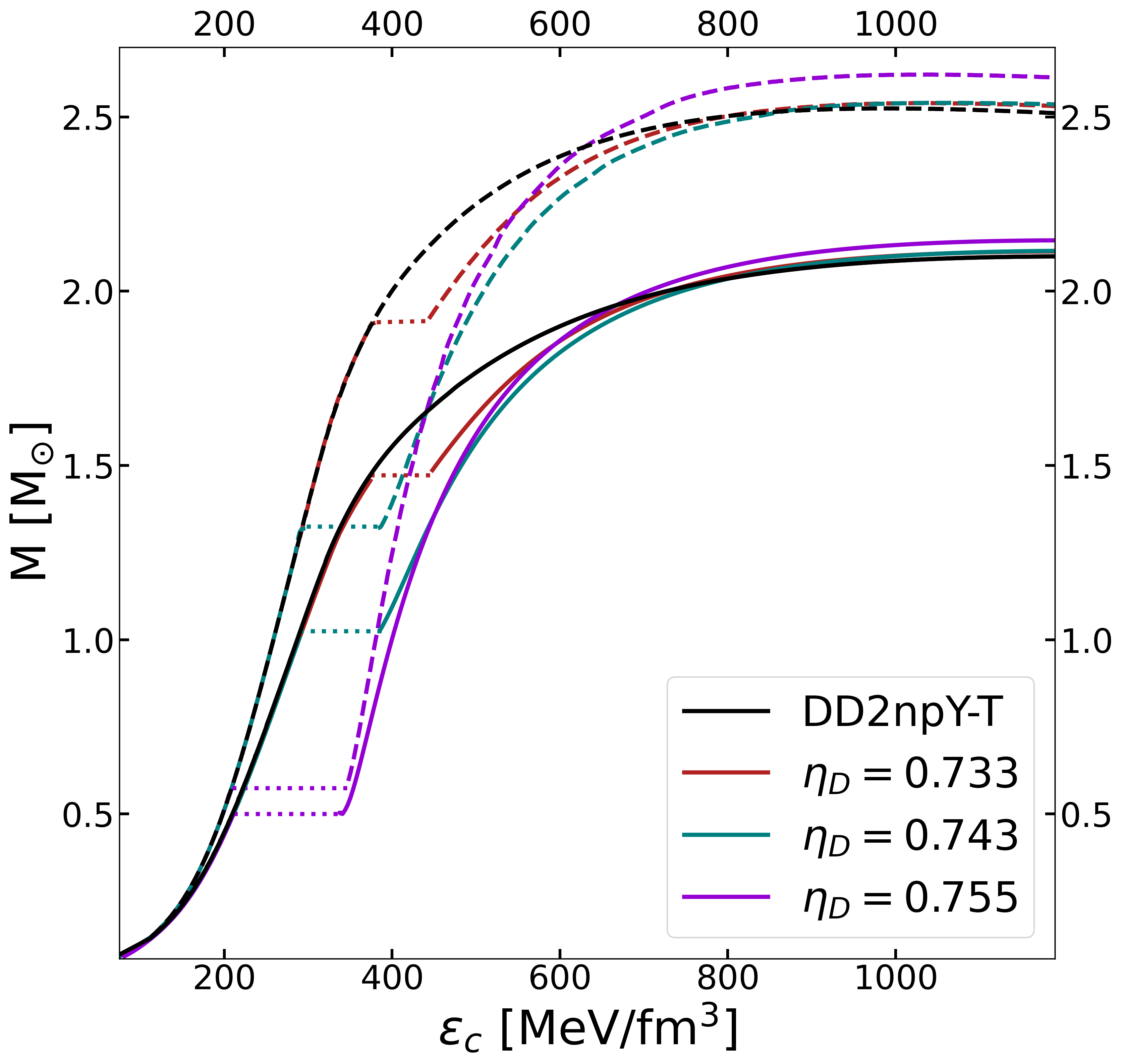}
\caption{The gravitational mass as a function of the central energy density for a set of static (solid curves) and rotating with the Kepler frequency (dashed curves) hybrid stars for the fixed vector coupling $\eta_V=0.3$ and different $\eta_D$ values listed in the legend (color curves). The dotted plateaus correspond to the first-order phase transition region.
The solid black curve represents the hadronic stars modelled within the DD2npY-T EoS.}
\label{fig4}
\end{figure}

\textit{Matter distribution.} 
As described in Section~\ref{eos} the two-phase hybrid star EoS is constructed by
matching its hadron and quark EoSs with the Maxwell construction. While the low-mass stars consist of hadronic matter, increasing the central energy density leads to the onset of the deconfinement phase transition. In Fig.~\ref{fig4}, the jump in energy density (horizontal dotted line) shows the transition region from hadronic to quark matter. No stable configurations are located along the horizontal dotted lines. Stars with a central energy density above the onset are hybrid stars with a quark core. As expected, an increasing diquark coupling triggers a much earlier onset of the deconfinement phase transition. We see that stars rotating at the Kepler frequency (dashed curves in Fig.~\ref{fig4}) show a similar structure to the non-rotating configurations (solid curves). The effect of rotation becomes more pronounced the later the phase transition occurs. Consequently, a higher value of $\eta_D$ leads to phase transitions at lower masses, and under rotation, the onset of these transitions shifts towards higher masses. In general, the more massive the object, the greater the shift towards higher masses. Therefore, an earlier phase transition will occur in low-mass stars, which can only remain stable under slow rotation. As a result, the shift due to the fastest possible spin is insignificant. In the case of $\eta_D = 0.733$, we clearly observe the impact of rapid rotation: the onset of the deconfinement phase transition shifts to higher masses, $M_{\rm onset}$, for the same central energy density, and the maximum mass increases by approximately $\approx 20\%$.

\section{The Kepler frequency}
\label{kepler}

The Kepler frequency plays an important role in understanding the interior composition of compact stars~\cite{1983ApJ...272..702S}. Any rotation at a faster speed will result in mass-shedding from the star's equator. \citet{1989Natur.340..451S} proposed the relativistic Roche model, for which the equality of the Kepler frequency with that of a particle orbiting at $r=R_K$ in the Schwarzschild spacetime around a point of mass $M$ at $r=0$ strictly holds~\cite{Chubarian:1999yn,Haensel:2009wa}
\begin{eqnarray}
\label{eqXVIII}
f_K^{\rm Roche}= \frac{1}{2\pi} \bigg{(}\frac{M}{R_{K}^3}\bigg{)}^{1/2},
\end{eqnarray}
where $R_K$ and $M$ are the circumferential equatorial radius $R_{\rm eq}$ and the gravitational mass of the Keplerian configuration.

A slightly different empirical relation can be written down between the Kepler frequency $f_K$, gravitational mass $M$ of the Keplerian configuration, and radius $R$ of the non-rotating star
\begin{eqnarray}
\label{eqXIX}
f_K= C \bigg{(}\frac{M}{M_{\odot}}\bigg{)}^{1/2}\bigg{(}\frac{R}{10\,{\rm km}}\bigg{)}^{-3/2}.
\end{eqnarray}
Relating two empirical formulas, the factor $C$ for the Roche model is equal to
\begin{eqnarray}
\label{eqXX}
C^{\rm Roche}&=& \frac{1}{2\pi} \bigg(\frac{2}{3}\bigg)^{3/2}
\bigg(\frac{M_{\odot}}{(10\,{\rm km})^3}\bigg)^{1/2}\\ \nonumber
&=&1.007 \, {\rm kHz}.
\end{eqnarray}
For realistic EoS, the factor $C$ is obtained by numerically calculating the Kepler frequency and comparing it with the empirical formula (\ref{eqXIX}). It is defined as
\begin{eqnarray}
\label{eqXXI}
C(M)&=& f_0
\bigg(\frac{R(M)}{R_K(M)}\bigg)^{3/2}~,
\end{eqnarray}
where $f_0=(3/2)^{3/2}C^{\rm Roche}=1.8335 \, {\rm kHz}$.
It was determined to be 1.04 kHz~\cite{Lattimer:2004pg} and further revised by~\citet{Haensel:2009wa} who found $C$=1.08 kHz for soft hadronic EoSs (e.g., APR, WFF, FPS) and  $C$=1.15 kHz for strange quark stars. An alternative polynomial function relating the Keplerian and static star properties is proposed by~\citet{Riahi:2019hmt}. One could, therefore, think of suggesting the value of $C$ as an indicator characterizing the star's interior. An empirical formula Eq. (\ref{eqXIX}) for the Kepler frequency is essential for constraining the EoS of strongly interacting matter with pulsar observations. Avoiding the computationally demanding numerical calculations Eq. (\ref{eqXIX}) provides a simple and universal relation between the quantities. 

While the values $C$=1.08 kHz and $C$=1.15 kHz are obtained for one-phase stars, i.e. hadronic and quark stars~\cite{Haensel:2009wa}, the deconfinement phase transition in the NS interior alters the obtained value of the factor $C$. Moreover, accommodation of the phase transition onset at various densities poses a complication. 

To obtain the behavior of $C(M)$ for the considered set of hybrid stars (see Table~\ref{tablemasses}), the masses of Keplerian $M^{K}$ and static $M^{TOV}$ configurations are matched following the relation of~\citet{Breu:2016ufb} suggesting a factor $M^{K}=bM^{TOV}$, where $b=1.204\pm0.002$. This result was recently revised in Ref. ~\cite{Musolino:2023edi} by analyzing a large number of generic EoSs and reported $b=1.255^{+0.047}_{-0.040}$. However, we found that for each set of the model parameters, $b$ lies in a slightly different range of $1.0-1.3$. Compared to the results of~\citet{Musolino:2023edi}, this range extends to smaller values since, unlike the study mentioned, our analysis also includes the NS masses in the interval $0.5-1~M_\odot$.

For most stars, the internal composition of both static and Keplerian configurations is the same, either hadronic or hybrid, resulting in similar values of $C$. However, within a narrow range of stellar masses, the two configurations differ, leading to a drop in $C(M)$ by approximately $80-100$ Hz. This behavior can be understood from the definition of $C(M)$ in Eq. (\ref{eqXXI}). In this mass region, the static configuration is already significantly compactified due to the deconfinement transition, while the Keplerian configuration at the same mass remains in the hadronic phase with a larger radius. To avoid compa\-ring hadronic stars with hybrid stars, the masses of static and rotating configurations are matched using a scaling factor $b=1.0-1.3$.

\begin{figure}
\centering
\includegraphics[width=\linewidth]{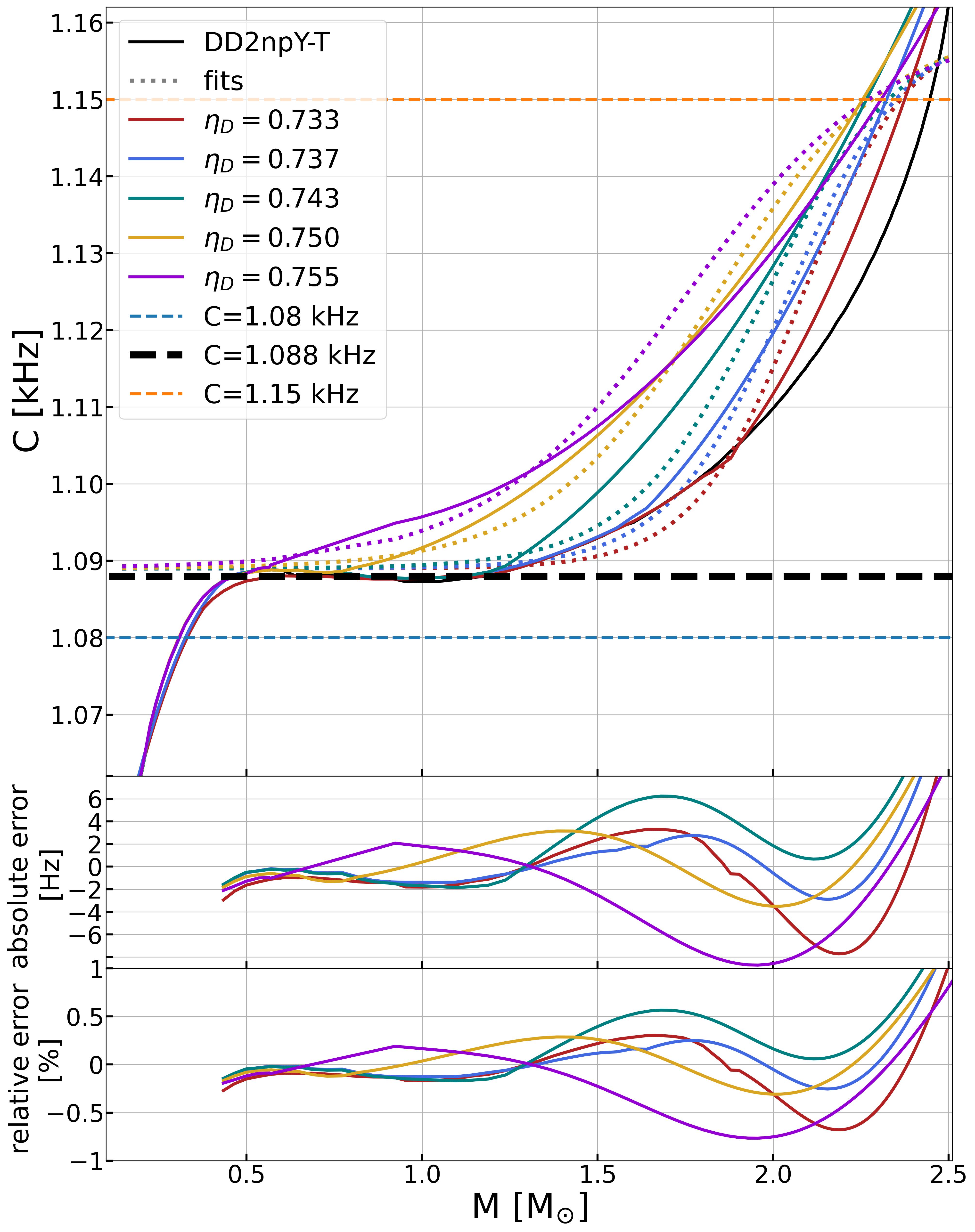}
\caption{Behavior of the empirical factor $C$ as a function of the gravitational mass $M$ of the Keplerian configuration for a set of hybrid stars. The solid curves illustrate the data, whereas the dotted curves show the fit with different deconfinement onset with Eq. (\ref{eqXXII}). The horizontal dashed lines indicate C=1.08 kHz (blue), C=1.088 kHz (black) obtained for hadronic EoSs, and C=1.15 kHz (orange) obtained for quark EoSs. The lower panels indicate the absolute and relative error of the fit.
}
\label{fig5}
\end{figure}

Fig.~\ref{fig5} illustrates the behavior of $C$ as a function of the gravitational mass of the Keplerian configuration $M^{K}$. We find the $C=1.088$ kHz value to be a better fit for the stiff DD2npY-T EoS in comparison to the $C=1.08$ kHz value obtained for a set of soft hadronic EoS~\cite{Haensel:2009wa}. In the range between 0.5 $M_{\odot}$ and the onset of quarks, the value of $C$ remains constant at 1.088 kHz, as shown by the black dashed line. The subsequent deviation from this constant value is associated with the deconfinement phase transition. For hybrid stars, $C$ increases after the phase transition, reaching a maximum of $C=1.15-1.16$ kHz. This result confirms that the hybrid EoS successfully reproduces the two limiting cases of hadronic and quark EoSs while incorporating the transition between them.

To account for different quark matter properties and the onset density of the deconfinement phase transition, we derived the parametrization of $C$ as a function of the onset mass $ M_{\rm onset}$ and gravitational mass at the Kepler frequency $M^{K}$
\begin{eqnarray}
\label{eqXXII}
 C=C_{\rm had}+\frac{\delta C}{1+5 \exp{(D(M_{\rm onset})-E(M_{\rm onset})M^{\it K})}}.
\end{eqnarray}
Here $\rm \delta C=0.072~kHz$ is defined as the difference between the value of $C$ in the quark and hadron limits $\rm \delta C=C_{quark}-C_{had}=1.16~kHz-1.088~kHz$. The obtained formula~(\ref{eqXXII}) corresponds to the logistic function with the inflection point equal to $ E(M_{\rm onset})\cdot M^{K}$, where the dimensionless parameter $E(M_{\rm onset})=3M_{\rm onset}+2$. The dimensionless function D is equal to $D(M_{\rm onset})=7.5M_{\rm onset}+0.7$.

This simple formula provides a good description of the $C$ factor in a wide range of the hybrid star properties and the onset of the phase transition with the maximal deviations of the order of $\rm \pm8~Hz$ and the relative error below $1\%$ (see the bottom part of Fig.~\ref{fig5}). The obtained fit is depicted as the dotted curves in Fig.~\ref{fig5}. The values of the onset mass were taken from Table~\ref{tablemasses}. Note that the fit was performed between 0.5 $M_{\odot}$ and the maximum mass.

The obtained parameterization reproduces two limits, i.e. hadronic stars with $ C_{\rm had}=1.088~ \rm kHz$ and quark stars with $ C_{\rm 
 quark}=1.16~\rm kHz$. The universal parameterization of $C$ considering the onset of the deconfinement phase transition at different densities is a useful tool to account for the general relativistic effects in the description of rapidly rotating stars with a different interior composition.

\begin{figure}[t]
\includegraphics[width=\columnwidth]{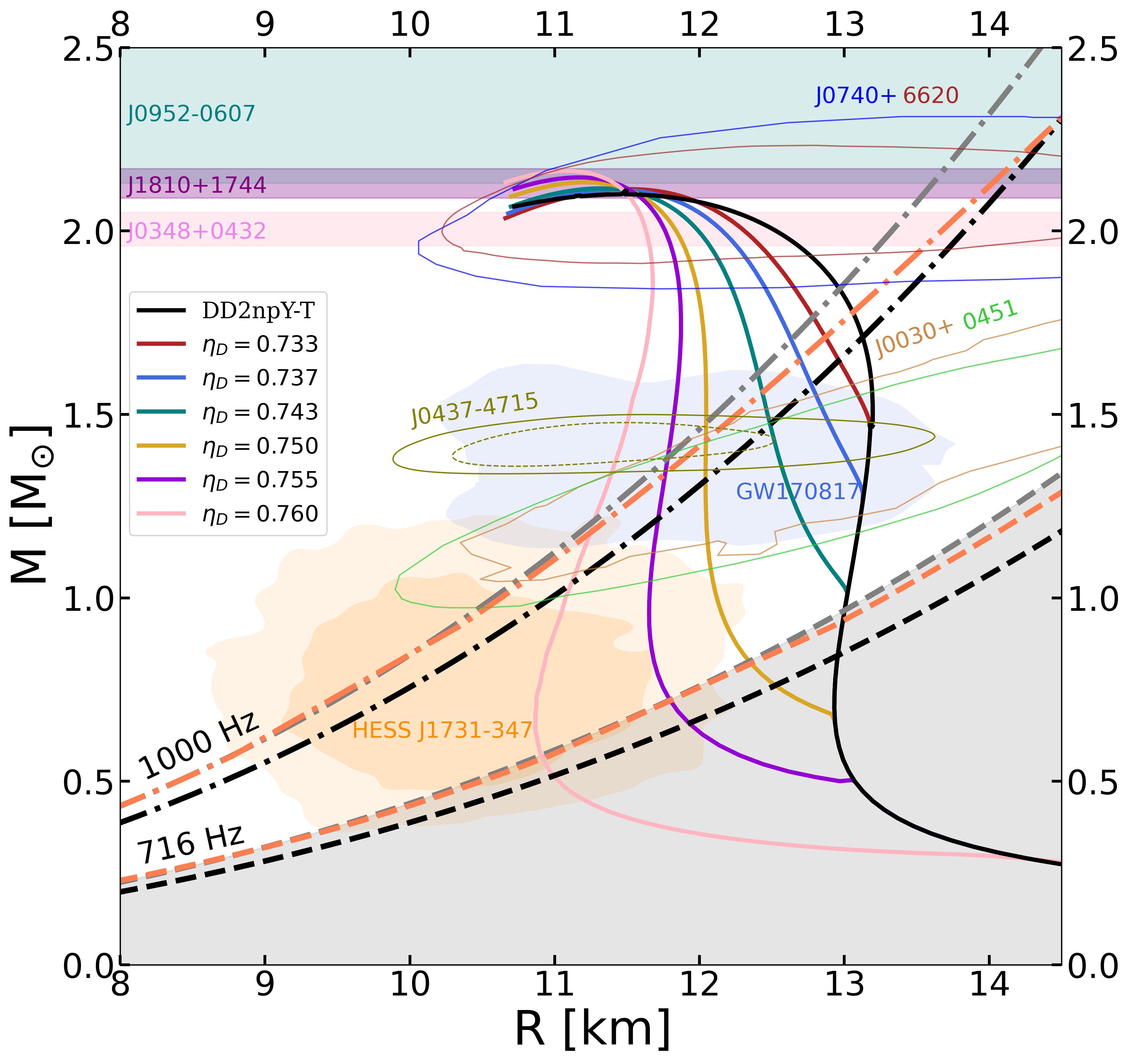}
\caption{Mass-radius relations for a set of static hybrid stars for the fixed vector coupling $\eta_V=0.30$ and different values of the diquark coupling $\eta_D$ as well as baryonic stars modeled within the DD2npY-T EoS (black curve). The teal, purple, and violet bands represent the 1$\sigma$ constraints on the mass of PSR J0952-0607~\citep{Romani:2022jhd}, PSR J1810+1744~\citep{Romani:2021xmb}, and PSR J0348+0432~\citep{Antoniadis:2013pzd}. The NICER measurement of PSR J0030+0451~\citep{Miller:2019cac,Riley:2019yda} is depicted with the light brown and lime green contours, while blue and brown contours represent the PSR J0740+6620 measurement~\citep{Miller:2021qha,Riley:2021pdl}. The olive solid (95\% CL) and dashed (68\% CL) contours represent the newly reported NICER measurement of PSR J0437-4715~\citep{Choudhury:2024xbk}. LIGO-Virgo detections of 
GW170817~\cite{LIGOScientific:2018cki} binary NS merger is shown in light blue. The 1$\sigma$ and 2$\sigma$ contours of HESS J1731-347~\cite{Doroshenko2022} are plotted in dark and light orange, respectively. The gray dashed curve and shaded gray region below are the lower boundary of the M-R relation obtained for the currently fastest spinning pulsar PSR J1748-2446ad~\cite{Hessels:2006ze} considering the value $C=1.08$~kHz found in~\cite{Haensel:2009wa}. The black and orange dashed curves depict the excluded region of the M-R diagram obtained for the quark EoS with $C=1.15$~kHz and hybrid EoS with the early deconfinement phase transition ($\eta_D=0.755$). The dash-dotted gray, black and orange curves illustrate the constraint coming from the detection of a pulsar with 1000 Hz spinning frequency considering the hadronic, quark and hybrid EoS, respectively.
}
\label{fig6}
\end{figure}

The fastest spinning pulsar, PSR J1748-2446ad, with a spin frequency of 716 Hz~\cite{Hessels:2006ze}, is used to put a limit on the NS properties and dense matter EoS. The lower limit on the M-R relation obtained by ~\citet{Haensel:2009wa} for a set of hadronic EoSs is depicted with the gray dashed curve in Fig.~\ref{fig6}. The gray-shaded area below shows an excluded region for NSs. On the other hand, the limits coming from quark EoSs with $C=1.15$~kHz and the hybrid EoS with the early deconfinement phase transition ($\eta_V=0.30$, $\eta_D=0.755$) are depicted by the black and orange dashed curves, respectively. As you can see in Fig.~\ref{fig6} hadronic EoSs provide a more stringent constraint on the mass and radius of low-mass NSs, while the hybrid EoS does not show a big difference from the former one. This insignificant deviation of the orange curve from the gray one could easily be understood from the be\-ha\-vior of the $C$ factor in Fig.~\ref{fig5}. Thus, in the middle-mass range, the deviation of $C$ for hybrid EoS from the hadronic $ C_{\rm had}=1.088$ kHz line does not exceed 20 Hz, while the difference becomes significant for higher masses. Therefore, detection of a pulsar with a 1000 Hz spinning frequency could lead to a much more stringent constraint on the compact star properties. The dash-dotted gray, black, and orange curves illustrate the lower bounds for the hadronic, quark, and hybrid EoS, respectively.

The inclusion of the hybrid EoS in the analysis of PSR J1748-2446ad resulted in a revised upper limit on the radius of a 1.4 solar mass NS. The limit extends from $R_{\rm 1.4}\leq$14.78~km (for the hadronic EoSs) to $R_{\rm 1.4}\leq$14.90~km (for the hybrid EoS with the early deconfinement phase transition $\eta_V=0.30$, $\eta_D=0.755$). 
While this limitation does not constrain the radius of a canonical NS compared to other available constraints,  
the detection of an even faster-rotating pulsar, e.g., with a spin frequency of 1000 Hz, would impose even tighter constraints on the properties of NSs and strongly interacting matter. 
It would also alter the upper limit from $R_{1.4}\leq$11.86~km (hadronic EoS) to $R_{1.4}\leq$11.90~km (hybrid EoS), providing significant insights into the EoS of dense matter.
On the other hand, confirmation of the existence of NSs with a mass compatible with the one reported for the HESS J1731-347 object~\cite{Doroshenko2022} would allow our results to significantly limit radii of such light NS.  
For example, for a 0.7 $M_\odot$ NS this would imply $R_{0.7}\le11.49$ km, which is more stringent than even the $2\sigma$ constraint from HESS J1731-347~\cite{Doroshenko2022}.

\section{Rapidly rotating hybrid stars}
\label{res}

\begin{figure*}[th!]
\centering
\setkeys{Gin}{width=0.498\linewidth}
\begin{tabularx}{\linewidth}{XX}
\includegraphics{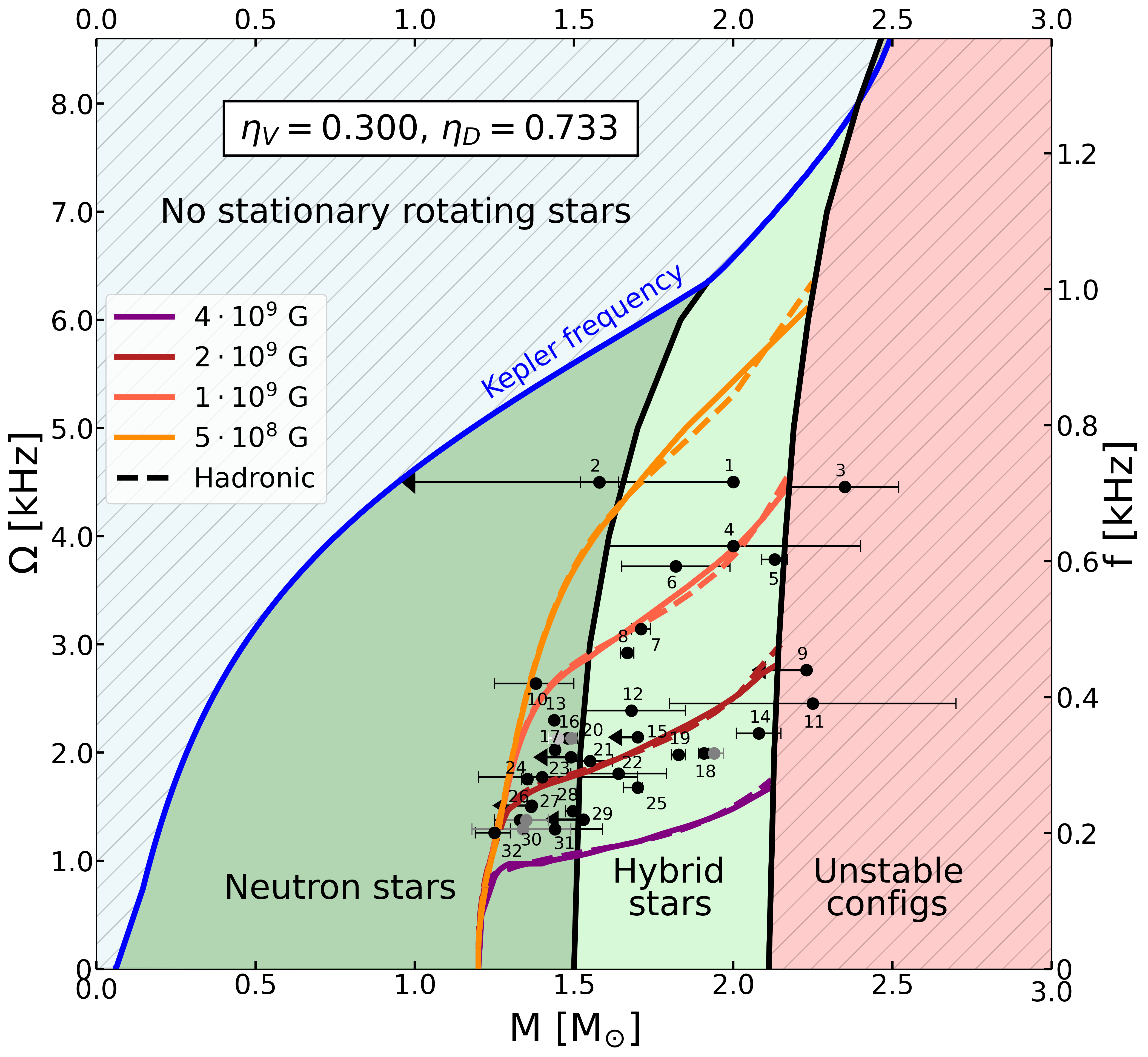}
\includegraphics{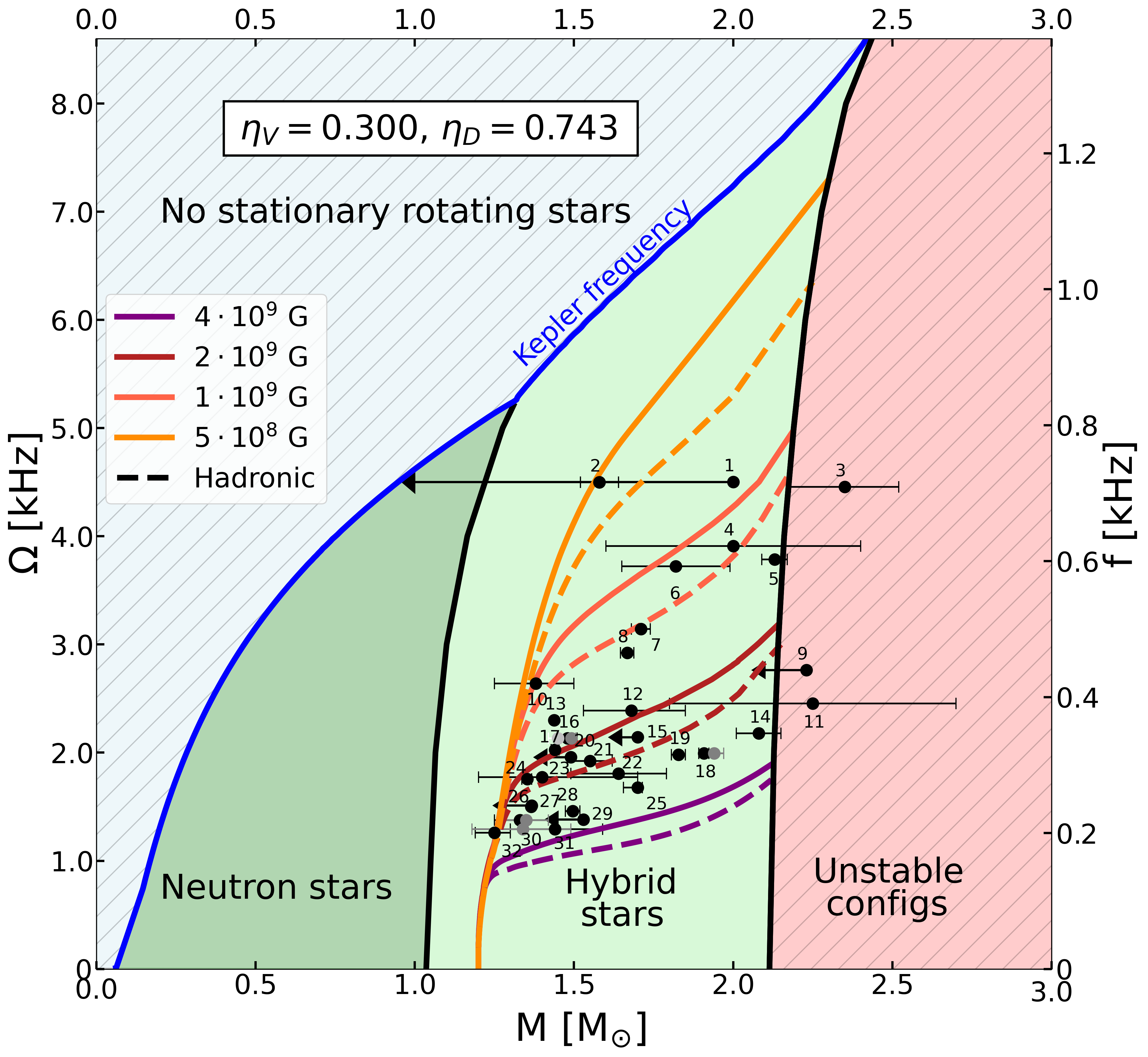}
\end{tabularx}
\begin{tabularx}{\linewidth}{XX}
\includegraphics{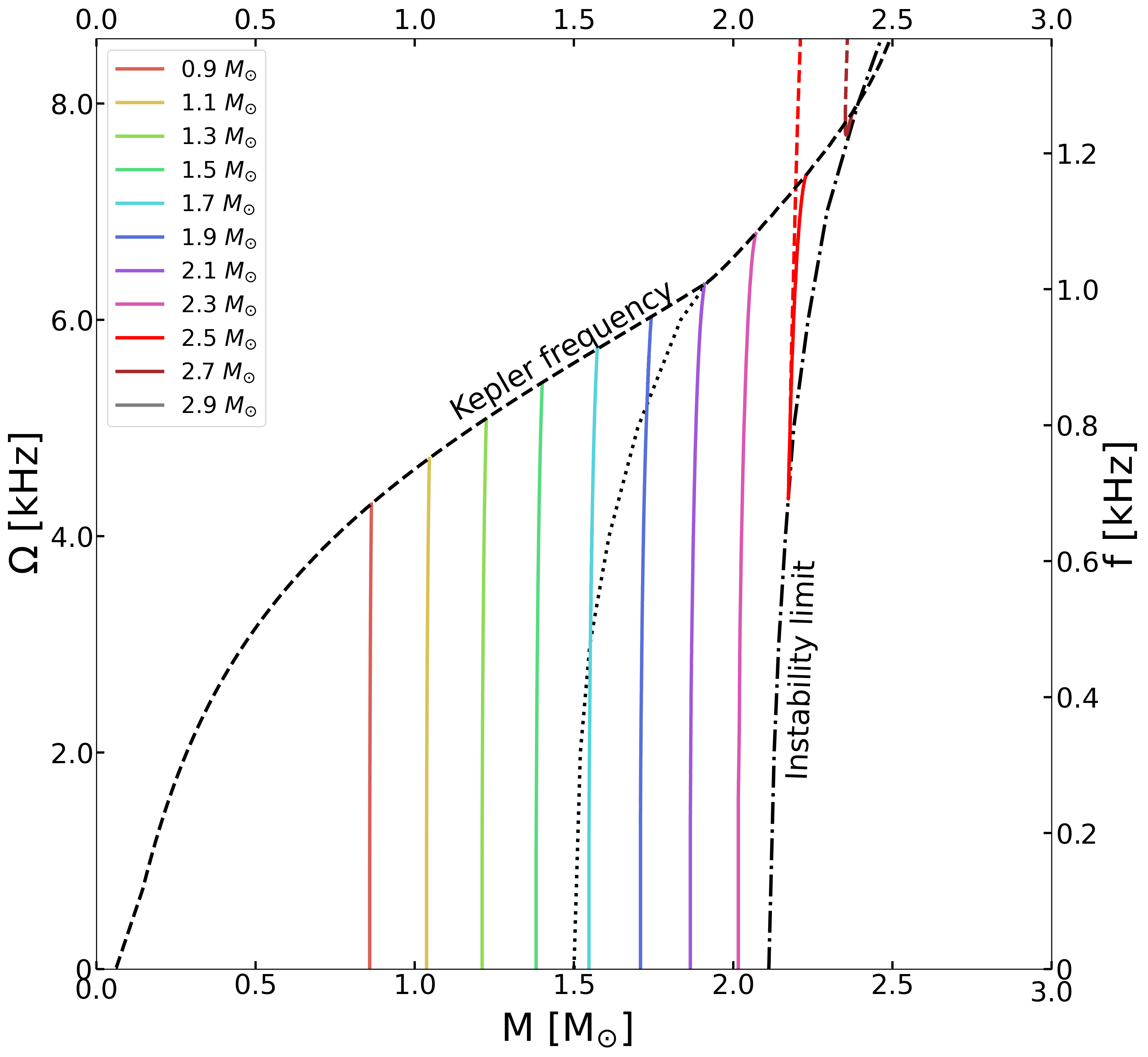}
\includegraphics{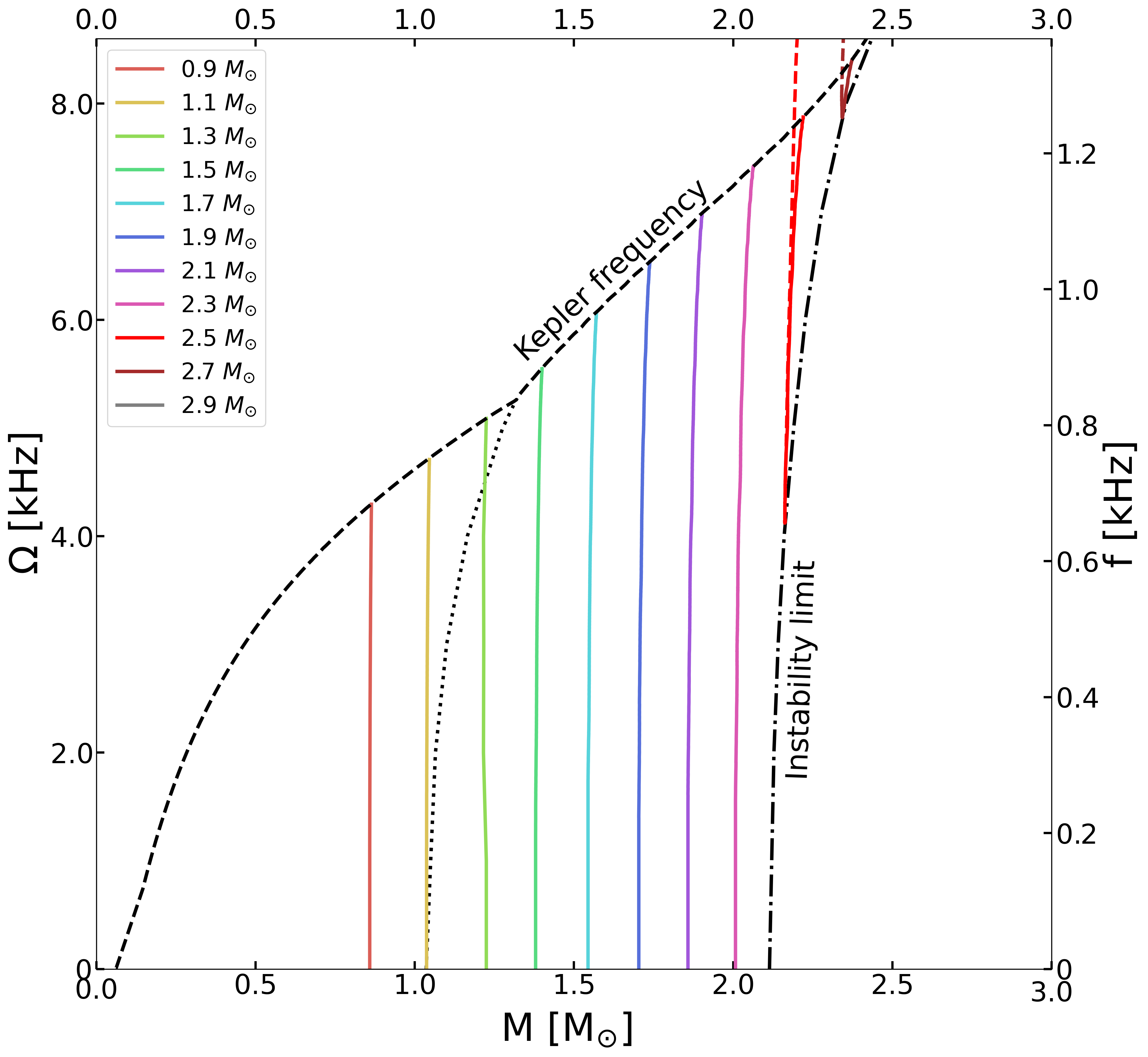}
\end{tabularx}
\caption{{\bf Upper panel:} The angular velocity as a function of the gravitational mass of hybrid stars. The solid blue curve represents the Kepler frequency, above which lies the gray-shaded area where no stationary rotating stars can be found. Black solid curves mark the separation between NSs (dark green area), hybrid stars (light green area), and black holes (pink beige area of unstable configurations). The circles with 1$\sigma$ confidence level error bars depict the measured mass and frequency of the fastest known MSPs with spin frequency f$>$200 Hz (the data are listed in Appendix~\ref{app:B}). The arrows, rather than error bars, represent the available upper limit on the mass estimate. Other independent mass measurements of the same object are depicted with gray dots. The color solid curves represent the evolution trajectories of the hybrid (solid) and hadronic (dashed) star of $1.2M_{\odot}$ with the corresponding strength of the magnetic field. 
{\bf Lower panel:} Evolutionary paths of non-accreting unmagnetized stars of the fixed constant rest mass depicted in different colors. The x-axis shows the gravitational mass. The dotted curve indicates the NS-hybrid star separation, similar to the upper panel. The results are obtained for the fixed value of the vector coupling $\eta_V=0.3$ and different values of the diquark coupling $\eta_D=0.733$ (left panels) and $\eta_D=0.743$ (right panels).}
\label{fig7}
\end{figure*}
\begin{figure*}[!ht]
\centering
\setkeys{Gin}{width=0.498\linewidth}
\begin{tabularx}{\linewidth}{XX}
\includegraphics{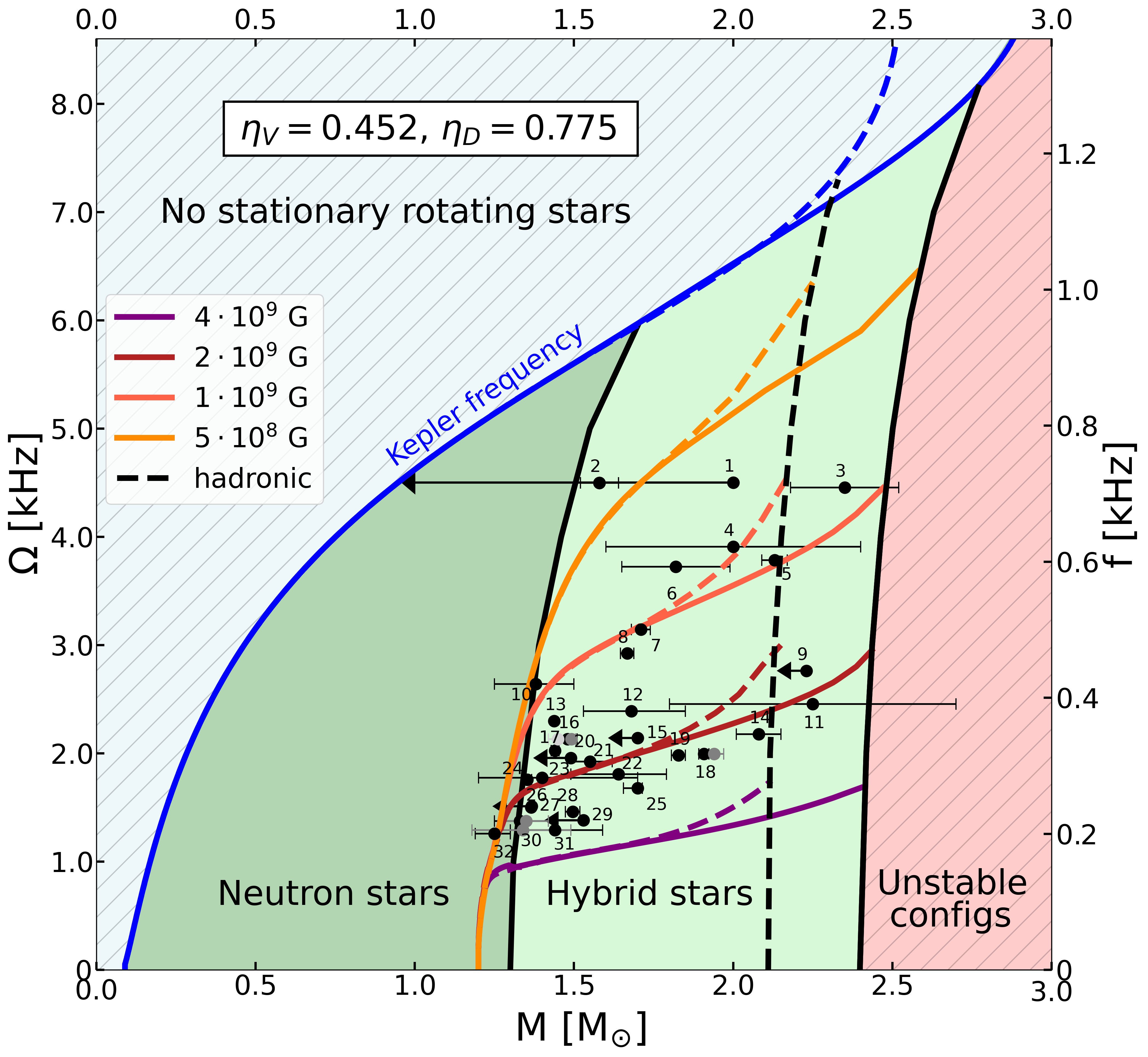}
\includegraphics{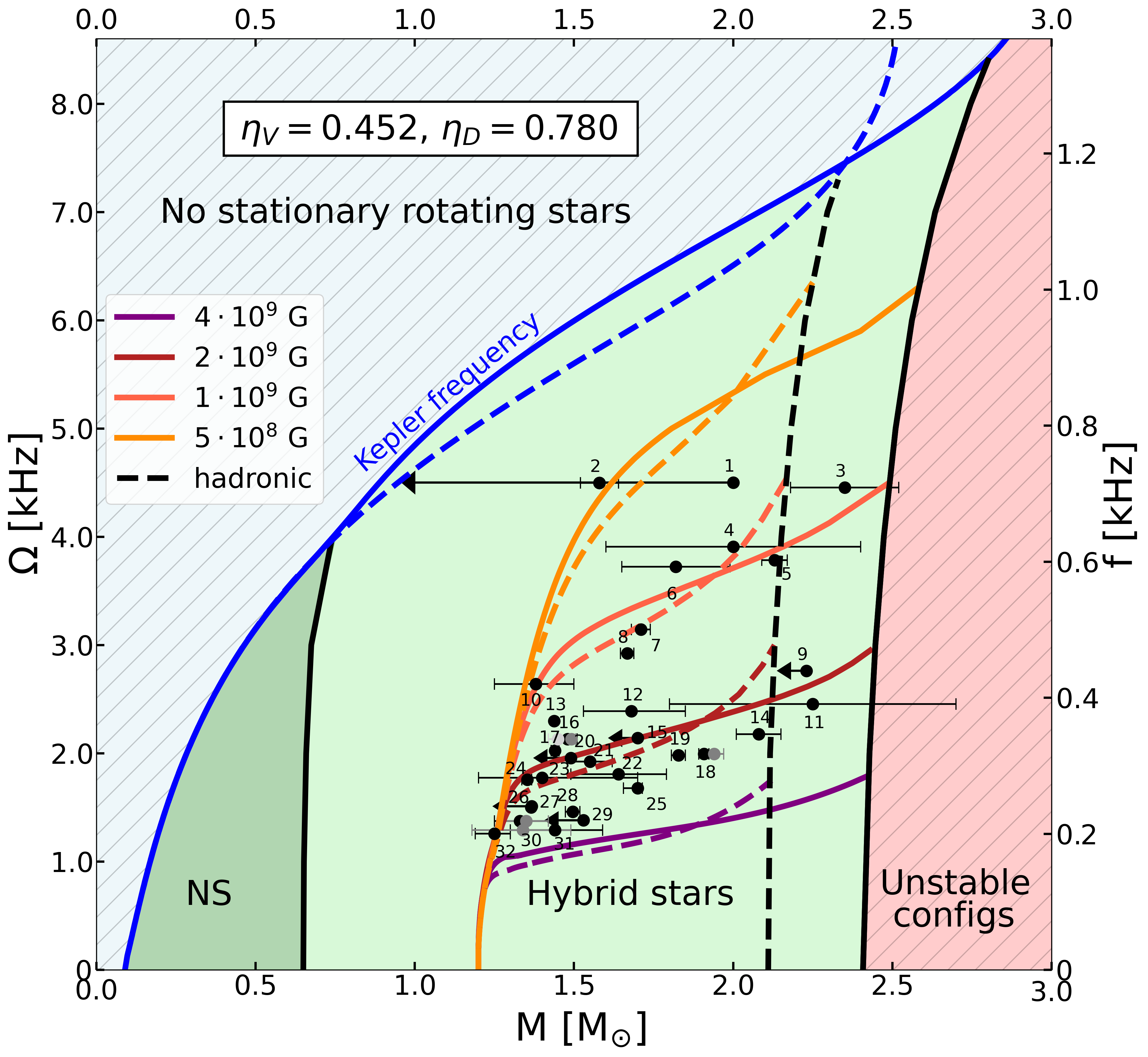}
\end{tabularx}
\caption{The same as the upper panel of Fig.~\ref{fig7} but for different values of model parameters: $\eta_V=0.452, \eta_D=0.775$ (left panel) and $\eta_V=0.452, \eta_D=0.780$ (right panel). The dashed color lines cease to exist where the hadronic stars reach the instability limit (black dashed curve). The Kepler limit for the hadronic DD2npY-T EoS is depicted with the dashed blue curve. 
}
\label{fig8}
\end{figure*}
For a better comparison, we selected hybrid star configurations for four pairs of couplings, namely ($\eta_V=0.3, \eta_D=0.733$), ($\eta_V=0.3, \eta_D=0.743$), ($\eta_V=0.452, \eta_D=0.775$) and ($\eta_V=0.452, \eta_D=0.780$). This choice is motivated by the correlation between the diquark coupling value and the onset mass of the deconfinement phase transition found in Ref.~\cite{Gartlein:2023vif}. Consequently, these confi\-gu\-ra\-tions represent two distinct cases of the model and hybrid star properties with the middle- and low-mass quark onset, and the maximum hybrid star mass exceeding that of the purely hadronic case and being below it.

Whether a star of a certain mass is a hybrid or a pure hadronic star is defined by the dense matter EoS. For the considered DD2npY-T -- RDF hybrid EoS we map the star's configurations in Fig.~\ref{fig7} in the plane of the angular velocity as a function of the star's gravitational mass. By solving the hydrostatic equations for equilibrated stars with different spin frequencies, we depict the region where the stars are purely hadronic (the dark green area in  Fig.~\ref{fig7}), hybrid stars with a quark core (the light green area) as well as unstable configurations (the pink beige area) separated by the solid black curves. The latter refers to configurations exceeding the maximum gravitational mass, where stellar oscillations would cause the star to collapse into a black hole. The blue curve above corresponds to the mass-shedding limit. Moreover, Fig.~\ref{fig7} depicts the observational data on the fastest MSPs for which the mass measurements are available. The data are listed in Appendix~\ref{app:B}. The error bars represent a 1$\sigma$ confidence interval, and for stars with only the upper estimate, the arrow indicates the range of masses. We observe that increasing the $\eta_D$ value from 0.733 (left panel) to 0.743 (right panel) enlarges the hybrid star region, encompassing all depicted MSPs within their error bars. Interestingly, for $\eta_V=0.3, \eta_D=0.743$ all the rapidly rotating MSPs are predicted to be hybrid stars, while for $\eta_V=0.3, \eta_D=0.733$ only the stars above $\sim 1.5~M_{\odot}$ are hybrid. 

Compact stars born with a rapid rotation evolve by emitting electromagnetic radiation and gravitational waves, eventually losing their angular momentum. As the baryon number is conserved the evolutionary path goes along the constant rest mass $M_0$ line shown in the lower panels of Fig.~\ref{fig7}. The color lines depict the evolutionary sequences (lines of NSs of constant rest mass, equal to baryon mass) as a function of the angular velocity $\Omega$. The x-axis displays the gravitational mass. The black dashed curve represents the rotational limit defined by the onset of mass-shedding from the equator (the Kepler frequency), while the dotted and dash-dotted curves indicate the onset of the deconfinement phase transition and configurations unstable due to radial oscillations, respectively. The cyan and blue lines in the lower left panel, as well as the yellow and light green lines in the lower right panel of Fig.~\ref{fig7}, illustrate the scenario where a hadronic star crosses the deconfinement onset as it spins down, further evolving as a hybrid star. In contrast, massive stars (represented by the red and brown lines) lack a static limit and will collapse into a black hole upon losing their angular velocity. Note that the evolutionary paths in the lower panels are obtained for unmagnetized non-accreting stars.

\begin{figure*}[ht!]
\centering
\setkeys{Gin}{width=0.498\linewidth}
\begin{tabularx}{\linewidth}{XX}
\includegraphics{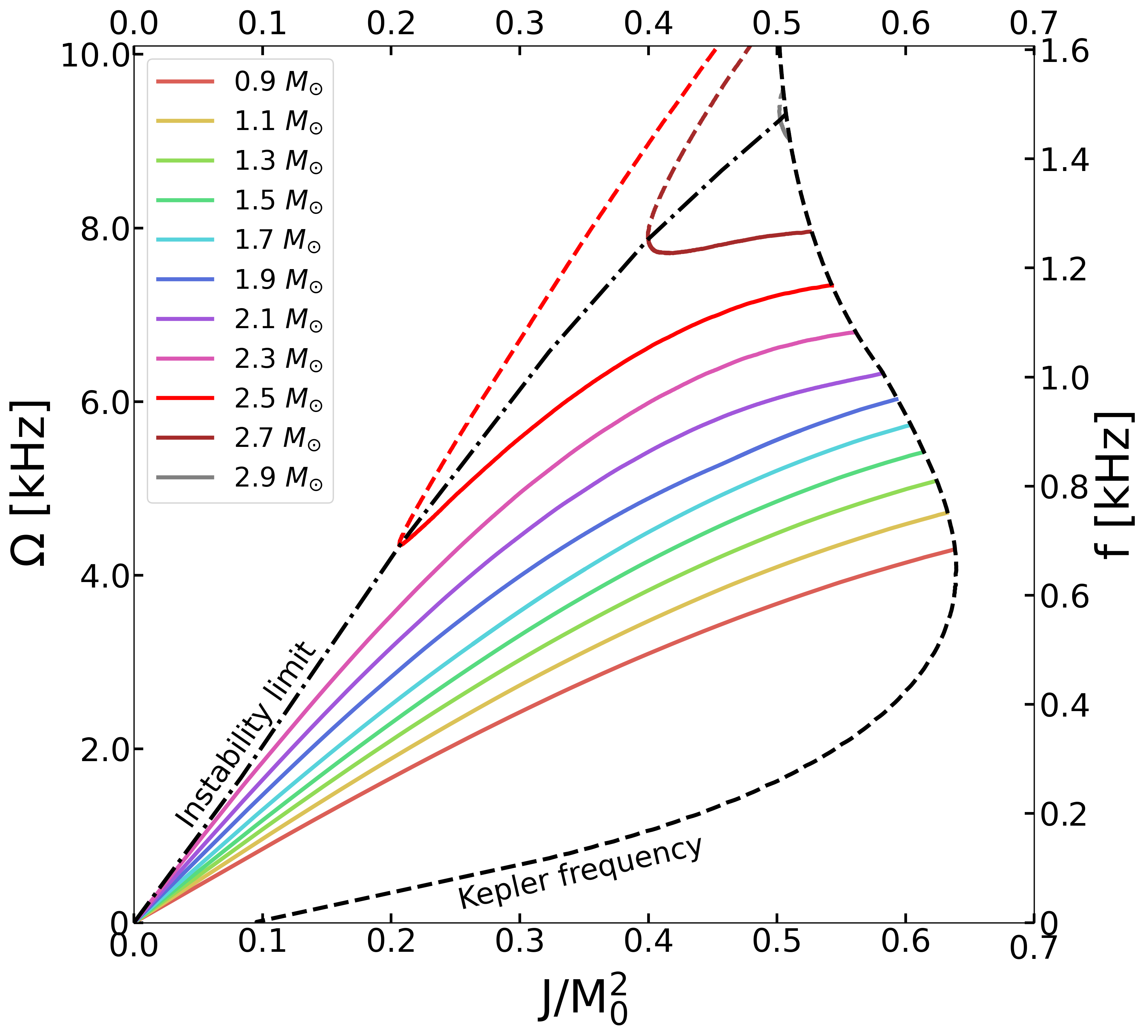}
\includegraphics{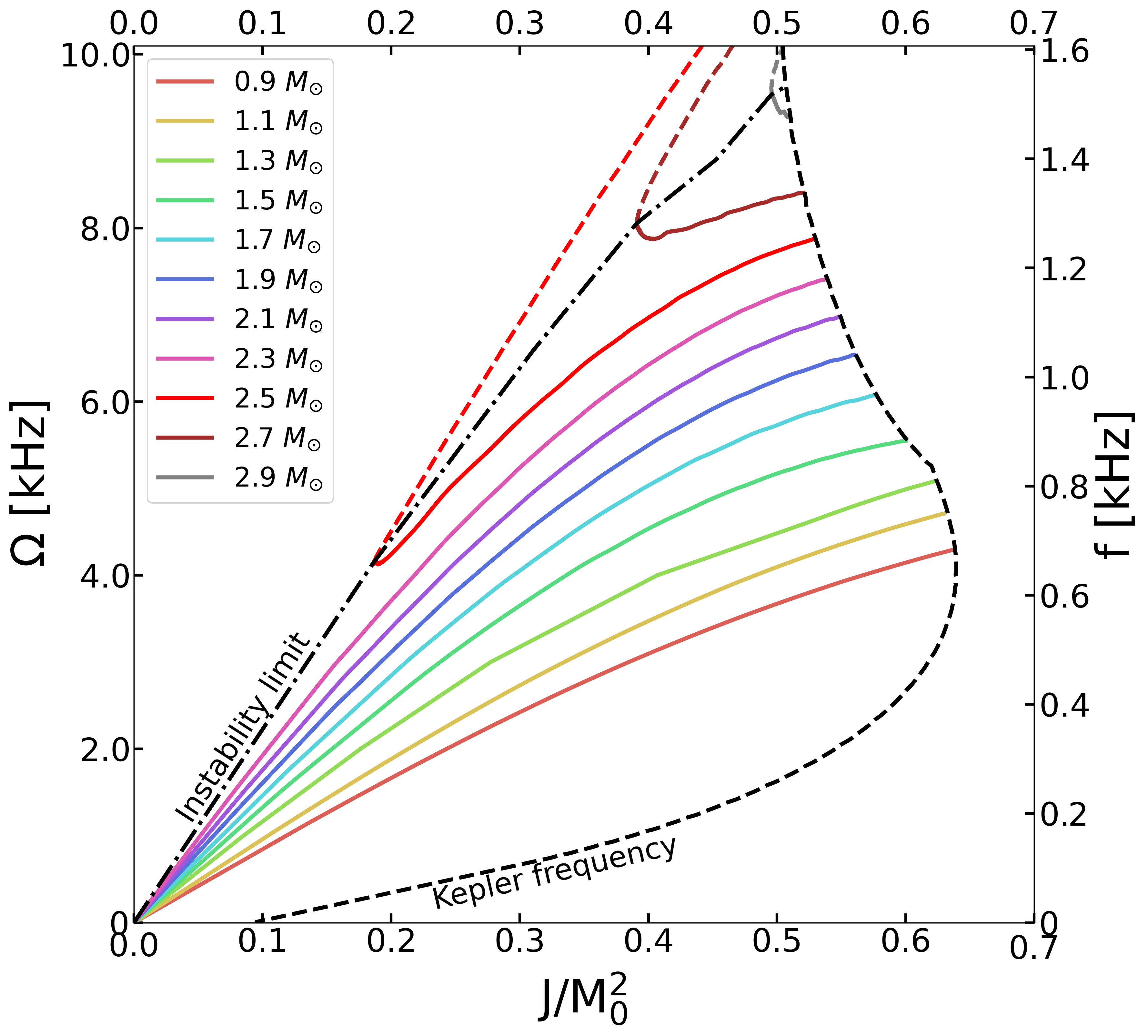}
\end{tabularx}
\caption{The angular velocity $\Omega$ as a function of the angular momentum $J$ divided by the rest mass squared for the fixed value of the vector coupling $\eta_V=0.3$ and different values of the diquark coupling $\eta_D=0.733$ (left panel) and $\eta_D=0.743$ (right panel).}
\label{fig9}
\end{figure*}

As you can see in Figs.~\ref{fig7}-\ref{fig8} even the fastest-spinning pulsars, J1748+2446ad~\cite{Hessels:2006ze}, 4U 1820–30 (J1820-30A)~\cite{Jaisawal_2024,Guver:2010td}, and J0952-0607~\cite{Romani:2022jhd} are far from the Kepler limit (for the mass of J1748+2446ad there is only an upper limit, but it is unlikely that its mass is below 1$M_{\odot}$). The fact of observing two of the fastest objects with very similar spin frequencies made us think that they corresponded to the cutoff limit. On the other hand, the MSPs depicted in Figs.~\ref{fig7}-\ref{fig8} show clustering in the frequency (angular velocity) interval $\rm 210 \leq f[Hz] \leq 340$ ($1300\leq \Omega[\rm Hz] \leq 2140$). The question arises as to whether any physical mechanisms affect spin evolution, e.g., binary evolution, accretion, magnetic field evolution, the appearance of the exotic degrees of freedom in the star interior, etc. 

A possible explanation for the population clustering caused by the waiting time in a certain frequency region is proposed by~\citet{Poghosyan:2000mr}. The onset of quarks in the interior of rapidly rotating stars entails a characteristic change in the spin evolution. The star's spin evolution is defined by the strength of the magnetic field, accretion rate, and the appearance of quark matter. The evolutionary tracks for the 1.2$M_{\odot}$ star with a fixed accretion rate of $\rm 10^{-7}~M_{\odot} yr^{-1}$, and initial magnetic field strength as indicated in the legend are shown in the upper panels of Fig.~\ref{fig7}. The solid curves depict the hybrid EoSs, while the dashed curves are obtained for the hadronic DD2npY-T EoS. As shown, the curves reproduce the MSP data well and explain a pulsar clustering. For further details about the model, see Appendix~\ref{app:C} and Ref.~\cite{Poghosyan:2000mr}.

\begin{figure*}[ht!]
\centering
\setkeys{Gin}{width=0.498\linewidth}
\begin{tabularx}{\linewidth}{XX}
\includegraphics{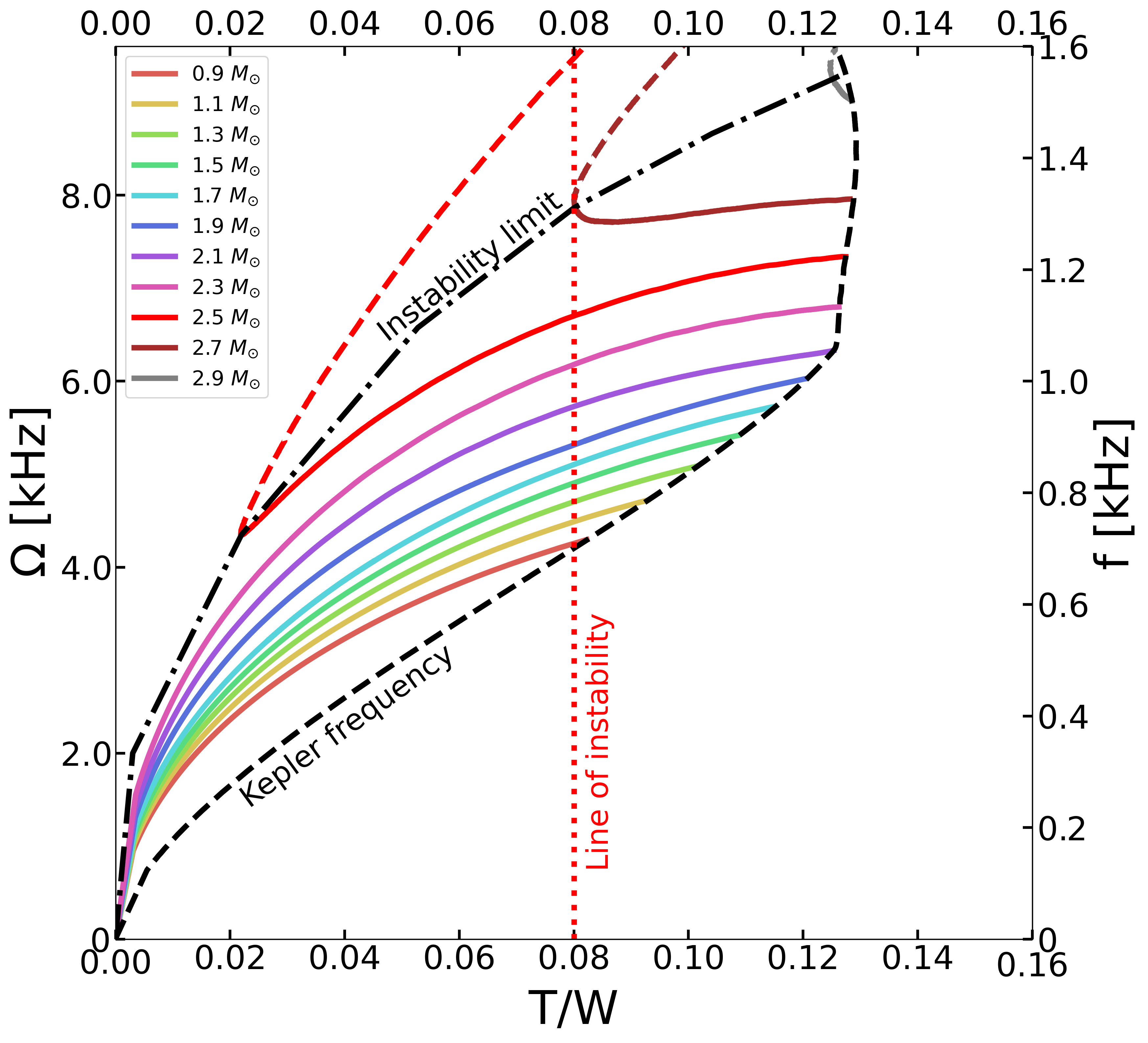}
\includegraphics{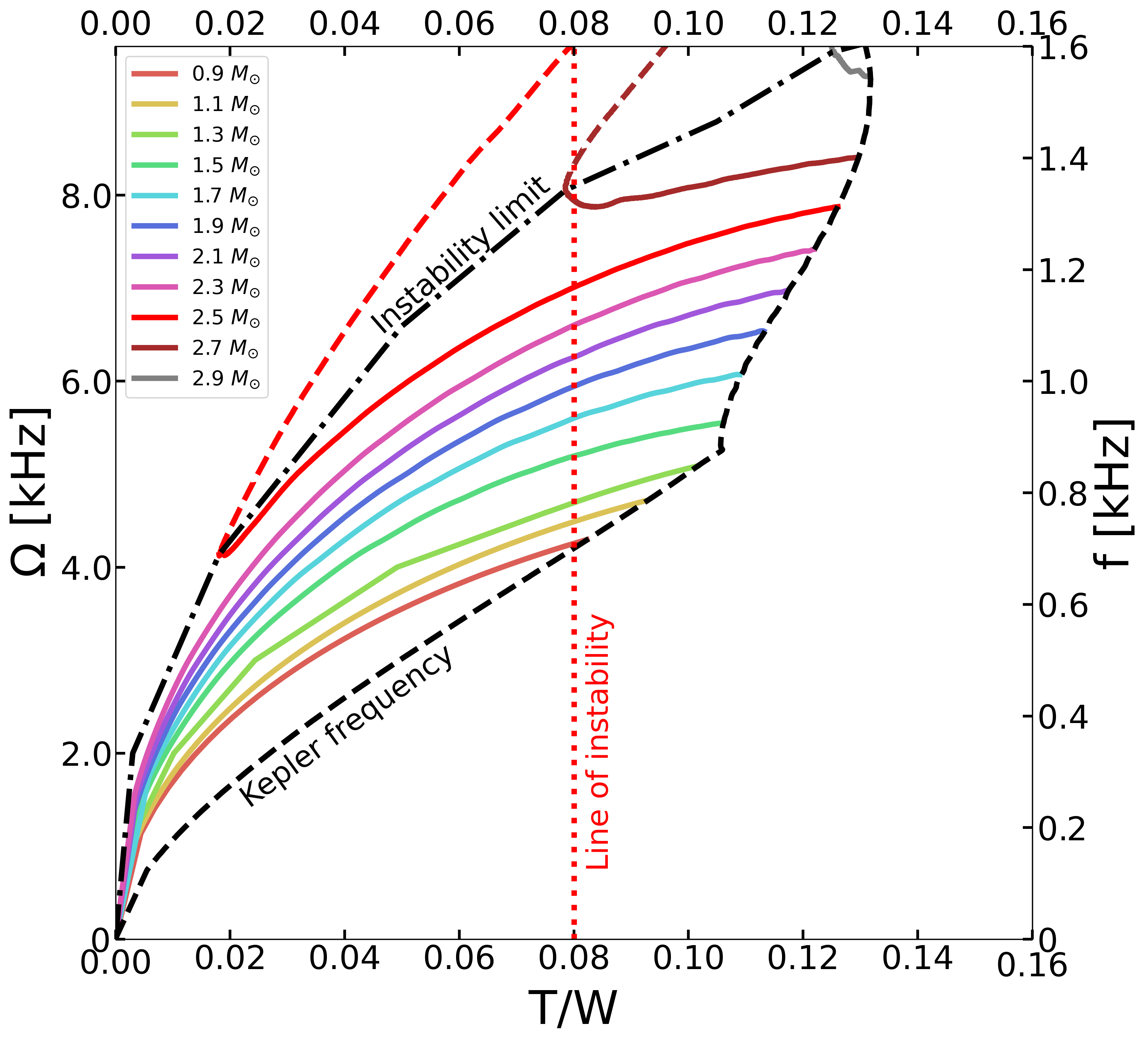}
\end{tabularx}
\begin{tabularx}{\linewidth}{XX}
\includegraphics{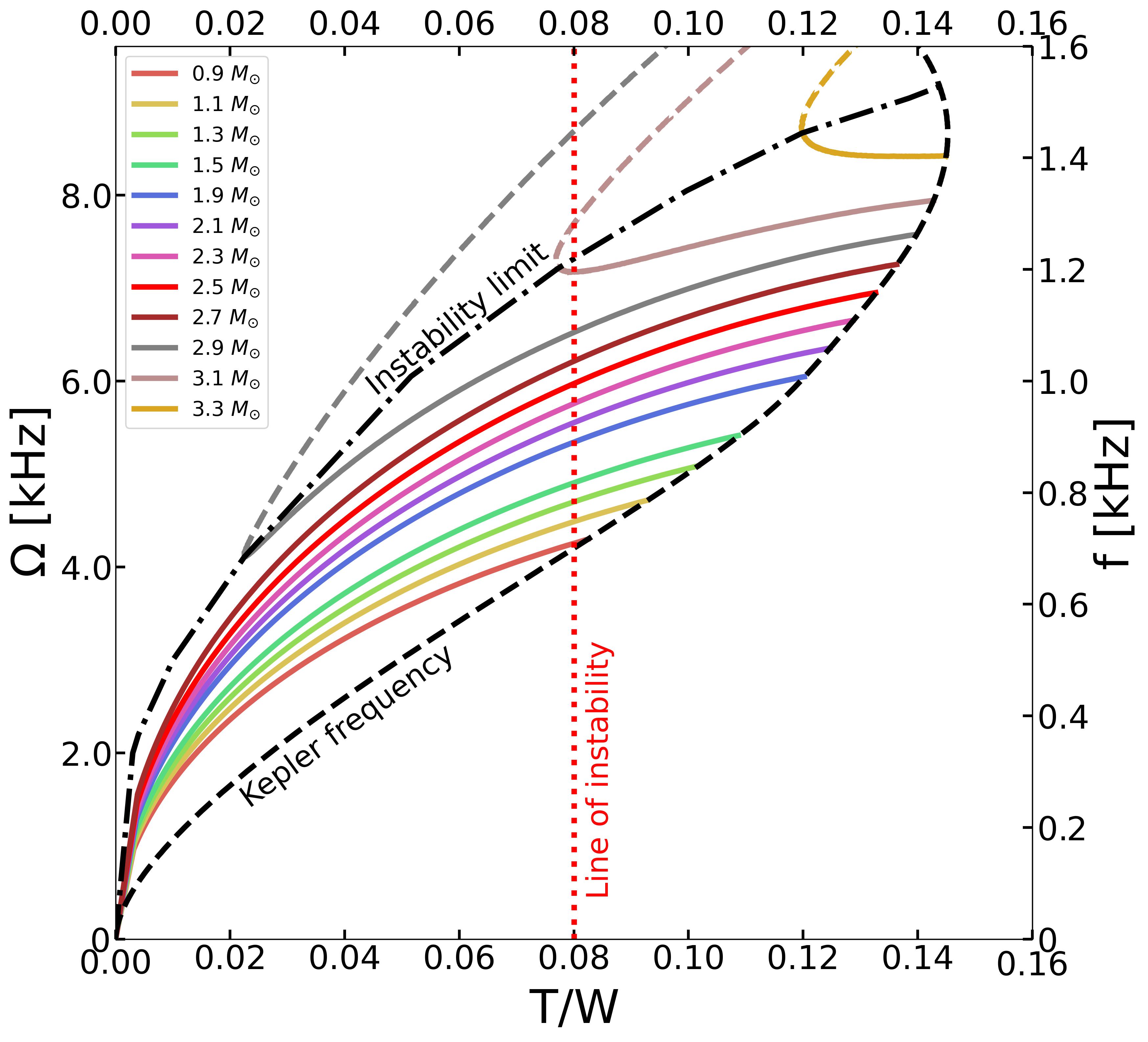}
\includegraphics{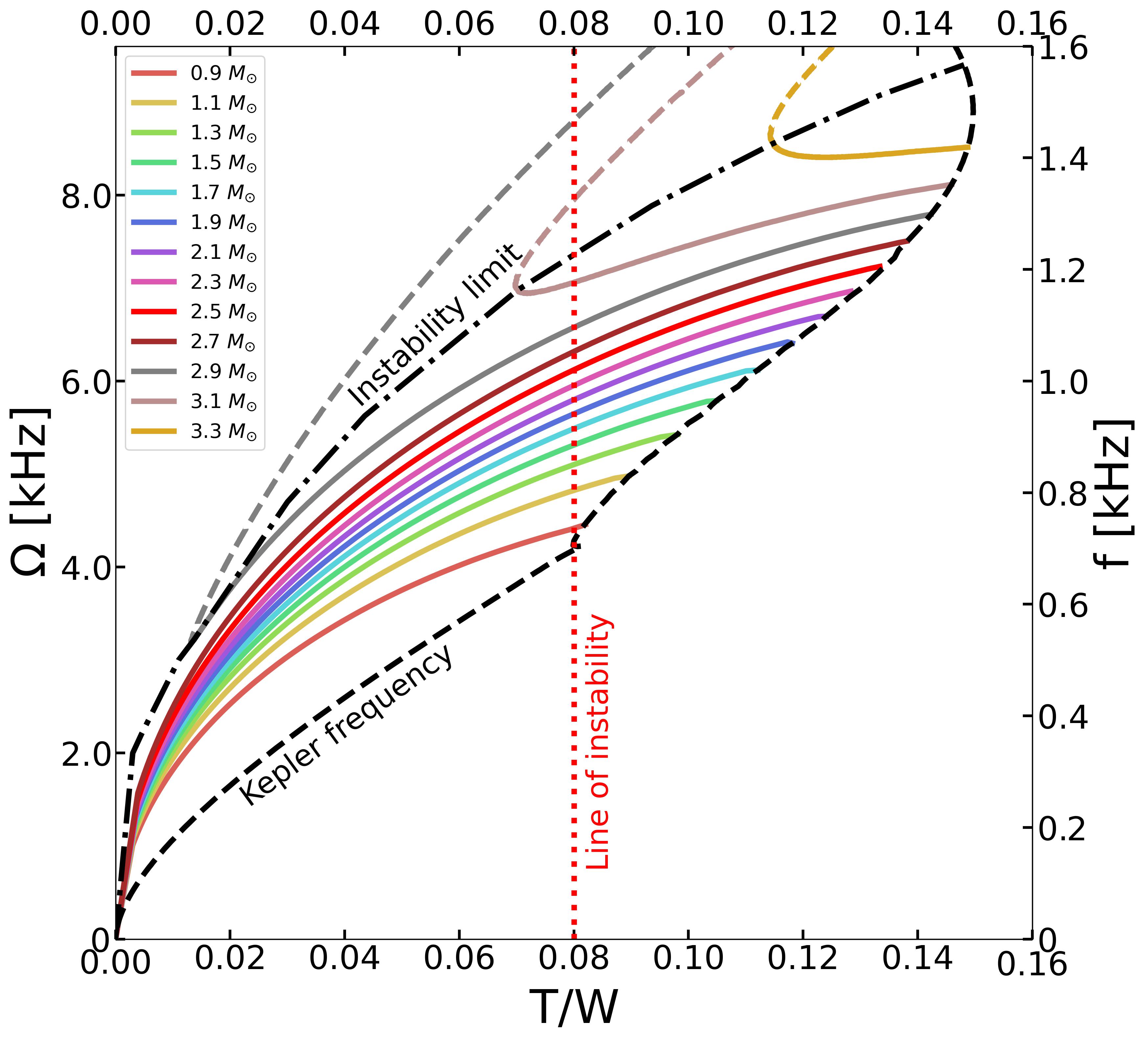}
\end{tabularx}
\caption{The angular velocity as a function of the ratio of rotational kinetic $T$ and gravitational potential $W$ energy. {\bf Upper panels:} the vector coupling is fixed to $\eta_V=0.3$ and the diquark coupling equals $\eta_D=0.733$ (left panel) and $\eta_D=0.743$ (right panel). {\bf Lower panels:} $\eta_V=0.452$ and the diquark coupling $\eta_D=0.775$ (left panel) and $\eta_D=0.780$ (right panel). The colorful curves represent the constant rest mass values listed in the legend. The black dashed and dash-dotted curves depict the mass-shedding limit and collapse to a black hole, respectively. The red dotted line corresponds to the onset of axisymmetric instability mentioned in the text.
}
\label{fig10}
\end{figure*}

Fig.~\ref{fig8} shows the modification of the angular velocity as a function of the gravitational mass for the model parameters ($\eta_V=0.452, \eta_D=0.775$) and ($\eta_V=0.452, \eta_D=0.780$)that resulted from the special point analysis of Ref. ~\cite{Gartlein:2023vif}. The value of $\eta_V=0.452$ was fixed by the $\omega$ meson mass and determines the maximum mass limit (dashed black lines) while the range of admissible values for the diquark coupling, $\eta_D=0.775$ and $\eta_D=0.780$ determines the upper (left panel) and lower (right panel) limits for the onset of deconfinement, respectively. Compared to Fig.~\ref{fig7}, the hybrid star regions in both panels are more extended, while the latter parameter set even encompasses all depicted MSPs.

Interestingly for these two pairs of $\eta_V, \eta_D$ values, we see a big difference in the behaviors of the evolutionary trajectories for the hybrid (solid) and hadronic (dashed) star of 1.2$M_{\odot}$. The dashed curves cease to exist where the hadronic stars reach the instability limit (dotted lines), while the solid curves continue up to the instability limit. Notably, the black-widow pulsar J0952-0607~\cite{Romani:2022jhd}, labeled `3' in Fig.~\ref{fig8} can only be explained as a hybrid star with the favorable strong vector coupling $\eta_V=0.452$, within the $1\sigma$ range of its mass because it lies beyond the limit of stability for the purely baryonic DD2npY-T EOS.

As was first classified by~\citet{Cook:1993qj} the evolu\-tio\-na\-ry sequences existing between the dashed and dashed-dotted black curves are the {\it normal} sequences that have a static limit (see the solid color lines in Fig.~\ref{fig9}). For all pairs of couplings, the normal sequences of hybrid stars characterized by the constant rest mass values show a shift towards the higher gravitational mass near the Kepler frequency. 

The second group of {\it supramassive} sequences does not have the static limit. Thus, the curves corresponding to the high rest mass values lack static or low-frequency counterparts. These stars while spinning down will collapse into black holes as the gravitational force becomes too strong, overcoming the repulsive forces. 

The evolutionary behavior for normal and supramassive sequences is well presented in the plane of the angular velocity $\Omega$ as a function of the angular momentum $J$ divided by the rest mass squared. The co\-lo\-rful dashed lines above the black dashed-dotted curve in Fig.~\ref{fig9} indicate the supramassive sequences. The solid color lines are the same as in Fig.~\ref{fig7} and obtained for the fixed values of the baryon mass.

For the normal NS sequences, the angular momentum $J$ consistently increases with $\Omega$, indicating that along the normal sequence, stars do not spin up as the angular momentum decreases. The supramassive stars exhibit the opposite behavior; with angular momentum lost during evolution, the angular velocity increases. The stability condition for the star of constant rest mass $M_0$ is defined as $\frac{\partial J}{\partial \epsilon_{c}}|_{M_0}<0$, where $\epsilon_{c}$ is the central energy density~\cite{Cook:1993qj}. The onset point of the quasi-radial oscillations occurs at the lowest $J$ value of each evolutionary sequence, i.e., on the dash-dotted curve dividing the two sequences. Furthermore, by comparing the two panels in Fig.~\ref{fig9}, we can conclude that for the higher value of the diquark coupling, the earlier onset makes hybrid stars more resistant to higher angular velocity and capable of sustaining a higher maximum mass.

\begin{figure*}[ht]
\centering
\setkeys{Gin}{width=0.498\linewidth}
\begin{tabularx}{\linewidth}{XX}
\includegraphics{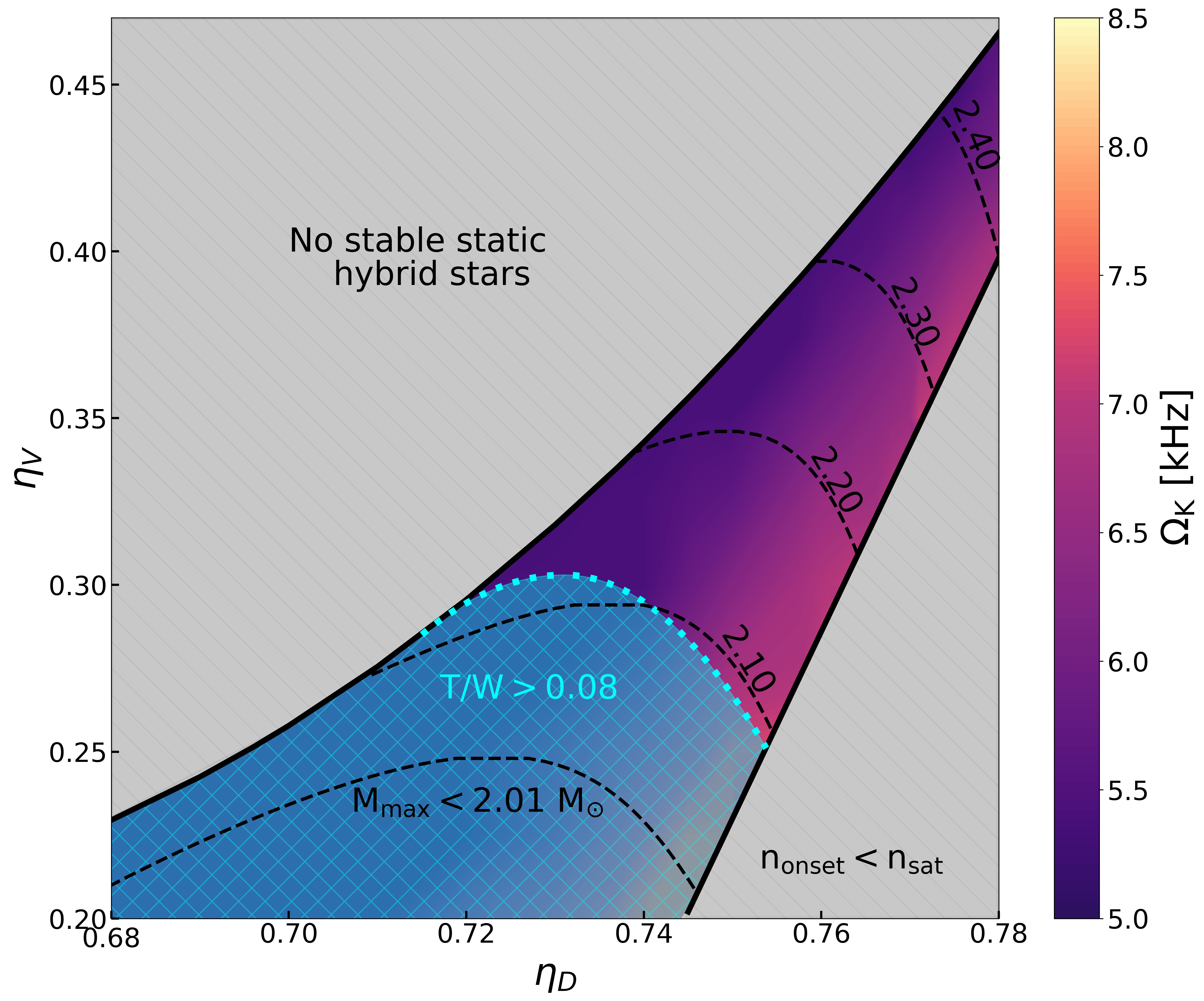} 
\includegraphics[width=0.498\linewidth]{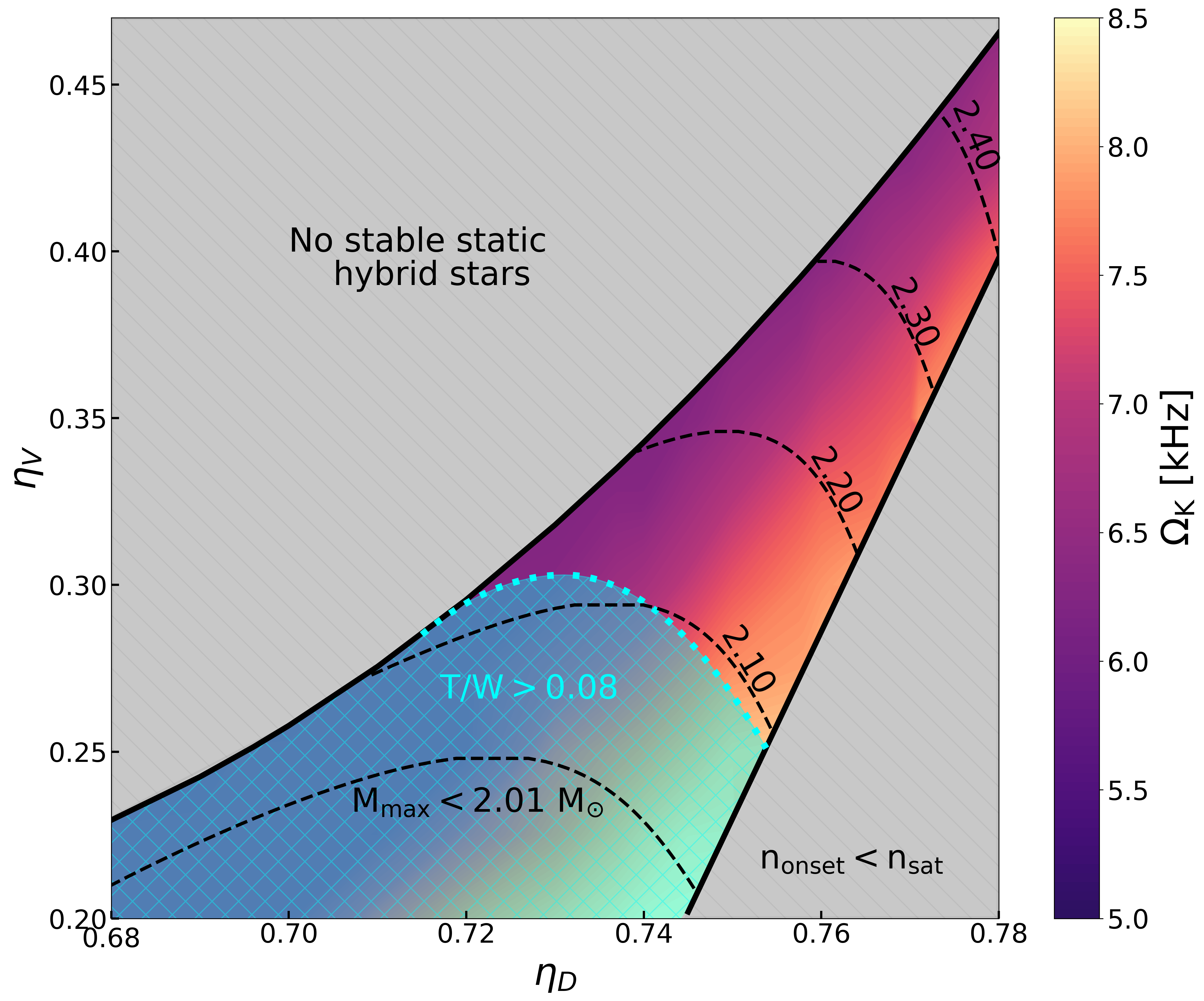}
\end{tabularx}
\caption{Parameter space in the $\eta_V$ - $\eta_D$ plane calculated for the  DD2npY-T - RDF hybrid EoS for the fixed value of the rest mass $1.5~M_\odot$ (left panel) and $2.1~M_\odot$ (right panel). The color represents the Kepler angular velocity of the rotating star in the region where the vector and diquark couplings are consistent with observational constraints on the maximum mass of NSs from~\cite{Fonseca:2021wxt}. The lavender region is excluded by the requirement of the stability of the quark branch and tidal deformability constraint from GW170817~\cite{LIGOScientific:2018cki}, while the light yellow region indicates the excluded region with the onset of the deconfinement phase transition below the saturation density. The dark green shaded area at the bottom corresponds to the excluded region of model parameters resulting in the maximum gravitational mass of the static stars below 
$M_{max}<2.01M_{\odot}$~\cite{Gartlein:2023vif}. The black dashed contour lines indicate the maximum achieved gravitational mass for those values of the parameters. The region excluded by the onset of gravitational radiation instabilities according to the criteria $T/W>$ 0.08 for the Black Widow Pulsar J0952$-$0607 is displayed with the cyan hatched region.}
\label{fig11}
\end{figure*}

A similar parameter, $K \equiv J/M^2$, where $M$ is the gravitational mass of a star, is known as the Kerr pa\-ra\-me\-ter. This dimensionless spin parameter, which reaches its maximum value at the mass-shedding limit, indicates the star's proximity to collapse into a black hole. As shown in Ref.~\cite{Koliogiannis:2020nhh}, a typical NS at the maximum mass reaches $K \sim$0.7, while for a black hole, it equals 1.0.

Although the configurations between the mass-shedding limit (the dashed black curve) and instability limit causing stars to collapse into black holes (the dash-dotted black curve) are considered stable, the ratio of rotational and gravitational energy, $T/W$, is strongly related to the onset of the gravitational radiation-driven instability that causes the star collapse into the Kerr black hole~\cite{Cook:1993qj,Paschalidis:2016vmz,Koliogiannis:2020nhh}. In Fig.~\ref{fig10} the angular velocity $\Omega$ as the function of the $T/W$ ratio is plotted for the sequence of hybrid stars of constant rest mass values (see solid curves of a different color). The vertical red dotted line indicates the onset of $l=m=2$ f-mode in\-sta\-bi\-li\-ty at $T/W \simeq 0.08$ for the star of 1.4~$M_{\odot}$ gravitational mass~\cite{1985ApJ...294..463M,1986ApJ...304..115F,Morsink:1998db}. Considering the weak dependence of the instability limit on the star's mass, we assume the same value for all stars is $T/W=0.08$. In the Newtonian limit, the onset of the non-axisymmetric perturbations would set in at $T/W \simeq 0.14$ for the 1.4~$M_{\odot}$ star. For all four different sets of model parameter sets ($\eta_V=0.3, \eta_D=0.733$), ($\eta_V=0.3, \eta_D=0.743$), ($\eta_V=0.452, \eta_D=0.775$) and ($\eta_V=0.452, \eta_D=0.780$) the non-axisymmetric instabilities will arise before the Kepler frequency is reached. Moreover, the heaviest star configurations are entirely ruled out by this limit. Consequently, observing rapidly rotating massive stars listed in Table~\ref{table2} allows us to constrain the allowed values of $\eta_V$ and $\eta_D$ within the f-mode instability window.

Fig.~\ref{fig11} shows the allowed range of the model parameters in the $\eta_V$ - $\eta_D$ plane that are consistent with the existing NS observations. For the fixed value of the rest mass $1.5~M_\odot$ (left panel) and $2.1~M_\odot$ (right panel), the color represents the mass-shedding angular velocity together with the excluded range of model parameters that lead to unstable static hybrid stars, excluded by the tidal deformability constraint from GW170817~\cite{LIGOScientific:2018cki}, and the negative squared frequency of the fundamental mode of radial oscillations~\cite{DiClemente:2020szl,Sagun:2020qvc} (see the shaded lavender region in Fig.~\ref{fig11}). Note, the considered rest mass  values $1.5~M_\odot$ (left panel) and $2.1~M_\odot$ (right panel) are equivalent to stars with the gravitational mass of $~\sim 1.4~M_\odot$ and $~\sim 1.9~M_\odot$, respectively. The gray region in the bottom right corner of both panels is excluded due to the criteria of the too early onset of the deconfinement phase transition. We require the phase transition to occur above the saturation density of normal nuclear matter, $n_{\rm sat}$, equivalent to the onset mass $M_{\rm onset}\geq 0.243~M_\odot$. Moreover, the excluded region of model parameters resulting in a maximum gravitational mass of static stars below $M_{\rm max}<2.01~M_{\odot}$, as found by~\citet{Gartlein:2023vif}, is indicated by a labelled dashed line at the bottom of both panels. The black dashed contour lines indicate the maximum gravitational mass of hybrid stars corresponding to those values of $\eta_V$, $\eta_D$. The cyan hatched region represents the limit for non-axisymmetric instabilities, determined by the criterion $T/W>$ 0.08, derived for the Black Widow Pulsar J0952$-$0607 with a mass of 2.35$~M_\odot$~\cite{Romani:2022jhd}. Particularly, the cyan hatched region depicts the excluded region where a star of 2.35$~M_\odot$ would have $T/W>$ 0.08. 


\section{Conclusions and perspectives}
\label{concl}

We have studied equilibrium configurations of rapidly ro\-ta\-ting hybrid stars, focusing on how the phase transition between hadronic and quark matter within their interiors modifies the limits of stability at high spin frequencies. Our analysis was based on a hybrid EoS with a hadronic phase of hypernuclear matter and a color-superconducting quark matter phase described by an RDF approach. By syste\-ma\-ti\-ca\-lly varying vector and diquark couplings in quark matter, we have explored the effects on the onset of the deconfinement and the maximum mass, ensuring consistency with astrophysical constraints, including the three NICER mass-radius measurements~\cite{Miller:2019cac,Riley:2019yda,Miller:2021qha,Riley:2021pdl,Choudhury:2024xbk}, and the tidal deformability measurement from the binary NS merger GW170817~\cite{LIGOScientific:2018cki}.

A novel aspect of this study is the investigation of rotating hybrid stars undergoing an early deconfinement phase transition, along with its previously unexamined impact on the empirical relation between the Kepler frequency, gravitational mass, and radius of non-rotating NSs.

Our results demonstrate the significant impact of increasing rotational frequencies on key parameters of compact stars, including the gravitational mass, energy density profile, and angular momentum. We have illustrated how rotation affects the structure of hybrid stars, showing that the onset of the deconfinement phase transition shifts to higher masses as the spin frequency increases. This effect is particularly pronounced for configurations with late phase transitions, as evidenced by our analysis of different diquark coupling values. Moreover, we demonstrated that stars characterized by the higher values of the diquark coupling $\eta_D$ can generally achieve higher rotation frequency or Kepler frequencies $f_K$, suggesting that more compact objects can sustain faster rotation.

A crucial aspect of our work is the revision of the empirical relation between the Kepler frequency, gravitational mass, and radius of non-rotating NSs. This relation was initially established under the assumption that all NSs, up to the heaviest, are either purely hadronic or quark stars. We proposed a parameterization for the $C$ factor that accounts for various scenarios of deconfinement onset while reproducing the two limiting cases of purely hadronic and quark EoSs. The $C$ factor is expressed as a function of the quark onset mass, $ M_{\rm onset}$, in the range between 0.5~$M_{\odot}$ and the maximum mass. By considering the currently fastest spinning pulsar, PSR J1748-2446ad, with its remarkable spin frequency of 716 Hz~\cite{Hessels:2006ze}, we have revised a lower bound on the mass-radius relation of compact stars. The concept of minimum mass is as important as the upper limit providing the lowest stable star configurations. The lower limit is especially interesting in the context of the lightest HESS J1731-347 compact object~\cite{Suwa:2018uni,Doroshenko2022,Sagun:2023rzp}. Our analysis reveals how the deconfinement phase transition in an NS core mo\-di\-fies the lower limit on the mass and radius, consequently affecting the constraints on the dense matter EoS. 
Notably, our findings place an upper limit on the radii of the 1.4 solar mass NS $R_{1.4} \le 14.90~{\rm km}$ and of the 0.7 solar mass NS $R_{0.7}<11.49$ km having important implications for the properties of dense matter at extreme conditions.

By analyzing the increase of the angular velocity as a function of the gravitational mass for hybrid stars with the different phase transition onset mass and quark matter properties governed by the $\eta_{V}$ and $\eta_{D}$ parameters we can map the region where the stars are hadronic or hybrid. These results are matched with observational data of the fastest-spinning MSPs with available mass measurements. We demonstrated the possibility of probing the interior composition of MSPs by tracing the evolutionary paths of unmagnetized stars with the same mass. We show that while spinning down with time, MSPs with hadronic matter in their interior could reach the conditions for the onset of the deconfinement phase transition and become hybrid stars. Similarly, fastly spinning heavy hybrid stars at their maximum mass limit could cease to exist due to reaching the instability line upon spinning down (see the red curve in the lower panel in Fig.~\ref{fig7}).

Although the mass-shedding and the quasi-radial oscillation limits define the boundaries for stationary rotating stars to exist, we show that the sta\-bi\-li\-ty condition against non-axisymmetric instabilities put an additional constraint narrowing down the region of admissible $\eta_{V}- \eta_{D}$ parameter values for hybrid star EoS. Particularly, the onset of the gravitational radiation-driven instability limits the higher frequencies for massive stars. The rotational and gravitational energy ratio $T/W\leq $0.08 defines the stable configurations. Thus, the observation of the rapidly-rotating Black Widow Pulsar J0952$-$0607 with the gravitational mass of 2.35$\pm$0.17 $M_{\odot}$~\cite{Romani:2022jhd} existing in the non-axisymmetric stability window provides a useful probe of the quark matter properties and the deconfinement onset, excluding $\eta_{V} \lesssim$ 0.27. This limit yields a new constraint on the microphysical parameters of the model and quark matter properties, which has a big implication for the dense matter EoS. 

As it is shown in Figs.~\ref{fig7} and ~\ref{fig8}, the fastest-rotating MSPs with the measured mass demonstrate a clear population clustering around 1.3-1.8$M_{\odot}$ and the frequency (angular velocity) of 210-340 Hz (1300-2140 Hz). Moreover, the detection of the two fastest spinning objects, i.e., J1748+2446ad and 4U 1820–30 (J1820-30A), with similar values of the spin frequency 716.36 Hz and 716 Hz, makes us think we might have approached the cutoff limit for NSs. Pulsar clustering and no pulsars found rotating faster than 716 Hz were suggested as a consequence of the deconfinement phase transition in the star interior and magnetic field evolution~\cite{Poghosyan:2000mr}. On the upper panels of Figs.~\ref{fig7} and \ref{fig8} we illustrate the spin evolution for the hybrid star of 1.2$M_{\odot}$ for different values of the magnetic field that explains well the population clustering due to the waiting time in a certain frequency region and the spin frequency cut off. This upper limit is also supported by the Bayesian analysis of observable accretion- and nuclear-powered pulsars that give $f_{\rm max}\sim$ 730 Hz~\cite{Jaisawal_2024}.

A striking result in favour of the hybrid NS case with color superconducting quark matter that we explored in this work is the fact that with the favoured parameter set, we can describe the Black Widow Pulsar J0952-0607 with its high mass and spin, while this star is out of the $1\sigma$ range for an explanation by the purely hadronic EoS.

Looking ahead, the potential detection of a pulsar with a spin frequency of 1000 Hz or higher would be a game-changer. Such a discovery would enable robust discri\-mi\-na\-tion between different EoS, significantly contributing to our understanding of NS interior composition and accretion. Our findings show that a pulsar with a spin frequency of 1000 Hz, would impose the upper limit $R_{1.4}\leq$11.90~km considering the hybrid EoS and $R_{1.4}\leq$11.86~km for the hadronic EoS. This underscores the importance of con\-ti\-nued observational efforts and theoretical modeling in this field.

It is worth noting that out of over 3,000 known pulsars, 560 are classified as MSPs\footnote{\url{https://www.atnf.csiro.au/research/pulsar/psrcat/}}. The field of pulsar astro\-no\-my is poised for significant advancements with upcoming and ongoing radio surveys. Initiatives such as the Five-hundred-meter Aperture Spherical radio Telescope (FAST)~\cite{Nan:2011um}, MeerKAT~\cite{Padmanabh:2023vma}, and the Square Kilometre Array (SKA)~\cite{Watts:2014tja}, are expected to substantially increase the number of detected pulsars. These surveys promise to provide crucial data on pulsar masses and frequencies, which will be instrumental in further constraining the EoS of strongly interacting matter.

On the other hand, future X-ray telescopes such as Advanced Telescope for High ENergy Astrophysics (Athena)~\cite{Barcons:2012zb}, STROBE-X~\cite{STROBE-XScienceWorkingGroup:2019cyd}, and enhanced X-ray Timing and Polarimetry (eXTP)~\cite{eXTP:2018anb} should be able to provide mass and radius measurement for even fainter pulsars.

Our study also highlights the complexity of modeling rapidly rotating compact stars. The framework we have employed, based on the RNS code~\cite{Paschalidis:2016vmz}, allows for the accurate calculation of stellar properties in the presence of strong gravitational fields and rapid rotation. This approach is instrumental in understanding the behavior of MSPs and their implications for fundamental physics.

In conclusion, our study underscores the importance of considering rotation and phase transitions in modeling compact stars. The interplay between these factors significantly influences our interpretation of observational data and understanding of the extreme state of matter within NSs. As we await more precise measurements from future surveys, the field of NS physics remains a vibrant area of research.

\subsection*{Acknowledgments}

The authors thank E. Giangrandi, H. Grigorian and L. Scurto for the useful comments. C.G. acknowledges the Funda\c c\~ao para a Ci\^encia e Tecnologia (FCT), Portugal, through the IDPASC PT-CERN program and support of the Center for Astrophysics and Gravitation (CENTRA/IST/ULisboa) through grant project No. UIDB/00099/2020 and grant No. PTDC/FIS-AST/28920/2017. V.S. gratefully acknowledges support from the UKRI-funded ``The next-generation gravitational-wave observatory network'' project (Grant No. ST/Y004248/1) and support from FCT – Fundação para a Ciência e a Tecnologia within the projects UIDP/\-04564/\-2020 and UIDB/\-04564/\-2020, respectively, with DOI identifiers 10.54499/UIDP/04564/2020 and 10.54499/UIDB/04564/2020. The work of O.I. was supported by the program Excellence Initiative--Research University of the University of Wrocław of the Ministry of Education and Science. D.B. and O.I. acknowledge the support from the Polish National Science Center under grant No. 2021/43/P/ST2/03319. I.L. would like to express his gratitude to the Funda\c c\~ao para a Ci\^encia e Tecnologia (FCT), Portugal, for providing financial support to the Center for Astrophysics and Gravitation (CENTRA/IST/ULisboa) through Grant Project No. UIDB/00099/2020 and Grant No. PTDC/FIS-AST/28920/2017. This work was produced with the support of INCD and funded by FCT I.P. under Advanced Computing Project 2023.10526.CPCA.A2 with DOI identifier 10.54499/2023.10526.CPCA.A2.

\subsection*{Data Availability Statement} 
The data that support the findings of this study are available from the corresponding author upon reasonable request.

\begin{appendix}
    
\section{ABPR parametrization of the RDF EOS}
\label{app:A}

We present the effective number of degrees of freedom, pairing gap, and bag constant of the ABPR parameterization of the RDF EoS from Ref.~\cite{Ivanytskyi:2022bjc}
\begin{eqnarray}
\label{eq_appa_I}    
A_4&=&a_1+b_1\eta_{V}+c_1\eta_{V}^{2}+\left(d_1+\frac{e_1}{\eta_{V}}\right)\eta_D,\\
\label{eq_appa_II}
\Delta&=&(a_2+b_2\eta_V+c_2\eta_V^2)\sqrt{d_2+e_2\eta_V+\eta_D},\\
\label{eq_appa_III}  
B&=&a_3+b_3\eta_V+c_3\eta_{V}^{2}+d_3\eta_{D}+e_3\eta_{D}^{2}.
\end{eqnarray}
The 15 parameters $a_i,b_i,c_i,d_i,e_i$ with $i=1,2,3$ were extracted from the fit of the original RDF EoS.
Their values are presented in Table~\ref{table1}.
For more details about the fit see~\cite{Gartlein:2023vif}.
\begin{table}[h]
\begin{tabular}{|c|c|c|c|c|c|c|c|c|c|c|c|}
\hline
i &       units      & $a_i$ & $b_i$ & $c_i$ & $d_i$ & $e_i$  \\ \hline
1 &      [~]            & 0.757 &-1.955 & 1.799 &-0.063 & 0.046  \\ \hline
2 &       [MeV]      & 300.7 & 8.534 &-308.2 &-0.235 & 1.458  \\ \hline
3 & $\rm[MeV/fm^3]$  & 72.018& 170.8 &-241.0 & 512.7 &-626.6  \\ \hline
\end{tabular}
\caption{Values of the parameters of Eqs. (\ref{eq_appa_I}-\ref{eq_appa_III}).}
\label{table1}
\end{table}

\section{Parameters of millisecond pulsars with known masses}
\label{app:B}

Table~\ref{table2} includes observational data of the fastest spinning MSPs, with spin frequency f$>$200 Hz, for which the mass measurements are available\footnote{The complete pulsar catalog with mass measurements can be accessed at \url{https://www3.mpifr-bonn.mpg.de/staff/pfreire/NS_masses.html}}. The pulsars are shown in Figs.~\ref{fig7}-\ref{fig8}. Number 2 corresponds to the recently reported observation by the NICER telescope of the second fastest spinning pulsar 4U 1820–30 (J1820$-$30A) with 716 Hz frequency~\cite{Jaisawal_2024}. For objects with several independent measurements, i.e., points 16,18,30,31, we plot them with a gray color.

\begin{table}[!t]
\begin{tabular}{|c|c|c|c|c|}
\hline
Number &       Name      & f [Hz] & $M/M_{\odot}$ & Ref.  \\ \hline
1 &    J1748$+$2446ad & 716.36 & $<$2 & \cite{Hessels:2006ze}\\
\hline
2  & J1820$-$30A & 716.00 & 1.58$\pm$0.06 & \cite{Jaisawal_2024,Guver:2010td}\\
\hline
3 &    J0952$-$0607 & 709.21 & 2.35$\pm$0.17 
& \cite{Romani:2022jhd}\\
\hline
4 &    B1957$+$20 & 622.15  & 1.81$\pm$0.07 &\cite{Clark:2023owb}\\
\hline
5 &  J1810$+$1744 & 602.41 & 2.13$\pm$0.04 
& \cite{Romani:2021xmb}\\
\hline
6 &   J1023$+$0038 & 592.42  &  1.82$\pm$0.17  & \cite{Strader:2018qbi}\\ 
\hline
7 &   J0955$-$6150 & 500.160 & 1.71$\pm$0.03  & \cite{Serylak:2022kna}\\ \hline
8 &   J1903$+$0327 & 465.135 & 1.667$\pm$0.021  & \cite{Freire:2010tf}\\ \hline
9 &   J1748$-$2446ao & 439.681 & $<$2.23  & \cite{Padmanabh:2024bsz}\\ \hline
10 &    J2043$+$1711 &  420.189  & $1.38^{+0.12}_{-0.13}$  & \cite{NANOGrav:2017wvv}\\ 
\hline
11 &    J1311$+$3430 & 390.57 & 1.8–2.7 & \cite{Romani:2015gaa}\\
\hline
12 &   J1125$-$6014 & 380.173  & $1.68^{+0.17}_{-0.15}$ & \cite{Shamohammadi:2022ttx}\\
\hline
13 &   J0337$+$1715 & 365.95 &1.4378$\pm$0.0013  & \cite{Ransom:2014xla}\\
\hline
14 &    J0740$+$6620 & 346.532 & 2.08$\pm$0.07 & \cite{Fonseca:2021wxt} \\ \hline
15 &  J1748$-$2446am & 340.853 & $<$1.7 & \cite{Andersen:2018nsx} \\ \hline
   &                &          & 1.48$\pm$0.03   &\cite{NANOGrav:2017wvv} \\ 
16 &   J1909$-$3744 & 339.316 & 1.492$\pm$0.014  &\cite{Liu:2020hkx} \\ 
   &               &         & 1.45$\pm$0.03     &\cite{Shamohammadi:2022ttx} \\ 
\hline
17 &   J1012$-$4235 & 322.462 & $1.44^{+0.013}_{-0.012}$ & \cite{Gautam:2023ref} \\ \hline
18 &  J1614$-$2230 &  317.379 & 1.908$\pm$0.016  & \cite{NANOGrav:2017wvv} \\
   &   &  & 1.94$\pm$0.03   & \cite{Shamohammadi:2022ttx} \\ \hline
19 &  J1946$+$3417 & 315.444 & 1.828$\pm$0.022  & \cite{Barr:2016vxv}\\ \hline
20 &  J0024$-$7204H & 311.493 &  $<$1.49  & \cite{Freire:2017mgu}\\ \hline
21 &  J1910$-$5958A & 306.167 & 1.55$\pm$0.07  & \cite{Corongiu:2023gft}\\ \hline
22 &  J0751$+$1807 & 287.458 & 1.64$\pm$0.15  & \cite{EPTA:2016ndq}\\ \hline
23 &  J1933$-$6211 &  282.212 & $1.4^{+0.3}_{-0.2}$  & \cite{Geyer:2023vng}\\ \hline
24 &    J2234$+$0611 & 279.597  & $1.353^{+0.014}_{-0.017}$ & \cite{Stovall:2018rvy}\\ \hline
25 &  J1748$-$2446ap &  267.044  & $1.700^{+0.015}_{-0.045}$ & \cite{Padmanabh:2024bsz}\\ \hline
26 &  J1824$-$2452C &  240.484  & $<$1.367 & \cite{Begin_2006}\\ \hline
27 &  J1807$-$2500B &  238.881  & 1.3655$\pm$0.0021 & \cite{Lynch:2011aa}\\ \hline
28 &  J1950$+$2414 & 232.300  & 1.496$\pm$0.023 & \cite{Zhu:2019oax}\\ \hline
29 &  J1748$-$2446au & 219.866 & $<$1.53 & \cite{Padmanabh:2024bsz}\\ \hline
30 & J1713$+$0747 & 218.811 & $1.33^{+0.09}_{-0.08}$ & \cite{EPTA:2016ndq}\\ 
   &             &          & 1.35$\pm$0.07 & \cite{NANOGrav:2017wvv}\\ \hline
31 & J0030$+$0451 & 205.53 & $1.44^{+0.15}_{-0.14}$ & \cite{Miller:2019cac}\\ 
  &  &  & $1.34^{+0.15}_{-0.16}$ & \cite{Riley:2019yda}\\ \hline
32 & J0514$-$4002A& 200.378 & $1.25^{+0.05}_{-0.06}$
& \cite{Ridolfi:2019wgs}\\
\hline
\end{tabular}
\caption{Data of the fastest MSPs with spin frequency f$>$200 Hz and available mass measurements. The columns include the pulsar's number in Figs.~\ref{fig7}-\ref{fig8}, the number in the catalog, spin frequency, gravitational mass, and the corresponding references. All mass measurements are listed within one standard deviation, while the J1903+0327 observed data are listed within three standard deviations. 
}
\label{table2}
\end{table}

\section{Spin evolution by mass accretion}
\label{app:C}

We present the underlying model for the evolutionary path of accreting and magnetized MSPs based on~\cite{Poghosyan:2000mr}. The differential equation we solve is given by
\begin{eqnarray}
\label{eq_appa_IV}    
\frac{d\Omega}{dt}=\frac{K_{\rm ext}-K_{\rm int}}{I(\Omega,N)+\Omega(\partial I(N,\Omega)/\partial \Omega)_N}.
\end{eqnarray}
The evolution of the angular velocity of a star depends on the external and internal torque, $K_{\rm ext}$ and $K_{\rm int}$, the accretion rate, and the decay of the magnetic field. The considered expression for the exponential decay of the magnetic field of the accretors and the time dependence of the baryon number for the constant accreting rate
$\dot{N}$ are given by
\begin{eqnarray}
\label{eq_appa_V}    
B(t)&=&[B(0)-B_{\infty}] \exp{(-t/\tau_B)}+B_{\infty}, \\
N(t)&=&N(0)+t\dot{N},
\end{eqnarray}
where $N(0)$ and $B(0)$ denote the baryon number initial values and the magnetic field strength, respectively. $\tau_B$ is the typical decay time of the magnetic field. The remnant magnetic field is chosen to be $B_{\infty}=10^{8}$ G. Additionally, the dependence on the EoS enters through the moment of inertia and its derivatives with respect to $N$ and $\Omega$. These terms enter the equation in the $K_{\rm int}$ as well as the denominator. For more details about the model see 
Ref.~\cite{Poghosyan:2000mr}.

\end{appendix}

%
\bibliography{references}

\begin{thebibliography}{130}%
\makeatletter
\providecommand \@ifxundefined [1]{%
 \@ifx{#1\undefined}
}%
\providecommand \@ifnum [1]{%
 \ifnum #1\expandafter \@firstoftwo
 \else \expandafter \@secondoftwo
 \fi
}%
\providecommand \@ifx [1]{%
 \ifx #1\expandafter \@firstoftwo
 \else \expandafter \@secondoftwo
 \fi
}%
\providecommand \natexlab [1]{#1}%
\providecommand \enquote  [1]{``#1''}%
\providecommand \bibnamefont  [1]{#1}%
\providecommand \bibfnamefont [1]{#1}%
\providecommand \citenamefont [1]{#1}%
\providecommand \href@noop [0]{\@secondoftwo}%
\providecommand \href [0]{\begingroup \@sanitize@url \@href}%
\providecommand \@href[1]{\@@startlink{#1}\@@href}%
\providecommand \@@href[1]{\endgroup#1\@@endlink}%
\providecommand \@sanitize@url [0]{\catcode `\\12\catcode `\$12\catcode `\&12\catcode `\#12\catcode `\^12\catcode `\_12\catcode `\%12\relax}%
\providecommand \@@startlink[1]{}%
\providecommand \@@endlink[0]{}%
\providecommand \url  [0]{\begingroup\@sanitize@url \@url }%
\providecommand \@url [1]{\endgroup\@href {#1}{\urlprefix }}%
\providecommand \urlprefix  [0]{URL }%
\providecommand \Eprint [0]{\href }%
\providecommand \doibase [0]{http://dx.doi.org/}%
\providecommand \selectlanguage [0]{\@gobble}%
\providecommand \bibinfo  [0]{\@secondoftwo}%
\providecommand \bibfield  [0]{\@secondoftwo}%
\providecommand \translation [1]{[#1]}%
\providecommand \BibitemOpen [0]{}%
\providecommand \bibitemStop [0]{}%
\providecommand \bibitemNoStop [0]{.\EOS\space}%
\providecommand \EOS [0]{\spacefactor3000\relax}%
\providecommand \BibitemShut  [1]{\csname bibitem#1\endcsname}%
\let\auto@bib@innerbib\@empty
\bibitem [{\citenamefont {Bhattacharya}\ and\ \citenamefont {van~den Heuvel}(1991)}]{1991PhR...203....1B}%
  \BibitemOpen
  \bibfield  {author} {\bibinfo {author} {\bibfnamefont {D.}~\bibnamefont {Bhattacharya}}\ and\ \bibinfo {author} {\bibfnamefont {E.~P.~J.}\ \bibnamefont {van~den Heuvel}},\ }\href {\doibase 10.1016/0370-1573(91)90064-S} {\bibfield  {journal} {\bibinfo  {journal} {Phys. Rep.}\ }\textbf {\bibinfo {volume} {203}},\ \bibinfo {pages} {1} (\bibinfo {year} {1991})}\BibitemShut {NoStop}%
\bibitem [{\citenamefont {Wijnands}\ and\ \citenamefont {van~der Klis}(1998)}]{1998Natur.394..344W}%
  \BibitemOpen
  \bibfield  {author} {\bibinfo {author} {\bibfnamefont {R.}~\bibnamefont {Wijnands}}\ and\ \bibinfo {author} {\bibfnamefont {M.}~\bibnamefont {van~der Klis}},\ }\href {\doibase 10.1038/28557} {\bibfield  {journal} {\bibinfo  {journal} {Nature}\ }\textbf {\bibinfo {volume} {394}},\ \bibinfo {pages} {344} (\bibinfo {year} {1998})}\BibitemShut {NoStop}%
\bibitem [{\citenamefont {Guillot}\ \emph {et~al.}(2019)\citenamefont {Guillot} \emph {et~al.}}]{Guillot:2019vqp}%
  \BibitemOpen
  \bibfield  {author} {\bibinfo {author} {\bibfnamefont {S.}~\bibnamefont {Guillot}} \emph {et~al.},\ }\href {\doibase 10.3847/2041-8213/ab511b} {\bibfield  {journal} {\bibinfo  {journal} {Astrophys. J. Lett.}\ }\textbf {\bibinfo {volume} {887}},\ \bibinfo {pages} {L27} (\bibinfo {year} {2019})},\ \Eprint {http://arxiv.org/abs/1912.05708} {arXiv:1912.05708 [astro-ph.HE]} \BibitemShut {NoStop}%
\bibitem [{\citenamefont {Romani}\ \emph {et~al.}(2022)\citenamefont {Romani}, \citenamefont {Kandel}, \citenamefont {Filippenko}, \citenamefont {Brink},\ and\ \citenamefont {Zheng}}]{Romani:2022jhd}%
  \BibitemOpen
  \bibfield  {author} {\bibinfo {author} {\bibfnamefont {R.~W.}\ \bibnamefont {Romani}}, \bibinfo {author} {\bibfnamefont {D.}~\bibnamefont {Kandel}}, \bibinfo {author} {\bibfnamefont {A.~V.}\ \bibnamefont {Filippenko}}, \bibinfo {author} {\bibfnamefont {T.~G.}\ \bibnamefont {Brink}}, \ and\ \bibinfo {author} {\bibfnamefont {W.}~\bibnamefont {Zheng}},\ }\href {\doibase 10.3847/2041-8213/ac8007} {\bibfield  {journal} {\bibinfo  {journal} {Astrophys. J. Lett.}\ }\textbf {\bibinfo {volume} {934}},\ \bibinfo {pages} {L17} (\bibinfo {year} {2022})},\ \Eprint {http://arxiv.org/abs/2207.05124} {arXiv:2207.05124 [astro-ph.HE]} \BibitemShut {NoStop}%
\bibitem [{\citenamefont {Alford}\ \emph {et~al.}(2007)\citenamefont {Alford}, \citenamefont {Blaschke}, \citenamefont {Drago}, \citenamefont {Klahn}, \citenamefont {Pagliara},\ and\ \citenamefont {Schaffner-Bielich}}]{Alford:2006vz}%
  \BibitemOpen
  \bibfield  {author} {\bibinfo {author} {\bibfnamefont {M.}~\bibnamefont {Alford}}, \bibinfo {author} {\bibfnamefont {D.}~\bibnamefont {Blaschke}}, \bibinfo {author} {\bibfnamefont {A.}~\bibnamefont {Drago}}, \bibinfo {author} {\bibfnamefont {T.}~\bibnamefont {Klahn}}, \bibinfo {author} {\bibfnamefont {G.}~\bibnamefont {Pagliara}}, \ and\ \bibinfo {author} {\bibfnamefont {J.}~\bibnamefont {Schaffner-Bielich}},\ }\href {\doibase 10.1038/nature05582} {\bibfield  {journal} {\bibinfo  {journal} {Nature}\ }\textbf {\bibinfo {volume} {445}},\ \bibinfo {pages} {E7} (\bibinfo {year} {2007})},\ \Eprint {http://arxiv.org/abs/astro-ph/0606524} {arXiv:astro-ph/0606524} \BibitemShut {NoStop}%
\bibitem [{\citenamefont {Koberlein}(2024)}]{Koberlein:2024}%
  \BibitemOpen
  \bibfield  {author} {\bibinfo {author} {\bibfnamefont {B.}~\bibnamefont {Koberlein}},\ }\href@noop {} {\enquote {\bibinfo {title} {Do the fastest-spinning pulsars contain quark matter?}}\ }\bibinfo {howpublished} {{https://phys.org/news/2024-12-fastest-pulsars-quark.html}} (\bibinfo {year} {2024}),\ \bibinfo {note} {accessed: 2025-02-13}\BibitemShut {NoStop}%
\bibitem [{\citenamefont {Rather}\ \emph {et~al.}(2021)\citenamefont {Rather}, \citenamefont {Rahaman}, \citenamefont {Imran}, \citenamefont {Das}, \citenamefont {Usmani},\ and\ \citenamefont {Patra}}]{Rather:2021yxo}%
  \BibitemOpen
  \bibfield  {author} {\bibinfo {author} {\bibfnamefont {I.~A.}\ \bibnamefont {Rather}}, \bibinfo {author} {\bibfnamefont {U.}~\bibnamefont {Rahaman}}, \bibinfo {author} {\bibfnamefont {M.}~\bibnamefont {Imran}}, \bibinfo {author} {\bibfnamefont {H.~C.}\ \bibnamefont {Das}}, \bibinfo {author} {\bibfnamefont {A.~A.}\ \bibnamefont {Usmani}}, \ and\ \bibinfo {author} {\bibfnamefont {S.~K.}\ \bibnamefont {Patra}},\ }\href {\doibase 10.1103/PhysRevC.103.055814} {\bibfield  {journal} {\bibinfo  {journal} {Phys. Rev. C}\ }\textbf {\bibinfo {volume} {103}},\ \bibinfo {pages} {055814} (\bibinfo {year} {2021})},\ \Eprint {http://arxiv.org/abs/2102.04067} {arXiv:2102.04067 [nucl-th]} \BibitemShut {NoStop}%
\bibitem [{\citenamefont {Espino}\ and\ \citenamefont {Paschalidis}(2022)}]{Espino:2021adh}%
  \BibitemOpen
  \bibfield  {author} {\bibinfo {author} {\bibfnamefont {P.~L.}\ \bibnamefont {Espino}}\ and\ \bibinfo {author} {\bibfnamefont {V.}~\bibnamefont {Paschalidis}},\ }\href {\doibase 10.1103/PhysRevD.105.043014} {\bibfield  {journal} {\bibinfo  {journal} {Phys. Rev. D}\ }\textbf {\bibinfo {volume} {105}},\ \bibinfo {pages} {043014} (\bibinfo {year} {2022})},\ \Eprint {http://arxiv.org/abs/2105.05269} {arXiv:2105.05269 [astro-ph.HE]} \BibitemShut {NoStop}%
\bibitem [{\citenamefont {Sen}\ and\ \citenamefont {Chaudhuri}(2022)}]{Sen:2022qol}%
  \BibitemOpen
  \bibfield  {author} {\bibinfo {author} {\bibfnamefont {D.}~\bibnamefont {Sen}}\ and\ \bibinfo {author} {\bibfnamefont {G.}~\bibnamefont {Chaudhuri}},\ }\href {\doibase 10.1088/1361-6471/ac6f14} {\bibfield  {journal} {\bibinfo  {journal} {J. Phys. G}\ }\textbf {\bibinfo {volume} {49}},\ \bibinfo {pages} {075201} (\bibinfo {year} {2022})},\ \Eprint {http://arxiv.org/abs/2205.00338} {arXiv:2205.00338 [nucl-th]} \BibitemShut {NoStop}%
\bibitem [{\citenamefont {Tsaloukidis}\ \emph {et~al.}(2023)\citenamefont {Tsaloukidis}, \citenamefont {Koliogiannis}, \citenamefont {Kanakis-Pegios},\ and\ \citenamefont {Moustakidis}}]{Tsaloukidis:2022rus}%
  \BibitemOpen
  \bibfield  {author} {\bibinfo {author} {\bibfnamefont {L.}~\bibnamefont {Tsaloukidis}}, \bibinfo {author} {\bibfnamefont {P.~S.}\ \bibnamefont {Koliogiannis}}, \bibinfo {author} {\bibfnamefont {A.}~\bibnamefont {Kanakis-Pegios}}, \ and\ \bibinfo {author} {\bibfnamefont {C.~C.}\ \bibnamefont {Moustakidis}},\ }\href {\doibase 10.1103/PhysRevD.107.023012} {\bibfield  {journal} {\bibinfo  {journal} {Phys. Rev. D}\ }\textbf {\bibinfo {volume} {107}},\ \bibinfo {pages} {023012} (\bibinfo {year} {2023})},\ \Eprint {http://arxiv.org/abs/2210.15644} {arXiv:2210.15644 [astro-ph.HE]} \BibitemShut {NoStop}%
\bibitem [{\citenamefont {Moreno}\ \emph {et~al.}(2023)\citenamefont {Moreno}, \citenamefont {Llanes-Estrada},\ and\ \citenamefont {Lope-Oter}}]{Moreno:2023xez}%
  \BibitemOpen
  \bibfield  {author} {\bibinfo {author} {\bibfnamefont {P.~N.}\ \bibnamefont {Moreno}}, \bibinfo {author} {\bibfnamefont {F.~J.}\ \bibnamefont {Llanes-Estrada}}, \ and\ \bibinfo {author} {\bibfnamefont {E.}~\bibnamefont {Lope-Oter}},\ }\href {\doibase 10.1016/j.aop.2023.169487} {\bibfield  {journal} {\bibinfo  {journal} {Annals Phys.}\ }\textbf {\bibinfo {volume} {459}},\ \bibinfo {pages} {169487} (\bibinfo {year} {2023})},\ \Eprint {http://arxiv.org/abs/2307.15366} {arXiv:2307.15366 [nucl-th]} \BibitemShut {NoStop}%
\bibitem [{\citenamefont {Kramer}\ \emph {et~al.}(2006)\citenamefont {Kramer} \emph {et~al.}}]{Kramer:2006nb}%
  \BibitemOpen
  \bibfield  {author} {\bibinfo {author} {\bibfnamefont {M.}~\bibnamefont {Kramer}} \emph {et~al.},\ }\href {\doibase 10.1126/science.1132305} {\bibfield  {journal} {\bibinfo  {journal} {Science}\ }\textbf {\bibinfo {volume} {314}},\ \bibinfo {pages} {97} (\bibinfo {year} {2006})},\ \Eprint {http://arxiv.org/abs/astro-ph/0609417} {arXiv:astro-ph/0609417} \BibitemShut {NoStop}%
\bibitem [{\citenamefont {Bejger}\ \emph {et~al.}(2005)\citenamefont {Bejger}, \citenamefont {Bulik},\ and\ \citenamefont {Haensel}}]{Bejger:2005jy}%
  \BibitemOpen
  \bibfield  {author} {\bibinfo {author} {\bibfnamefont {M.}~\bibnamefont {Bejger}}, \bibinfo {author} {\bibfnamefont {T.}~\bibnamefont {Bulik}}, \ and\ \bibinfo {author} {\bibfnamefont {P.}~\bibnamefont {Haensel}},\ }\href {\doibase 10.1111/j.1365-2966.2005.09575.x} {\bibfield  {journal} {\bibinfo  {journal} {Mon. Not. Roy. Astron. Soc.}\ }\textbf {\bibinfo {volume} {364}},\ \bibinfo {pages} {635} (\bibinfo {year} {2005})},\ \Eprint {http://arxiv.org/abs/astro-ph/0508105} {arXiv:astro-ph/0508105} \BibitemShut {NoStop}%
\bibitem [{\citenamefont {Kramer}\ and\ \citenamefont {Wex}(2009)}]{Kramer:2009zza}%
  \BibitemOpen
  \bibfield  {author} {\bibinfo {author} {\bibfnamefont {M.}~\bibnamefont {Kramer}}\ and\ \bibinfo {author} {\bibfnamefont {N.}~\bibnamefont {Wex}},\ }\href {\doibase 10.1088/0264-9381/26/7/073001} {\bibfield  {journal} {\bibinfo  {journal} {Class. Quant. Grav.}\ }\textbf {\bibinfo {volume} {26}},\ \bibinfo {pages} {073001} (\bibinfo {year} {2009})}\BibitemShut {NoStop}%
\bibitem [{\citenamefont {Raaijmakers}\ \emph {et~al.}(2019)\citenamefont {Raaijmakers} \emph {et~al.}}]{Raaijmakers:2019qny}%
  \BibitemOpen
  \bibfield  {author} {\bibinfo {author} {\bibfnamefont {G.}~\bibnamefont {Raaijmakers}} \emph {et~al.},\ }\href {\doibase 10.3847/2041-8213/ab451a} {\bibfield  {journal} {\bibinfo  {journal} {Astrophys. J. Lett.}\ }\textbf {\bibinfo {volume} {887}},\ \bibinfo {pages} {L22} (\bibinfo {year} {2019})},\ \Eprint {http://arxiv.org/abs/1912.05703} {arXiv:1912.05703 [astro-ph.HE]} \BibitemShut {NoStop}%
\bibitem [{\citenamefont {Romani}\ \emph {et~al.}(2021)\citenamefont {Romani}, \citenamefont {Kandel}, \citenamefont {Filippenko}, \citenamefont {Brink},\ and\ \citenamefont {Zheng}}]{Romani:2021xmb}%
  \BibitemOpen
  \bibfield  {author} {\bibinfo {author} {\bibfnamefont {R.~W.}\ \bibnamefont {Romani}}, \bibinfo {author} {\bibfnamefont {D.}~\bibnamefont {Kandel}}, \bibinfo {author} {\bibfnamefont {A.~V.}\ \bibnamefont {Filippenko}}, \bibinfo {author} {\bibfnamefont {T.~G.}\ \bibnamefont {Brink}}, \ and\ \bibinfo {author} {\bibfnamefont {W.}~\bibnamefont {Zheng}},\ }\href {\doibase 10.3847/2041-8213/abe2b4} {\bibfield  {journal} {\bibinfo  {journal} {Astrophys. J. Lett.}\ }\textbf {\bibinfo {volume} {908}},\ \bibinfo {pages} {L46} (\bibinfo {year} {2021})},\ \Eprint {http://arxiv.org/abs/2101.09822} {arXiv:2101.09822 [astro-ph.HE]} \BibitemShut {NoStop}%
\bibitem [{\citenamefont {Hessels}\ \emph {et~al.}(2006)\citenamefont {Hessels}, \citenamefont {Ransom}, \citenamefont {Stairs}, \citenamefont {Freire}, \citenamefont {Kaspi},\ and\ \citenamefont {Camilo}}]{Hessels:2006ze}%
  \BibitemOpen
  \bibfield  {author} {\bibinfo {author} {\bibfnamefont {J.~W.~T.}\ \bibnamefont {Hessels}}, \bibinfo {author} {\bibfnamefont {S.~M.}\ \bibnamefont {Ransom}}, \bibinfo {author} {\bibfnamefont {I.~H.}\ \bibnamefont {Stairs}}, \bibinfo {author} {\bibfnamefont {P.~C.~C.}\ \bibnamefont {Freire}}, \bibinfo {author} {\bibfnamefont {V.~M.}\ \bibnamefont {Kaspi}}, \ and\ \bibinfo {author} {\bibfnamefont {F.}~\bibnamefont {Camilo}},\ }\href {\doibase 10.1126/science.1123430} {\bibfield  {journal} {\bibinfo  {journal} {Science}\ }\textbf {\bibinfo {volume} {311}},\ \bibinfo {pages} {1901} (\bibinfo {year} {2006})},\ \Eprint {http://arxiv.org/abs/astro-ph/0601337} {arXiv:astro-ph/0601337} \BibitemShut {NoStop}%
\bibitem [{\citenamefont {Haskell}\ \emph {et~al.}(2018)\citenamefont {Haskell}, \citenamefont {Zdunik}, \citenamefont {Fortin}, \citenamefont {Bejger}, \citenamefont {Wijnands},\ and\ \citenamefont {Patruno}}]{Haskell:2018nlh}%
  \BibitemOpen
  \bibfield  {author} {\bibinfo {author} {\bibfnamefont {B.}~\bibnamefont {Haskell}}, \bibinfo {author} {\bibfnamefont {J.~L.}\ \bibnamefont {Zdunik}}, \bibinfo {author} {\bibfnamefont {M.}~\bibnamefont {Fortin}}, \bibinfo {author} {\bibfnamefont {M.}~\bibnamefont {Bejger}}, \bibinfo {author} {\bibfnamefont {R.}~\bibnamefont {Wijnands}}, \ and\ \bibinfo {author} {\bibfnamefont {A.}~\bibnamefont {Patruno}},\ }\href {\doibase 10.1051/0004-6361/201833521} {\bibfield  {journal} {\bibinfo  {journal} {Astron. Astrophys.}\ }\textbf {\bibinfo {volume} {620}},\ \bibinfo {pages} {A69} (\bibinfo {year} {2018})},\ \Eprint {http://arxiv.org/abs/1805.11277} {arXiv:1805.11277 [astro-ph.HE]} \BibitemShut {NoStop}%
\bibitem [{\citenamefont {Patruno}\ \emph {et~al.}(2017)\citenamefont {Patruno}, \citenamefont {Haskell},\ and\ \citenamefont {Andersson}}]{Patruno:2017oum}%
  \BibitemOpen
  \bibfield  {author} {\bibinfo {author} {\bibfnamefont {A.}~\bibnamefont {Patruno}}, \bibinfo {author} {\bibfnamefont {B.}~\bibnamefont {Haskell}}, \ and\ \bibinfo {author} {\bibfnamefont {N.}~\bibnamefont {Andersson}},\ }\href {\doibase 10.3847/1538-4357/aa927a} {\bibfield  {journal} {\bibinfo  {journal} {Astrophys. J.}\ }\textbf {\bibinfo {volume} {850}},\ \bibinfo {pages} {106} (\bibinfo {year} {2017})},\ \Eprint {http://arxiv.org/abs/1705.07669} {arXiv:1705.07669 [astro-ph.HE]} \BibitemShut {NoStop}%
\bibitem [{\citenamefont {Glendenning}\ and\ \citenamefont {Weber}(2001)}]{Glendenning:2000zz}%
  \BibitemOpen
  \bibfield  {author} {\bibinfo {author} {\bibfnamefont {N.~K.}\ \bibnamefont {Glendenning}}\ and\ \bibinfo {author} {\bibfnamefont {F.}~\bibnamefont {Weber}},\ }\href {\doibase 10.1086/323972} {\bibfield  {journal} {\bibinfo  {journal} {Astrophys. J. Lett.}\ }\textbf {\bibinfo {volume} {559}},\ \bibinfo {pages} {L119} (\bibinfo {year} {2001})},\ \Eprint {http://arxiv.org/abs/astro-ph/0003426} {arXiv:astro-ph/0003426} \BibitemShut {NoStop}%
\bibitem [{\citenamefont {Bejger}\ \emph {et~al.}(2017)\citenamefont {Bejger}, \citenamefont {Blaschke}, \citenamefont {Haensel}, \citenamefont {Zdunik},\ and\ \citenamefont {Fortin}}]{Bejger:2016emu}%
  \BibitemOpen
  \bibfield  {author} {\bibinfo {author} {\bibfnamefont {M.}~\bibnamefont {Bejger}}, \bibinfo {author} {\bibfnamefont {D.}~\bibnamefont {Blaschke}}, \bibinfo {author} {\bibfnamefont {P.}~\bibnamefont {Haensel}}, \bibinfo {author} {\bibfnamefont {J.~L.}\ \bibnamefont {Zdunik}}, \ and\ \bibinfo {author} {\bibfnamefont {M.}~\bibnamefont {Fortin}},\ }\href {\doibase 10.1051/0004-6361/201629580} {\bibfield  {journal} {\bibinfo  {journal} {Astron. Astrophys.}\ }\textbf {\bibinfo {volume} {600}},\ \bibinfo {pages} {A39} (\bibinfo {year} {2017})},\ \Eprint {http://arxiv.org/abs/1608.07049} {arXiv:1608.07049 [astro-ph.HE]} \BibitemShut {NoStop}%
\bibitem [{\citenamefont {Friedman}\ \emph {et~al.}(1986{\natexlab{a}})\citenamefont {Friedman}, \citenamefont {Parker},\ and\ \citenamefont {Ipser}}]{Friedman:1986tx}%
  \BibitemOpen
  \bibfield  {author} {\bibinfo {author} {\bibfnamefont {J.~L.}\ \bibnamefont {Friedman}}, \bibinfo {author} {\bibfnamefont {L.}~\bibnamefont {Parker}}, \ and\ \bibinfo {author} {\bibfnamefont {J.~R.}\ \bibnamefont {Ipser}},\ }\href {\doibase 10.1086/164149} {\bibfield  {journal} {\bibinfo  {journal} {Astrophys. J.}\ }\textbf {\bibinfo {volume} {304}},\ \bibinfo {pages} {115} (\bibinfo {year} {1986}{\natexlab{a}})}\BibitemShut {NoStop}%
\bibitem [{\citenamefont {Komatsu}\ \emph {et~al.}(1989)\citenamefont {Komatsu}, \citenamefont {Eriguchi},\ and\ \citenamefont {Hachisu}}]{Komatsu:1989zz}%
  \BibitemOpen
  \bibfield  {author} {\bibinfo {author} {\bibfnamefont {H.}~\bibnamefont {Komatsu}}, \bibinfo {author} {\bibfnamefont {Y.}~\bibnamefont {Eriguchi}}, \ and\ \bibinfo {author} {\bibfnamefont {I.}~\bibnamefont {Hachisu}},\ }\href@noop {} {\bibfield  {journal} {\bibinfo  {journal} {Mon. Not. Roy. Astron. Soc.}\ }\textbf {\bibinfo {volume} {237}},\ \bibinfo {pages} {355} (\bibinfo {year} {1989})}\BibitemShut {NoStop}%
\bibitem [{\citenamefont {Cook}\ \emph {et~al.}(1994{\natexlab{a}})\citenamefont {Cook}, \citenamefont {Shapiro},\ and\ \citenamefont {Teukolsky}}]{Cook:1993qj}%
  \BibitemOpen
  \bibfield  {author} {\bibinfo {author} {\bibfnamefont {G.~B.}\ \bibnamefont {Cook}}, \bibinfo {author} {\bibfnamefont {S.~L.}\ \bibnamefont {Shapiro}}, \ and\ \bibinfo {author} {\bibfnamefont {S.~A.}\ \bibnamefont {Teukolsky}},\ }\href@noop {} {\bibfield  {journal} {\bibinfo  {journal} {Astrophys. J.}\ }\textbf {\bibinfo {volume} {422}},\ \bibinfo {pages} {227} (\bibinfo {year} {1994}{\natexlab{a}})}\BibitemShut {NoStop}%
\bibitem [{\citenamefont {Cook}\ \emph {et~al.}(1994{\natexlab{b}})\citenamefont {Cook}, \citenamefont {Shapiro},\ and\ \citenamefont {Teukolsky}}]{Cook:1993qr}%
  \BibitemOpen
  \bibfield  {author} {\bibinfo {author} {\bibfnamefont {G.~B.}\ \bibnamefont {Cook}}, \bibinfo {author} {\bibfnamefont {S.~L.}\ \bibnamefont {Shapiro}}, \ and\ \bibinfo {author} {\bibfnamefont {S.~A.}\ \bibnamefont {Teukolsky}},\ }\href {\doibase 10.1086/173934} {\bibfield  {journal} {\bibinfo  {journal} {Astrophys. J.}\ }\textbf {\bibinfo {volume} {424}},\ \bibinfo {pages} {823} (\bibinfo {year} {1994}{\natexlab{b}})}\BibitemShut {NoStop}%
\bibitem [{\citenamefont {Stergioulas}\ and\ \citenamefont {Friedman}(1995)}]{Stergioulas:1994ea}%
  \BibitemOpen
  \bibfield  {author} {\bibinfo {author} {\bibfnamefont {N.}~\bibnamefont {Stergioulas}}\ and\ \bibinfo {author} {\bibfnamefont {J.~L.}\ \bibnamefont {Friedman}},\ }\href {\doibase 10.1086/175605} {\bibfield  {journal} {\bibinfo  {journal} {Astrophys. J.}\ }\textbf {\bibinfo {volume} {444}},\ \bibinfo {pages} {306} (\bibinfo {year} {1995})},\ \Eprint {http://arxiv.org/abs/astro-ph/9411032} {arXiv:astro-ph/9411032} \BibitemShut {NoStop}%
\bibitem [{\citenamefont {Kr\"uger}\ \emph {et~al.}(2021)\citenamefont {Kr\"uger}, \citenamefont {Kokkotas}, \citenamefont {Manoharan},\ and\ \citenamefont {V\"olkel}}]{Kruger:2021zta}%
  \BibitemOpen
  \bibfield  {author} {\bibinfo {author} {\bibfnamefont {C.~J.}\ \bibnamefont {Kr\"uger}}, \bibinfo {author} {\bibfnamefont {K.~D.}\ \bibnamefont {Kokkotas}}, \bibinfo {author} {\bibfnamefont {P.}~\bibnamefont {Manoharan}}, \ and\ \bibinfo {author} {\bibfnamefont {S.~H.}\ \bibnamefont {V\"olkel}},\ }\href {\doibase 10.3389/fspas.2021.736918} {\bibfield  {journal} {\bibinfo  {journal} {Front. Astron. Space Sci.}\ }\textbf {\bibinfo {volume} {8}},\ \bibinfo {pages} {736918} (\bibinfo {year} {2021})},\ \Eprint {http://arxiv.org/abs/2110.00393} {arXiv:2110.00393 [gr-qc]} \BibitemShut {NoStop}%
\bibitem [{\citenamefont {Weber}\ \emph {et~al.}(2013)\citenamefont {Weber}, \citenamefont {Orsaria},\ and\ \citenamefont {Negreiros}}]{Weber:2013uja}%
  \BibitemOpen
  \bibfield  {author} {\bibinfo {author} {\bibfnamefont {F.}~\bibnamefont {Weber}}, \bibinfo {author} {\bibfnamefont {M.}~\bibnamefont {Orsaria}}, \ and\ \bibinfo {author} {\bibfnamefont {R.}~\bibnamefont {Negreiros}},\ }in\ \href@noop {} {\emph {\bibinfo {booktitle} {{Compact Stars in the QCD Phase Diagram III}}}}\ (\bibinfo {year} {2013})\ \Eprint {http://arxiv.org/abs/1307.1103} {arXiv:1307.1103 [astro-ph.SR]} \BibitemShut {NoStop}%
\bibitem [{\citenamefont {Beznogov}\ \emph {et~al.}(2023)\citenamefont {Beznogov}, \citenamefont {Novak}, \citenamefont {Page},\ and\ \citenamefont {Raduta}}]{Beznogov:2022wae}%
  \BibitemOpen
  \bibfield  {author} {\bibinfo {author} {\bibfnamefont {M.~V.}\ \bibnamefont {Beznogov}}, \bibinfo {author} {\bibfnamefont {J.}~\bibnamefont {Novak}}, \bibinfo {author} {\bibfnamefont {D.}~\bibnamefont {Page}}, \ and\ \bibinfo {author} {\bibfnamefont {A.~R.}\ \bibnamefont {Raduta}},\ }\href {\doibase 10.3847/1538-4357/ac9eb7} {\bibfield  {journal} {\bibinfo  {journal} {Astrophys. J.}\ }\textbf {\bibinfo {volume} {942}},\ \bibinfo {pages} {72} (\bibinfo {year} {2023})},\ \Eprint {http://arxiv.org/abs/2206.04539} {arXiv:2206.04539 [astro-ph.HE]} \BibitemShut {NoStop}%
\bibitem [{\citenamefont {Krastev}\ \emph {et~al.}(2008)\citenamefont {Krastev}, \citenamefont {Li},\ and\ \citenamefont {Worley}}]{Krastev:2007wh}%
  \BibitemOpen
  \bibfield  {author} {\bibinfo {author} {\bibfnamefont {P.~G.}\ \bibnamefont {Krastev}}, \bibinfo {author} {\bibfnamefont {B.-A.}\ \bibnamefont {Li}}, \ and\ \bibinfo {author} {\bibfnamefont {A.}~\bibnamefont {Worley}},\ }\href {\doibase 10.1086/528736} {\bibfield  {journal} {\bibinfo  {journal} {Astrophys. J.}\ }\textbf {\bibinfo {volume} {676}},\ \bibinfo {pages} {1170} (\bibinfo {year} {2008})},\ \Eprint {http://arxiv.org/abs/0709.3621} {arXiv:0709.3621 [astro-ph]} \BibitemShut {NoStop}%
\bibitem [{\citenamefont {Largani}\ \emph {et~al.}(2022)\citenamefont {Largani}, \citenamefont {Fischer}, \citenamefont {Sedrakian}, \citenamefont {Cierniak}, \citenamefont {Alvarez-Castillo},\ and\ \citenamefont {Blaschke}}]{Largani:2021hjo}%
  \BibitemOpen
  \bibfield  {author} {\bibinfo {author} {\bibfnamefont {N.~K.}\ \bibnamefont {Largani}}, \bibinfo {author} {\bibfnamefont {T.}~\bibnamefont {Fischer}}, \bibinfo {author} {\bibfnamefont {A.}~\bibnamefont {Sedrakian}}, \bibinfo {author} {\bibfnamefont {M.}~\bibnamefont {Cierniak}}, \bibinfo {author} {\bibfnamefont {D.~E.}\ \bibnamefont {Alvarez-Castillo}}, \ and\ \bibinfo {author} {\bibfnamefont {D.~B.}\ \bibnamefont {Blaschke}},\ }\href {\doibase 10.1093/mnras/stac1916} {\bibfield  {journal} {\bibinfo  {journal} {Mon. Not. Roy. Astron. Soc.}\ }\textbf {\bibinfo {volume} {515}},\ \bibinfo {pages} {3539} (\bibinfo {year} {2022})},\ \Eprint {http://arxiv.org/abs/2112.10439} {arXiv:2112.10439 [astro-ph.HE]} \BibitemShut {NoStop}%
\bibitem [{\citenamefont {Kr\"uger}\ and\ \citenamefont {V\"olkel}(2023)}]{Kruger:2023olj}%
  \BibitemOpen
  \bibfield  {author} {\bibinfo {author} {\bibfnamefont {C.~J.}\ \bibnamefont {Kr\"uger}}\ and\ \bibinfo {author} {\bibfnamefont {S.~H.}\ \bibnamefont {V\"olkel}},\ }\href {\doibase 10.1103/PhysRevD.108.124056} {\bibfield  {journal} {\bibinfo  {journal} {Phys. Rev. D}\ }\textbf {\bibinfo {volume} {108}},\ \bibinfo {pages} {124056} (\bibinfo {year} {2023})},\ \Eprint {http://arxiv.org/abs/2309.05643} {arXiv:2309.05643 [gr-qc]} \BibitemShut {NoStop}%
\bibitem [{\citenamefont {Zdunik}\ \emph {et~al.}(2006)\citenamefont {Zdunik}, \citenamefont {Bejger}, \citenamefont {Haensel},\ and\ \citenamefont {Gourgoulhon}}]{Zdunik:2005kh}%
  \BibitemOpen
  \bibfield  {author} {\bibinfo {author} {\bibfnamefont {J.~L.}\ \bibnamefont {Zdunik}}, \bibinfo {author} {\bibfnamefont {M.}~\bibnamefont {Bejger}}, \bibinfo {author} {\bibfnamefont {P.}~\bibnamefont {Haensel}}, \ and\ \bibinfo {author} {\bibfnamefont {E.}~\bibnamefont {Gourgoulhon}},\ }\href {\doibase 10.1051/0004-6361:20054260} {\bibfield  {journal} {\bibinfo  {journal} {Astron. Astrophys.}\ }\textbf {\bibinfo {volume} {450}},\ \bibinfo {pages} {747} (\bibinfo {year} {2006})},\ \Eprint {http://arxiv.org/abs/astro-ph/0509806} {arXiv:astro-ph/0509806} \BibitemShut {NoStop}%
\bibitem [{\citenamefont {Dimmelmeier}\ \emph {et~al.}(2009)\citenamefont {Dimmelmeier}, \citenamefont {Bejger}, \citenamefont {Haensel},\ and\ \citenamefont {Zdunik}}]{Dimmelmeier:2009vw}%
  \BibitemOpen
  \bibfield  {author} {\bibinfo {author} {\bibfnamefont {H.}~\bibnamefont {Dimmelmeier}}, \bibinfo {author} {\bibfnamefont {M.}~\bibnamefont {Bejger}}, \bibinfo {author} {\bibfnamefont {P.}~\bibnamefont {Haensel}}, \ and\ \bibinfo {author} {\bibfnamefont {J.~L.}\ \bibnamefont {Zdunik}},\ }\href {\doibase 10.1111/j.1365-2966.2009.14891.x} {\bibfield  {journal} {\bibinfo  {journal} {Mon. Not. Roy. Astron. Soc.}\ }\textbf {\bibinfo {volume} {396}},\ \bibinfo {pages} {2269} (\bibinfo {year} {2009})},\ \Eprint {http://arxiv.org/abs/0901.3819} {arXiv:0901.3819 [astro-ph.SR]} \BibitemShut {NoStop}%
\bibitem [{\citenamefont {Ippolito}\ \emph {et~al.}(2008)\citenamefont {Ippolito}, \citenamefont {Ruggieri}, \citenamefont {Rischke}, \citenamefont {Sedrakian},\ and\ \citenamefont {Weber}}]{Ippolito:2007hn}%
  \BibitemOpen
  \bibfield  {author} {\bibinfo {author} {\bibfnamefont {N.}~\bibnamefont {Ippolito}}, \bibinfo {author} {\bibfnamefont {M.}~\bibnamefont {Ruggieri}}, \bibinfo {author} {\bibfnamefont {D.}~\bibnamefont {Rischke}}, \bibinfo {author} {\bibfnamefont {A.}~\bibnamefont {Sedrakian}}, \ and\ \bibinfo {author} {\bibfnamefont {F.}~\bibnamefont {Weber}},\ }\href {\doibase 10.1103/PhysRevD.77.023004} {\bibfield  {journal} {\bibinfo  {journal} {Phys. Rev. D}\ }\textbf {\bibinfo {volume} {77}},\ \bibinfo {pages} {023004} (\bibinfo {year} {2008})},\ \Eprint {http://arxiv.org/abs/0710.3874} {arXiv:0710.3874 [astro-ph]} \BibitemShut {NoStop}%
\bibitem [{\citenamefont {Dhiman}\ \emph {et~al.}(2010)\citenamefont {Dhiman}, \citenamefont {Mahajan},\ and\ \citenamefont {Agrawal}}]{Dhiman:2010imr}%
  \BibitemOpen
  \bibfield  {author} {\bibinfo {author} {\bibfnamefont {S.~K.}\ \bibnamefont {Dhiman}}, \bibinfo {author} {\bibfnamefont {G.}~\bibnamefont {Mahajan}}, \ and\ \bibinfo {author} {\bibfnamefont {B.~K.}\ \bibnamefont {Agrawal}},\ }\href {\doibase 10.1016/j.nuclphysa.2009.12.063} {\bibfield  {journal} {\bibinfo  {journal} {Nucl. Phys. A}\ }\textbf {\bibinfo {volume} {836}},\ \bibinfo {pages} {183} (\bibinfo {year} {2010})}\BibitemShut {NoStop}%
\bibitem [{\citenamefont {Ayvazyan}\ \emph {et~al.}(2013)\citenamefont {Ayvazyan}, \citenamefont {Colucci}, \citenamefont {Rischke},\ and\ \citenamefont {Sedrakian}}]{Ayvazyan:2013cva}%
  \BibitemOpen
  \bibfield  {author} {\bibinfo {author} {\bibfnamefont {N.~S.}\ \bibnamefont {Ayvazyan}}, \bibinfo {author} {\bibfnamefont {G.}~\bibnamefont {Colucci}}, \bibinfo {author} {\bibfnamefont {D.~H.}\ \bibnamefont {Rischke}}, \ and\ \bibinfo {author} {\bibfnamefont {A.}~\bibnamefont {Sedrakian}},\ }\href {\doibase 10.1051/0004-6361/201322484} {\bibfield  {journal} {\bibinfo  {journal} {Astron. Astrophys.}\ }\textbf {\bibinfo {volume} {559}},\ \bibinfo {pages} {A118} (\bibinfo {year} {2013})},\ \Eprint {http://arxiv.org/abs/1308.3053} {arXiv:1308.3053 [astro-ph.SR]} \BibitemShut {NoStop}%
\bibitem [{\citenamefont {Bhattacharyya}\ \emph {et~al.}(2017)\citenamefont {Bhattacharyya}, \citenamefont {Bombaci}, \citenamefont {Bandyopadhyay}, \citenamefont {Thampan},\ and\ \citenamefont {Logoteta}}]{Bhattacharyya:2017tos}%
  \BibitemOpen
  \bibfield  {author} {\bibinfo {author} {\bibfnamefont {S.}~\bibnamefont {Bhattacharyya}}, \bibinfo {author} {\bibfnamefont {I.}~\bibnamefont {Bombaci}}, \bibinfo {author} {\bibfnamefont {D.}~\bibnamefont {Bandyopadhyay}}, \bibinfo {author} {\bibfnamefont {A.~V.}\ \bibnamefont {Thampan}}, \ and\ \bibinfo {author} {\bibfnamefont {D.}~\bibnamefont {Logoteta}},\ }\href {\doibase 10.1016/j.newast.2017.01.008} {\bibfield  {journal} {\bibinfo  {journal} {New Astron.}\ }\textbf {\bibinfo {volume} {54}},\ \bibinfo {pages} {61} (\bibinfo {year} {2017})},\ \Eprint {http://arxiv.org/abs/1701.03489} {arXiv:1701.03489 [astro-ph.HE]} \BibitemShut {NoStop}%
\bibitem [{\citenamefont {Shapiro}\ and\ \citenamefont {Teukolsky}(1983)}]{Shapiro:1983du}%
  \BibitemOpen
  \bibfield  {author} {\bibinfo {author} {\bibfnamefont {S.~L.}\ \bibnamefont {Shapiro}}\ and\ \bibinfo {author} {\bibfnamefont {S.~A.}\ \bibnamefont {Teukolsky}},\ }\href {\doibase 10.1002/9783527617661} {\emph {\bibinfo {title} {{Black holes, white dwarfs, and neutron stars: The physics of compact objects}}}}\ (\bibinfo {year} {1983})\BibitemShut {NoStop}%
\bibitem [{\citenamefont {Miller}\ \emph {et~al.}(2019)\citenamefont {Miller} \emph {et~al.}}]{Miller:2019cac}%
  \BibitemOpen
  \bibfield  {author} {\bibinfo {author} {\bibfnamefont {M.~C.}\ \bibnamefont {Miller}} \emph {et~al.},\ }\href {\doibase 10.3847/2041-8213/ab50c5} {\bibfield  {journal} {\bibinfo  {journal} {Astrophys. J. Lett.}\ }\textbf {\bibinfo {volume} {887}},\ \bibinfo {pages} {L24} (\bibinfo {year} {2019})},\ \Eprint {http://arxiv.org/abs/1912.05705} {arXiv:1912.05705 [astro-ph.HE]} \BibitemShut {NoStop}%
\bibitem [{\citenamefont {Riley}\ \emph {et~al.}(2019)\citenamefont {Riley} \emph {et~al.}}]{Riley:2019yda}%
  \BibitemOpen
  \bibfield  {author} {\bibinfo {author} {\bibfnamefont {T.~E.}\ \bibnamefont {Riley}} \emph {et~al.},\ }\href {\doibase 10.3847/2041-8213/ab481c} {\bibfield  {journal} {\bibinfo  {journal} {Astrophys. J. Lett.}\ }\textbf {\bibinfo {volume} {887}},\ \bibinfo {pages} {L21} (\bibinfo {year} {2019})},\ \Eprint {http://arxiv.org/abs/1912.05702} {arXiv:1912.05702 [astro-ph.HE]} \BibitemShut {NoStop}%
\bibitem [{\citenamefont {Miller}\ \emph {et~al.}(2021)\citenamefont {Miller} \emph {et~al.}}]{Miller:2021qha}%
  \BibitemOpen
  \bibfield  {author} {\bibinfo {author} {\bibfnamefont {M.~C.}\ \bibnamefont {Miller}} \emph {et~al.},\ }\href {\doibase 10.3847/2041-8213/ac089b} {\bibfield  {journal} {\bibinfo  {journal} {Astrophys. J. Lett.}\ }\textbf {\bibinfo {volume} {918}},\ \bibinfo {pages} {L28} (\bibinfo {year} {2021})},\ \Eprint {http://arxiv.org/abs/2105.06979} {arXiv:2105.06979 [astro-ph.HE]} \BibitemShut {NoStop}%
\bibitem [{\citenamefont {Riley}\ \emph {et~al.}(2021)\citenamefont {Riley} \emph {et~al.}}]{Riley:2021pdl}%
  \BibitemOpen
  \bibfield  {author} {\bibinfo {author} {\bibfnamefont {T.~E.}\ \bibnamefont {Riley}} \emph {et~al.},\ }\href {\doibase 10.3847/2041-8213/ac0a81} {\bibfield  {journal} {\bibinfo  {journal} {Astrophys. J. Lett.}\ }\textbf {\bibinfo {volume} {918}},\ \bibinfo {pages} {L27} (\bibinfo {year} {2021})},\ \Eprint {http://arxiv.org/abs/2105.06980} {arXiv:2105.06980 [astro-ph.HE]} \BibitemShut {NoStop}%
\bibitem [{\citenamefont {Choudhury}\ \emph {et~al.}(2024)\citenamefont {Choudhury} \emph {et~al.}}]{Choudhury:2024xbk}%
  \BibitemOpen
  \bibfield  {author} {\bibinfo {author} {\bibfnamefont {D.}~\bibnamefont {Choudhury}} \emph {et~al.},\ }\href@noop {} {\  (\bibinfo {year} {2024})},\ \Eprint {http://arxiv.org/abs/2407.06789} {arXiv:2407.06789 [astro-ph.HE]} \BibitemShut {NoStop}%
\bibitem [{\citenamefont {Antoniadis}\ \emph {et~al.}(2013)\citenamefont {Antoniadis} \emph {et~al.}}]{Antoniadis:2013pzd}%
  \BibitemOpen
  \bibfield  {author} {\bibinfo {author} {\bibfnamefont {J.}~\bibnamefont {Antoniadis}} \emph {et~al.},\ }\href {\doibase 10.1126/science.1233232} {\bibfield  {journal} {\bibinfo  {journal} {Science}\ }\textbf {\bibinfo {volume} {340}},\ \bibinfo {pages} {6131} (\bibinfo {year} {2013})},\ \Eprint {http://arxiv.org/abs/1304.6875} {arXiv:1304.6875 [astro-ph.HE]} \BibitemShut {NoStop}%
\bibitem [{\citenamefont {Ivanytskyi}\ and\ \citenamefont {Blaschke}(2022{\natexlab{a}})}]{Ivanytskyi:2022oxv}%
  \BibitemOpen
  \bibfield  {author} {\bibinfo {author} {\bibfnamefont {O.}~\bibnamefont {Ivanytskyi}}\ and\ \bibinfo {author} {\bibfnamefont {D.}~\bibnamefont {Blaschke}},\ }\href {\doibase 10.1103/PhysRevD.105.114042} {\bibfield  {journal} {\bibinfo  {journal} {Phys. Rev. D}\ }\textbf {\bibinfo {volume} {105}},\ \bibinfo {pages} {114042} (\bibinfo {year} {2022}{\natexlab{a}})},\ \Eprint {http://arxiv.org/abs/2204.03611} {arXiv:2204.03611 [nucl-th]} \BibitemShut {NoStop}%
\bibitem [{\citenamefont {G\"artlein}\ \emph {et~al.}(2023)\citenamefont {G\"artlein}, \citenamefont {Ivanytskyi}, \citenamefont {Sagun},\ and\ \citenamefont {Blaschke}}]{Gartlein:2023vif}%
  \BibitemOpen
  \bibfield  {author} {\bibinfo {author} {\bibfnamefont {C.}~\bibnamefont {G\"artlein}}, \bibinfo {author} {\bibfnamefont {O.}~\bibnamefont {Ivanytskyi}}, \bibinfo {author} {\bibfnamefont {V.}~\bibnamefont {Sagun}}, \ and\ \bibinfo {author} {\bibfnamefont {D.}~\bibnamefont {Blaschke}},\ }\href {\doibase 10.1103/PhysRevD.108.114028} {\bibfield  {journal} {\bibinfo  {journal} {Phys. Rev. D}\ }\textbf {\bibinfo {volume} {108}},\ \bibinfo {pages} {114028} (\bibinfo {year} {2023})},\ \Eprint {http://arxiv.org/abs/2301.10765} {arXiv:2301.10765 [nucl-th]} \BibitemShut {NoStop}%
\bibitem [{\citenamefont {Alford}\ \emph {et~al.}(2005)\citenamefont {Alford}, \citenamefont {Braby}, \citenamefont {Paris},\ and\ \citenamefont {Reddy}}]{Alford:2004pf}%
  \BibitemOpen
  \bibfield  {author} {\bibinfo {author} {\bibfnamefont {M.}~\bibnamefont {Alford}}, \bibinfo {author} {\bibfnamefont {M.}~\bibnamefont {Braby}}, \bibinfo {author} {\bibfnamefont {M.~W.}\ \bibnamefont {Paris}}, \ and\ \bibinfo {author} {\bibfnamefont {S.}~\bibnamefont {Reddy}},\ }\href {\doibase 10.1086/430902} {\bibfield  {journal} {\bibinfo  {journal} {Astrophys. J.}\ }\textbf {\bibinfo {volume} {629}},\ \bibinfo {pages} {969} (\bibinfo {year} {2005})},\ \Eprint {http://arxiv.org/abs/nucl-th/0411016} {arXiv:nucl-th/0411016} \BibitemShut {NoStop}%
\bibitem [{\citenamefont {Shahrbaf}\ \emph {et~al.}(2022)\citenamefont {Shahrbaf}, \citenamefont {Blaschke}, \citenamefont {Typel}, \citenamefont {Farrar},\ and\ \citenamefont {Alvarez-Castillo}}]{Shahrbaf:2022upc}%
  \BibitemOpen
  \bibfield  {author} {\bibinfo {author} {\bibfnamefont {M.}~\bibnamefont {Shahrbaf}}, \bibinfo {author} {\bibfnamefont {D.}~\bibnamefont {Blaschke}}, \bibinfo {author} {\bibfnamefont {S.}~\bibnamefont {Typel}}, \bibinfo {author} {\bibfnamefont {G.~R.}\ \bibnamefont {Farrar}}, \ and\ \bibinfo {author} {\bibfnamefont {D.~E.}\ \bibnamefont {Alvarez-Castillo}},\ }\href {\doibase 10.1103/PhysRevD.105.103005} {\bibfield  {journal} {\bibinfo  {journal} {Phys. Rev. D}\ }\textbf {\bibinfo {volume} {105}},\ \bibinfo {pages} {103005} (\bibinfo {year} {2022})},\ \Eprint {http://arxiv.org/abs/2202.00652} {arXiv:2202.00652 [nucl-th]} \BibitemShut {NoStop}%
\bibitem [{\citenamefont {Abbott}\ \emph {et~al.}(2018)\citenamefont {Abbott} \emph {et~al.}}]{LIGOScientific:2018cki}%
  \BibitemOpen
  \bibfield  {author} {\bibinfo {author} {\bibfnamefont {B.~P.}\ \bibnamefont {Abbott}} \emph {et~al.} (\bibinfo {collaboration} {LIGO Scientific, Virgo}),\ }\href {\doibase 10.1103/PhysRevLett.121.161101} {\bibfield  {journal} {\bibinfo  {journal} {Phys. Rev. Lett.}\ }\textbf {\bibinfo {volume} {121}},\ \bibinfo {pages} {161101} (\bibinfo {year} {2018})},\ \Eprint {http://arxiv.org/abs/1805.11581} {arXiv:1805.11581 [gr-qc]} \BibitemShut {NoStop}%
\bibitem [{\citenamefont {Typel}(2018)}]{Typel:2018wmm}%
  \BibitemOpen
  \bibfield  {author} {\bibinfo {author} {\bibfnamefont {S.}~\bibnamefont {Typel}},\ }\href {\doibase 10.1088/1361-6471/aadea5} {\bibfield  {journal} {\bibinfo  {journal} {J. Phys. G}\ }\textbf {\bibinfo {volume} {45}},\ \bibinfo {pages} {114001} (\bibinfo {year} {2018})}\BibitemShut {NoStop}%
\bibitem [{\citenamefont {Ivanytskyi}\ and\ \citenamefont {Blaschke}(2022{\natexlab{b}})}]{Ivanytskyi:2022bjc}%
  \BibitemOpen
  \bibfield  {author} {\bibinfo {author} {\bibfnamefont {O.}~\bibnamefont {Ivanytskyi}}\ and\ \bibinfo {author} {\bibfnamefont {D.~B.}\ \bibnamefont {Blaschke}},\ }\href {\doibase 10.3390/particles5040038} {\bibfield  {journal} {\bibinfo  {journal} {Particles}\ }\textbf {\bibinfo {volume} {5}},\ \bibinfo {pages} {514} (\bibinfo {year} {2022}{\natexlab{b}})},\ \Eprint {http://arxiv.org/abs/2209.02050} {arXiv:2209.02050 [nucl-th]} \BibitemShut {NoStop}%
\bibitem [{\citenamefont {Glendenning}(1992)}]{Glendenning:1992vb}%
  \BibitemOpen
  \bibfield  {author} {\bibinfo {author} {\bibfnamefont {N.~K.}\ \bibnamefont {Glendenning}},\ }\href {\doibase 10.1103/PhysRevD.46.1274} {\bibfield  {journal} {\bibinfo  {journal} {Phys. Rev. D}\ }\textbf {\bibinfo {volume} {46}},\ \bibinfo {pages} {1274} (\bibinfo {year} {1992})}\BibitemShut {NoStop}%
\bibitem [{\citenamefont {Maslov}\ \emph {et~al.}(2019)\citenamefont {Maslov}, \citenamefont {Yasutake}, \citenamefont {Ayriyan}, \citenamefont {Blaschke}, \citenamefont {Grigorian}, \citenamefont {Maruyama}, \citenamefont {Tatsumi},\ and\ \citenamefont {Voskresensky}}]{Maslov:2018ghi}%
  \BibitemOpen
  \bibfield  {author} {\bibinfo {author} {\bibfnamefont {K.}~\bibnamefont {Maslov}}, \bibinfo {author} {\bibfnamefont {N.}~\bibnamefont {Yasutake}}, \bibinfo {author} {\bibfnamefont {A.}~\bibnamefont {Ayriyan}}, \bibinfo {author} {\bibfnamefont {D.}~\bibnamefont {Blaschke}}, \bibinfo {author} {\bibfnamefont {H.}~\bibnamefont {Grigorian}}, \bibinfo {author} {\bibfnamefont {T.}~\bibnamefont {Maruyama}}, \bibinfo {author} {\bibfnamefont {T.}~\bibnamefont {Tatsumi}}, \ and\ \bibinfo {author} {\bibfnamefont {D.~N.}\ \bibnamefont {Voskresensky}},\ }\href {\doibase 10.1103/PhysRevC.100.025802} {\bibfield  {journal} {\bibinfo  {journal} {Phys. Rev. C}\ }\textbf {\bibinfo {volume} {100}},\ \bibinfo {pages} {025802} (\bibinfo {year} {2019})},\ \Eprint {http://arxiv.org/abs/1812.11889} {arXiv:1812.11889 [nucl-th]} \BibitemShut {NoStop}%
\bibitem [{\citenamefont {Voskresensky}\ \emph {et~al.}(2002)\citenamefont {Voskresensky}, \citenamefont {Yasuhira},\ and\ \citenamefont {Tatsumi}}]{Voskresensky:2001jq}%
  \BibitemOpen
  \bibfield  {author} {\bibinfo {author} {\bibfnamefont {D.~N.}\ \bibnamefont {Voskresensky}}, \bibinfo {author} {\bibfnamefont {M.}~\bibnamefont {Yasuhira}}, \ and\ \bibinfo {author} {\bibfnamefont {T.}~\bibnamefont {Tatsumi}},\ }\href {\doibase 10.1016/S0370-2693(02)02186-X} {\bibfield  {journal} {\bibinfo  {journal} {Phys. Lett. B}\ }\textbf {\bibinfo {volume} {541}},\ \bibinfo {pages} {93} (\bibinfo {year} {2002})},\ \Eprint {http://arxiv.org/abs/nucl-th/0109009} {arXiv:nucl-th/0109009} \BibitemShut {NoStop}%
\bibitem [{\citenamefont {Voskresensky}\ \emph {et~al.}(2003)\citenamefont {Voskresensky}, \citenamefont {Yasuhira},\ and\ \citenamefont {Tatsumi}}]{Voskresensky:2002hu}%
  \BibitemOpen
  \bibfield  {author} {\bibinfo {author} {\bibfnamefont {D.~N.}\ \bibnamefont {Voskresensky}}, \bibinfo {author} {\bibfnamefont {M.}~\bibnamefont {Yasuhira}}, \ and\ \bibinfo {author} {\bibfnamefont {T.}~\bibnamefont {Tatsumi}},\ }\href {\doibase 10.1016/S0375-9474(03)01313-7} {\bibfield  {journal} {\bibinfo  {journal} {Nucl. Phys. A}\ }\textbf {\bibinfo {volume} {723}},\ \bibinfo {pages} {291} (\bibinfo {year} {2003})},\ \Eprint {http://arxiv.org/abs/nucl-th/0208067} {arXiv:nucl-th/0208067} \BibitemShut {NoStop}%
\bibitem [{\citenamefont {Abgaryan}\ \emph {et~al.}(2018)\citenamefont {Abgaryan}, \citenamefont {Alvarez-Castillo}, \citenamefont {Ayriyan}, \citenamefont {Blaschke},\ and\ \citenamefont {Grigorian}}]{Abgaryan:2018gqp}%
  \BibitemOpen
  \bibfield  {author} {\bibinfo {author} {\bibfnamefont {V.}~\bibnamefont {Abgaryan}}, \bibinfo {author} {\bibfnamefont {D.}~\bibnamefont {Alvarez-Castillo}}, \bibinfo {author} {\bibfnamefont {A.}~\bibnamefont {Ayriyan}}, \bibinfo {author} {\bibfnamefont {D.}~\bibnamefont {Blaschke}}, \ and\ \bibinfo {author} {\bibfnamefont {H.}~\bibnamefont {Grigorian}},\ }\href {\doibase 10.3390/universe4090094} {\bibfield  {journal} {\bibinfo  {journal} {Universe}\ }\textbf {\bibinfo {volume} {4}},\ \bibinfo {pages} {94} (\bibinfo {year} {2018})},\ \Eprint {http://arxiv.org/abs/1807.08034} {arXiv:1807.08034 [astro-ph.HE]} \BibitemShut {NoStop}%
\bibitem [{\citenamefont {Cierniak}\ and\ \citenamefont {Blaschke}(2021)}]{Cierniak:2021knt}%
  \BibitemOpen
  \bibfield  {author} {\bibinfo {author} {\bibfnamefont {M.}~\bibnamefont {Cierniak}}\ and\ \bibinfo {author} {\bibfnamefont {D.}~\bibnamefont {Blaschke}},\ }\href {\doibase 10.1002/asna.202114000} {\bibfield  {journal} {\bibinfo  {journal} {Astron. Nachr.}\ }\textbf {\bibinfo {volume} {342}},\ \bibinfo {pages} {819} (\bibinfo {year} {2021})},\ \Eprint {http://arxiv.org/abs/2106.06986} {arXiv:2106.06986 [nucl-th]} \BibitemShut {NoStop}%
\bibitem [{\citenamefont {Ivanytskyi}\ and\ \citenamefont {Blaschke}(2022{\natexlab{c}})}]{Ivanytskyi:2022wln}%
  \BibitemOpen
  \bibfield  {author} {\bibinfo {author} {\bibfnamefont {O.}~\bibnamefont {Ivanytskyi}}\ and\ \bibinfo {author} {\bibfnamefont {D.}~\bibnamefont {Blaschke}},\ }\href {\doibase 10.1140/epja/s10050-022-00808-5} {\bibfield  {journal} {\bibinfo  {journal} {Eur. Phys. J. A}\ }\textbf {\bibinfo {volume} {58}},\ \bibinfo {pages} {152} (\bibinfo {year} {2022}{\natexlab{c}})},\ \Eprint {http://arxiv.org/abs/2205.03455} {arXiv:2205.03455 [nucl-th]} \BibitemShut {NoStop}%
\bibitem [{\citenamefont {{Seidov}}(1971)}]{1971SvA15347S}%
  \BibitemOpen
  \bibfield  {author} {\bibinfo {author} {\bibfnamefont {Z.~F.}\ \bibnamefont {{Seidov}}},\ }\href@noop {} {\bibfield  {journal} {\bibinfo  {journal} {Soviet Astronomy}\ }\textbf {\bibinfo {volume} {15}},\ \bibinfo {pages} {347} (\bibinfo {year} {1971})}\BibitemShut {NoStop}%
\bibitem [{\citenamefont {Blaschke}\ \emph {et~al.}(2020)\citenamefont {Blaschke}, \citenamefont {Alvarez-Castillo}, \citenamefont {Ayriyan}, \citenamefont {Grigorian}, \citenamefont {Largani},\ and\ \citenamefont {Weber}}]{Blaschke:2019tbh}%
  \BibitemOpen
  \bibfield  {author} {\bibinfo {author} {\bibfnamefont {D.}~\bibnamefont {Blaschke}}, \bibinfo {author} {\bibfnamefont {D.~E.}\ \bibnamefont {Alvarez-Castillo}}, \bibinfo {author} {\bibfnamefont {A.}~\bibnamefont {Ayriyan}}, \bibinfo {author} {\bibfnamefont {H.}~\bibnamefont {Grigorian}}, \bibinfo {author} {\bibfnamefont {N.~K.}\ \bibnamefont {Largani}}, \ and\ \bibinfo {author} {\bibfnamefont {F.}~\bibnamefont {Weber}},\ }\enquote {\bibinfo {title} {{Astrophysical aspects of general relativistic mass twin stars}},}\ \ (\bibinfo {year} {2020})\ pp.\ \bibinfo {pages} {207--256},\ \Eprint {http://arxiv.org/abs/1906.02522} {arXiv:1906.02522 [astro-ph.HE]} \BibitemShut {NoStop}%
\bibitem [{\citenamefont {Goncalves}\ \emph {et~al.}(2022)\citenamefont {Goncalves}, \citenamefont {Jim\'enez},\ and\ \citenamefont {Lazzari}}]{Goncalves:2022phg}%
  \BibitemOpen
  \bibfield  {author} {\bibinfo {author} {\bibfnamefont {V.~P.}\ \bibnamefont {Goncalves}}, \bibinfo {author} {\bibfnamefont {J.~C.}\ \bibnamefont {Jim\'enez}}, \ and\ \bibinfo {author} {\bibfnamefont {L.}~\bibnamefont {Lazzari}},\ }\href {\doibase 10.1140/epjc/s10052-022-11115-0} {\bibfield  {journal} {\bibinfo  {journal} {Eur. Phys. J. C}\ }\textbf {\bibinfo {volume} {82}},\ \bibinfo {pages} {1117} (\bibinfo {year} {2022})},\ \Eprint {http://arxiv.org/abs/2206.10513} {arXiv:2206.10513 [nucl-th]} \BibitemShut {NoStop}%
\bibitem [{\citenamefont {Chanlaridis}\ \emph {et~al.}(2024)\citenamefont {Chanlaridis}, \citenamefont {Ohse}, \citenamefont {Alvarez-Castillo}, \citenamefont {Antoniadis}, \citenamefont {Blaschke}, \citenamefont {Danchev}, \citenamefont {Langer},\ and\ \citenamefont {Misra}}]{Chanlaridis:2024rov}%
  \BibitemOpen
  \bibfield  {author} {\bibinfo {author} {\bibfnamefont {S.}~\bibnamefont {Chanlaridis}}, \bibinfo {author} {\bibfnamefont {D.}~\bibnamefont {Ohse}}, \bibinfo {author} {\bibfnamefont {D.~E.}\ \bibnamefont {Alvarez-Castillo}}, \bibinfo {author} {\bibfnamefont {J.}~\bibnamefont {Antoniadis}}, \bibinfo {author} {\bibfnamefont {D.}~\bibnamefont {Blaschke}}, \bibinfo {author} {\bibfnamefont {V.}~\bibnamefont {Danchev}}, \bibinfo {author} {\bibfnamefont {N.}~\bibnamefont {Langer}}, \ and\ \bibinfo {author} {\bibfnamefont {D.}~\bibnamefont {Misra}},\ }\href@noop {} {\  (\bibinfo {year} {2024})},\ \Eprint {http://arxiv.org/abs/2409.04755} {arXiv:2409.04755 [astro-ph.HE]} \BibitemShut {NoStop}%
\bibitem [{\citenamefont {Glendenning}\ \emph {et~al.}(1997)\citenamefont {Glendenning}, \citenamefont {Pei},\ and\ \citenamefont {Weber}}]{Glendenning:1997fy}%
  \BibitemOpen
  \bibfield  {author} {\bibinfo {author} {\bibfnamefont {N.~K.}\ \bibnamefont {Glendenning}}, \bibinfo {author} {\bibfnamefont {S.}~\bibnamefont {Pei}}, \ and\ \bibinfo {author} {\bibfnamefont {F.}~\bibnamefont {Weber}},\ }\href {\doibase 10.1103/PhysRevLett.79.1603} {\bibfield  {journal} {\bibinfo  {journal} {Phys. Rev. Lett.}\ }\textbf {\bibinfo {volume} {79}},\ \bibinfo {pages} {1603} (\bibinfo {year} {1997})},\ \Eprint {http://arxiv.org/abs/astro-ph/9705235} {arXiv:astro-ph/9705235} \BibitemShut {NoStop}%
\bibitem [{\citenamefont {Sagert}\ \emph {et~al.}(2009)\citenamefont {Sagert}, \citenamefont {Fischer}, \citenamefont {Hempel}, \citenamefont {Pagliara}, \citenamefont {Schaffner-Bielich}, \citenamefont {Mezzacappa}, \citenamefont {Thielemann},\ and\ \citenamefont {Liebendorfer}}]{Sagert:2008ka}%
  \BibitemOpen
  \bibfield  {author} {\bibinfo {author} {\bibfnamefont {I.}~\bibnamefont {Sagert}}, \bibinfo {author} {\bibfnamefont {T.}~\bibnamefont {Fischer}}, \bibinfo {author} {\bibfnamefont {M.}~\bibnamefont {Hempel}}, \bibinfo {author} {\bibfnamefont {G.}~\bibnamefont {Pagliara}}, \bibinfo {author} {\bibfnamefont {J.}~\bibnamefont {Schaffner-Bielich}}, \bibinfo {author} {\bibfnamefont {A.}~\bibnamefont {Mezzacappa}}, \bibinfo {author} {\bibfnamefont {F.~K.}\ \bibnamefont {Thielemann}}, \ and\ \bibinfo {author} {\bibfnamefont {M.}~\bibnamefont {Liebendorfer}},\ }\href {\doibase 10.1103/PhysRevLett.102.081101} {\bibfield  {journal} {\bibinfo  {journal} {Phys. Rev. Lett.}\ }\textbf {\bibinfo {volume} {102}},\ \bibinfo {pages} {081101} (\bibinfo {year} {2009})},\ \Eprint {http://arxiv.org/abs/0809.4225} {arXiv:0809.4225 [astro-ph]} \BibitemShut {NoStop}%
\bibitem [{\citenamefont {Cipolletta}\ \emph {et~al.}(2015)\citenamefont {Cipolletta}, \citenamefont {Cherubini}, \citenamefont {Filippi}, \citenamefont {Rueda},\ and\ \citenamefont {Ruffini}}]{Cipolletta_2015}%
  \BibitemOpen
  \bibfield  {author} {\bibinfo {author} {\bibfnamefont {F.}~\bibnamefont {Cipolletta}}, \bibinfo {author} {\bibfnamefont {C.}~\bibnamefont {Cherubini}}, \bibinfo {author} {\bibfnamefont {S.}~\bibnamefont {Filippi}}, \bibinfo {author} {\bibfnamefont {J.}~\bibnamefont {Rueda}}, \ and\ \bibinfo {author} {\bibfnamefont {R.}~\bibnamefont {Ruffini}},\ }\href {\doibase 10.1103/physrevd.92.023007} {\bibfield  {journal} {\bibinfo  {journal} {Physical Review D}\ }\textbf {\bibinfo {volume} {92}} (\bibinfo {year} {2015}),\ 10.1103/physrevd.92.023007}\BibitemShut {NoStop}%
\bibitem [{\citenamefont {Chakraborty}\ \emph {et~al.}(2014)\citenamefont {Chakraborty}, \citenamefont {Modak},\ and\ \citenamefont {Bandyopadhyay}}]{Chakraborty:2014qba}%
  \BibitemOpen
  \bibfield  {author} {\bibinfo {author} {\bibfnamefont {C.}~\bibnamefont {Chakraborty}}, \bibinfo {author} {\bibfnamefont {K.~P.}\ \bibnamefont {Modak}}, \ and\ \bibinfo {author} {\bibfnamefont {D.}~\bibnamefont {Bandyopadhyay}},\ }\href {\doibase 10.1088/0004-637X/790/1/2} {\bibfield  {journal} {\bibinfo  {journal} {Astrophys. J.}\ }\textbf {\bibinfo {volume} {790}},\ \bibinfo {pages} {2} (\bibinfo {year} {2014})},\ \Eprint {http://arxiv.org/abs/1402.6108} {arXiv:1402.6108 [astro-ph.HE]} \BibitemShut {NoStop}%
\bibitem [{\citenamefont {Paschalidis}\ and\ \citenamefont {Stergioulas}(2017)}]{Paschalidis:2016vmz}%
  \BibitemOpen
  \bibfield  {author} {\bibinfo {author} {\bibfnamefont {V.}~\bibnamefont {Paschalidis}}\ and\ \bibinfo {author} {\bibfnamefont {N.}~\bibnamefont {Stergioulas}},\ }\href {\doibase 10.1007/s41114-017-0008-x} {\bibfield  {journal} {\bibinfo  {journal} {Living Rev. Rel.}\ }\textbf {\bibinfo {volume} {20}},\ \bibinfo {pages} {7} (\bibinfo {year} {2017})},\ \Eprint {http://arxiv.org/abs/1612.03050} {arXiv:1612.03050 [astro-ph.HE]} \BibitemShut {NoStop}%
\bibitem [{\citenamefont {Poisson}\ and\ \citenamefont {Doucot}(2017)}]{Poisson:2016wtv}%
  \BibitemOpen
  \bibfield  {author} {\bibinfo {author} {\bibfnamefont {E.}~\bibnamefont {Poisson}}\ and\ \bibinfo {author} {\bibfnamefont {J.}~\bibnamefont {Doucot}},\ }\href {\doibase 10.1103/PhysRevD.95.044023} {\bibfield  {journal} {\bibinfo  {journal} {Phys. Rev. D}\ }\textbf {\bibinfo {volume} {95}},\ \bibinfo {pages} {044023} (\bibinfo {year} {2017})},\ \Eprint {http://arxiv.org/abs/1612.04255} {arXiv:1612.04255 [gr-qc]} \BibitemShut {NoStop}%
\bibitem [{\citenamefont {Baarden}(1973)}]{Baarden:1973ula}%
  \BibitemOpen
  \bibfield  {author} {\bibinfo {author} {\bibfnamefont {J.~M.}\ \bibnamefont {Baarden}},\ }in\ \href@noop {} {\emph {\bibinfo {booktitle} {{Les Houches Summer School of Theoretical Physics}: {Black Holes}}}}\ (\bibinfo {year} {1973})\ pp.\ \bibinfo {pages} {241--290}\BibitemShut {NoStop}%
\bibitem [{\citenamefont {Tolman}(1939)}]{Tolman:1939jz}%
  \BibitemOpen
  \bibfield  {author} {\bibinfo {author} {\bibfnamefont {R.~C.}\ \bibnamefont {Tolman}},\ }\href {\doibase 10.1103/PhysRev.55.364} {\bibfield  {journal} {\bibinfo  {journal} {Phys. Rev.}\ }\textbf {\bibinfo {volume} {55}},\ \bibinfo {pages} {364} (\bibinfo {year} {1939})}\BibitemShut {NoStop}%
\bibitem [{\citenamefont {Oppenheimer}\ and\ \citenamefont {Volkoff}(1939)}]{Oppenheimer:1939ne}%
  \BibitemOpen
  \bibfield  {author} {\bibinfo {author} {\bibfnamefont {J.~R.}\ \bibnamefont {Oppenheimer}}\ and\ \bibinfo {author} {\bibfnamefont {G.~M.}\ \bibnamefont {Volkoff}},\ }\href {\doibase 10.1103/PhysRev.55.374} {\bibfield  {journal} {\bibinfo  {journal} {Phys. Rev.}\ }\textbf {\bibinfo {volume} {55}},\ \bibinfo {pages} {374} (\bibinfo {year} {1939})}\BibitemShut {NoStop}%
\bibitem [{\citenamefont {Stergioulas}(1998)}]{Stergioulas:1998hx}%
  \BibitemOpen
  \bibfield  {author} {\bibinfo {author} {\bibfnamefont {N.}~\bibnamefont {Stergioulas}},\ }\href {\doibase 10.12942/lrr-1998-8} {\bibfield  {journal} {\bibinfo  {journal} {Living Rev. Rel.}\ }\textbf {\bibinfo {volume} {1}},\ \bibinfo {pages} {8} (\bibinfo {year} {1998})},\ \Eprint {http://arxiv.org/abs/gr-qc/9805012} {arXiv:gr-qc/9805012} \BibitemShut {NoStop}%
\bibitem [{\citenamefont {Cierniak}\ and\ \citenamefont {Blaschke}(2020)}]{Cierniak:2020eyh}%
  \BibitemOpen
  \bibfield  {author} {\bibinfo {author} {\bibfnamefont {M.}~\bibnamefont {Cierniak}}\ and\ \bibinfo {author} {\bibfnamefont {D.}~\bibnamefont {Blaschke}},\ }\href {\doibase 10.1140/epjst/e2020-000235-5} {\bibfield  {journal} {\bibinfo  {journal} {Eur. Phys. J. ST}\ }\textbf {\bibinfo {volume} {229}},\ \bibinfo {pages} {3663} (\bibinfo {year} {2020})},\ \Eprint {http://arxiv.org/abs/2009.12353} {arXiv:2009.12353 [astro-ph.HE]} \BibitemShut {NoStop}%
\bibitem [{\citenamefont {Harrison}\ \emph {et~al.}(1965)\citenamefont {Harrison}, \citenamefont {Thorne}, \citenamefont {Wakano},\ and\ \citenamefont {Wheeler}}]{1965gtgc.book.....H}%
  \BibitemOpen
  \bibfield  {author} {\bibinfo {author} {\bibfnamefont {B.~K.}\ \bibnamefont {Harrison}}, \bibinfo {author} {\bibfnamefont {K.~S.}\ \bibnamefont {Thorne}}, \bibinfo {author} {\bibfnamefont {M.}~\bibnamefont {Wakano}}, \ and\ \bibinfo {author} {\bibfnamefont {J.~A.}\ \bibnamefont {Wheeler}},\ }\href@noop {} {\enquote {\bibinfo {title} {Gravitation theory and gravitational collapse},}\ } (\bibinfo {year} {1965})\BibitemShut {NoStop}%
\bibitem [{\citenamefont {Gourgoulhon}\ \emph {et~al.}(1995)\citenamefont {Gourgoulhon}, \citenamefont {Haensel},\ and\ \citenamefont {Gondek}}]{1995A&A...294..747G}%
  \BibitemOpen
  \bibfield  {author} {\bibinfo {author} {\bibfnamefont {E.}~\bibnamefont {Gourgoulhon}}, \bibinfo {author} {\bibfnamefont {P.}~\bibnamefont {Haensel}}, \ and\ \bibinfo {author} {\bibfnamefont {D.}~\bibnamefont {Gondek}},\ }\href@noop {} {\bibfield  {journal} {\bibinfo  {journal} {Astron. Astrophys.}\ }\textbf {\bibinfo {volume} {294}},\ \bibinfo {pages} {747} (\bibinfo {year} {1995})}\BibitemShut {NoStop}%
\bibitem [{\citenamefont {Sagun}\ \emph {et~al.}(2020)\citenamefont {Sagun}, \citenamefont {Panotopoulos},\ and\ \citenamefont {Lopes}}]{Sagun:2020qvc}%
  \BibitemOpen
  \bibfield  {author} {\bibinfo {author} {\bibfnamefont {V.}~\bibnamefont {Sagun}}, \bibinfo {author} {\bibfnamefont {G.}~\bibnamefont {Panotopoulos}}, \ and\ \bibinfo {author} {\bibfnamefont {I.}~\bibnamefont {Lopes}},\ }\href {\doibase 10.1103/PhysRevD.101.063025} {\bibfield  {journal} {\bibinfo  {journal} {Phys. Rev. D}\ }\textbf {\bibinfo {volume} {101}},\ \bibinfo {pages} {063025} (\bibinfo {year} {2020})},\ \Eprint {http://arxiv.org/abs/2002.12209} {arXiv:2002.12209 [astro-ph.HE]} \BibitemShut {NoStop}%
\bibitem [{\citenamefont {{Sedrakyan}}\ and\ \citenamefont {{Chubaryan}}(1968{\natexlab{a}})}]{1968Ap......4...87S}%
  \BibitemOpen
  \bibfield  {author} {\bibinfo {author} {\bibfnamefont {D.~M.}\ \bibnamefont {{Sedrakyan}}}\ and\ \bibinfo {author} {\bibfnamefont {E.~V.}\ \bibnamefont {{Chubaryan}}},\ }\href {\doibase 10.1007/BF01020005} {\bibfield  {journal} {\bibinfo  {journal} {Astrophysics}\ }\textbf {\bibinfo {volume} {4}},\ \bibinfo {pages} {87} (\bibinfo {year} {1968}{\natexlab{a}})}\BibitemShut {NoStop}%
\bibitem [{\citenamefont {{Sedrakyan}}\ and\ \citenamefont {{Chubaryan}}(1968{\natexlab{b}})}]{1968Ap......4..227S}%
  \BibitemOpen
  \bibfield  {author} {\bibinfo {author} {\bibfnamefont {D.~M.}\ \bibnamefont {{Sedrakyan}}}\ and\ \bibinfo {author} {\bibfnamefont {E.~V.}\ \bibnamefont {{Chubaryan}}},\ }\href {\doibase 10.1007/BF01013134} {\bibfield  {journal} {\bibinfo  {journal} {Astrophysics}\ }\textbf {\bibinfo {volume} {4}},\ \bibinfo {pages} {227} (\bibinfo {year} {1968}{\natexlab{b}})}\BibitemShut {NoStop}%
\bibitem [{\citenamefont {Chubarian}\ \emph {et~al.}(2000)\citenamefont {Chubarian}, \citenamefont {Grigorian}, \citenamefont {Poghosyan},\ and\ \citenamefont {Blaschke}}]{Chubarian:1999yn}%
  \BibitemOpen
  \bibfield  {author} {\bibinfo {author} {\bibfnamefont {E.}~\bibnamefont {Chubarian}}, \bibinfo {author} {\bibfnamefont {H.}~\bibnamefont {Grigorian}}, \bibinfo {author} {\bibfnamefont {G.~S.}\ \bibnamefont {Poghosyan}}, \ and\ \bibinfo {author} {\bibfnamefont {D.}~\bibnamefont {Blaschke}},\ }\href@noop {} {\bibfield  {journal} {\bibinfo  {journal} {Astron. Astrophys.}\ }\textbf {\bibinfo {volume} {357}},\ \bibinfo {pages} {968} (\bibinfo {year} {2000})},\ \Eprint {http://arxiv.org/abs/astro-ph/9903489} {arXiv:astro-ph/9903489} \BibitemShut {NoStop}%
\bibitem [{\citenamefont {Konstantinou}\ and\ \citenamefont {Morsink}(2022)}]{Konstantinou:2022vkr}%
  \BibitemOpen
  \bibfield  {author} {\bibinfo {author} {\bibfnamefont {A.}~\bibnamefont {Konstantinou}}\ and\ \bibinfo {author} {\bibfnamefont {S.~M.}\ \bibnamefont {Morsink}},\ }\href {\doibase 10.3847/1538-4357/ac7b86} {\bibfield  {journal} {\bibinfo  {journal} {Astrophys. J.}\ }\textbf {\bibinfo {volume} {934}},\ \bibinfo {pages} {139} (\bibinfo {year} {2022})},\ \Eprint {http://arxiv.org/abs/2206.12515} {arXiv:2206.12515 [astro-ph.HE]} \BibitemShut {NoStop}%
\bibitem [{\citenamefont {Hartle}(1967)}]{Hartle:1967he}%
  \BibitemOpen
  \bibfield  {author} {\bibinfo {author} {\bibfnamefont {J.~B.}\ \bibnamefont {Hartle}},\ }\href {\doibase 10.1086/149400} {\bibfield  {journal} {\bibinfo  {journal} {Astrophys. J.}\ }\textbf {\bibinfo {volume} {150}},\ \bibinfo {pages} {1005} (\bibinfo {year} {1967})}\BibitemShut {NoStop}%
\bibitem [{\citenamefont {Hartle}\ and\ \citenamefont {Thorne}(1968)}]{Hartle:1968si}%
  \BibitemOpen
  \bibfield  {author} {\bibinfo {author} {\bibfnamefont {J.~B.}\ \bibnamefont {Hartle}}\ and\ \bibinfo {author} {\bibfnamefont {K.~S.}\ \bibnamefont {Thorne}},\ }\href {\doibase 10.1086/149707} {\bibfield  {journal} {\bibinfo  {journal} {Astrophys. J.}\ }\textbf {\bibinfo {volume} {153}},\ \bibinfo {pages} {807} (\bibinfo {year} {1968})}\BibitemShut {NoStop}%
\bibitem [{\citenamefont {Shapiro}\ \emph {et~al.}(1983)\citenamefont {Shapiro}, \citenamefont {Teukolsky},\ and\ \citenamefont {Wasserman}}]{1983ApJ...272..702S}%
  \BibitemOpen
  \bibfield  {author} {\bibinfo {author} {\bibfnamefont {S.~L.}\ \bibnamefont {Shapiro}}, \bibinfo {author} {\bibfnamefont {S.~A.}\ \bibnamefont {Teukolsky}}, \ and\ \bibinfo {author} {\bibfnamefont {I.}~\bibnamefont {Wasserman}},\ }\href {\doibase 10.1086/161332} {\bibfield  {journal} {\bibinfo  {journal} {Astroph. J.}\ }\textbf {\bibinfo {volume} {272}},\ \bibinfo {pages} {702} (\bibinfo {year} {1983})}\BibitemShut {NoStop}%
\bibitem [{\citenamefont {Shapiro}\ \emph {et~al.}(1989)\citenamefont {Shapiro}, \citenamefont {Teukolsky},\ and\ \citenamefont {Wasserman}}]{1989Natur.340..451S}%
  \BibitemOpen
  \bibfield  {author} {\bibinfo {author} {\bibfnamefont {S.~L.}\ \bibnamefont {Shapiro}}, \bibinfo {author} {\bibfnamefont {S.~A.}\ \bibnamefont {Teukolsky}}, \ and\ \bibinfo {author} {\bibfnamefont {I.}~\bibnamefont {Wasserman}},\ }\href {\doibase 10.1038/340451a0} {\bibfield  {journal} {\bibinfo  {journal} {Nature}\ }\textbf {\bibinfo {volume} {340}},\ \bibinfo {pages} {451} (\bibinfo {year} {1989})}\BibitemShut {NoStop}%
\bibitem [{\citenamefont {Haensel}\ \emph {et~al.}(2009)\citenamefont {Haensel}, \citenamefont {Zdunik}, \citenamefont {Bejger},\ and\ \citenamefont {Lattimer}}]{Haensel:2009wa}%
  \BibitemOpen
  \bibfield  {author} {\bibinfo {author} {\bibfnamefont {P.}~\bibnamefont {Haensel}}, \bibinfo {author} {\bibfnamefont {J.~L.}\ \bibnamefont {Zdunik}}, \bibinfo {author} {\bibfnamefont {M.}~\bibnamefont {Bejger}}, \ and\ \bibinfo {author} {\bibfnamefont {J.~M.}\ \bibnamefont {Lattimer}},\ }\href {\doibase 10.1051/0004-6361/200811605} {\bibfield  {journal} {\bibinfo  {journal} {Astron. Astrophys.}\ }\textbf {\bibinfo {volume} {502}},\ \bibinfo {pages} {605} (\bibinfo {year} {2009})},\ \Eprint {http://arxiv.org/abs/0901.1268} {arXiv:0901.1268 [astro-ph.SR]} \BibitemShut {NoStop}%
\bibitem [{\citenamefont {Lattimer}\ and\ \citenamefont {Prakash}(2004)}]{Lattimer:2004pg}%
  \BibitemOpen
  \bibfield  {author} {\bibinfo {author} {\bibfnamefont {J.~M.}\ \bibnamefont {Lattimer}}\ and\ \bibinfo {author} {\bibfnamefont {M.}~\bibnamefont {Prakash}},\ }\href {\doibase 10.1126/science.1090720} {\bibfield  {journal} {\bibinfo  {journal} {Science}\ }\textbf {\bibinfo {volume} {304}},\ \bibinfo {pages} {536} (\bibinfo {year} {2004})},\ \Eprint {http://arxiv.org/abs/astro-ph/0405262} {arXiv:astro-ph/0405262} \BibitemShut {NoStop}%
\bibitem [{\citenamefont {Riahi}\ \emph {et~al.}(2019)\citenamefont {Riahi}, \citenamefont {Kalantari},\ and\ \citenamefont {Rueda~Hernandez}}]{Riahi:2019hmt}%
  \BibitemOpen
  \bibfield  {author} {\bibinfo {author} {\bibfnamefont {R.}~\bibnamefont {Riahi}}, \bibinfo {author} {\bibfnamefont {S.~Z.}\ \bibnamefont {Kalantari}}, \ and\ \bibinfo {author} {\bibfnamefont {J.~A.}\ \bibnamefont {Rueda~Hernandez}},\ }\href {\doibase 10.1103/PhysRevD.99.043004} {\bibfield  {journal} {\bibinfo  {journal} {Phys. Rev. D}\ }\textbf {\bibinfo {volume} {99}},\ \bibinfo {pages} {043004} (\bibinfo {year} {2019})},\ \Eprint {http://arxiv.org/abs/1902.00349} {arXiv:1902.00349 [astro-ph.HE]} \BibitemShut {NoStop}%
\bibitem [{\citenamefont {Breu}\ and\ \citenamefont {Rezzolla}(2016)}]{Breu:2016ufb}%
  \BibitemOpen
  \bibfield  {author} {\bibinfo {author} {\bibfnamefont {C.}~\bibnamefont {Breu}}\ and\ \bibinfo {author} {\bibfnamefont {L.}~\bibnamefont {Rezzolla}},\ }\href {\doibase 10.1093/mnras/stw575} {\bibfield  {journal} {\bibinfo  {journal} {Mon. Not. Roy. Astron. Soc.}\ }\textbf {\bibinfo {volume} {459}},\ \bibinfo {pages} {646} (\bibinfo {year} {2016})},\ \Eprint {http://arxiv.org/abs/1601.06083} {arXiv:1601.06083 [gr-qc]} \BibitemShut {NoStop}%
\bibitem [{\citenamefont {Musolino}\ \emph {et~al.}(2024)\citenamefont {Musolino}, \citenamefont {Ecker},\ and\ \citenamefont {Rezzolla}}]{Musolino:2023edi}%
  \BibitemOpen
  \bibfield  {author} {\bibinfo {author} {\bibfnamefont {C.}~\bibnamefont {Musolino}}, \bibinfo {author} {\bibfnamefont {C.}~\bibnamefont {Ecker}}, \ and\ \bibinfo {author} {\bibfnamefont {L.}~\bibnamefont {Rezzolla}},\ }\href {\doibase 10.3847/1538-4357/ad1758} {\bibfield  {journal} {\bibinfo  {journal} {Astrophys. J.}\ }\textbf {\bibinfo {volume} {962}},\ \bibinfo {pages} {61} (\bibinfo {year} {2024})},\ \Eprint {http://arxiv.org/abs/2307.03225} {arXiv:2307.03225 [gr-qc]} \BibitemShut {NoStop}%
\bibitem [{\citenamefont {Doroshenko}\ \emph {et~al.}(2022)\citenamefont {Doroshenko}, \citenamefont {Suleimanov}, \citenamefont {Pühlhofer},\ and\ \citenamefont {et~al.}}]{Doroshenko2022}%
  \BibitemOpen
  \bibfield  {author} {\bibinfo {author} {\bibfnamefont {V.}~\bibnamefont {Doroshenko}}, \bibinfo {author} {\bibfnamefont {V.}~\bibnamefont {Suleimanov}}, \bibinfo {author} {\bibfnamefont {G.}~\bibnamefont {Pühlhofer}}, \ and\ \bibinfo {author} {\bibnamefont {et~al.}},\ }\href {\doibase 10.1038/s41550-022-01800-1} {\bibfield  {journal} {\bibinfo  {journal} {Nature Astronomy}\ }\textbf {\bibinfo {volume} {6}},\ \bibinfo {pages} {1444} (\bibinfo {year} {2022})}\BibitemShut {NoStop}%
\bibitem [{\citenamefont {Jaisawal}\ \emph {et~al.}(2024)\citenamefont {Jaisawal}, \citenamefont {Bostancım}, \citenamefont {Boztepe}, \citenamefont {Güver}, \citenamefont {Strohmayer}, \citenamefont {Ballantyne}, \citenamefont {Beck}, \citenamefont {Göğüş}, \citenamefont {Altamirano}, \citenamefont {Arzoumanian}, \citenamefont {Chakrabarty}, \citenamefont {Gendreau}, \citenamefont {Guillot}, \citenamefont {Ludlam}, \citenamefont {Ng}, \citenamefont {Sanna},\ and\ \citenamefont {Chenevez}}]{Jaisawal_2024}%
  \BibitemOpen
  \bibfield  {author} {\bibinfo {author} {\bibfnamefont {G.~K.}\ \bibnamefont {Jaisawal}}, \bibinfo {author} {\bibfnamefont {Z.~F.}\ \bibnamefont {Bostancım}}, \bibinfo {author} {\bibfnamefont {T.}~\bibnamefont {Boztepe}}, \bibinfo {author} {\bibfnamefont {T.}~\bibnamefont {Güver}}, \bibinfo {author} {\bibfnamefont {T.~E.}\ \bibnamefont {Strohmayer}}, \bibinfo {author} {\bibfnamefont {D.~R.}\ \bibnamefont {Ballantyne}}, \bibinfo {author} {\bibfnamefont {J.~H.}\ \bibnamefont {Beck}}, \bibinfo {author} {\bibfnamefont {E.}~\bibnamefont {Göğüş}}, \bibinfo {author} {\bibfnamefont {D.}~\bibnamefont {Altamirano}}, \bibinfo {author} {\bibfnamefont {Z.}~\bibnamefont {Arzoumanian}}, \bibinfo {author} {\bibfnamefont {D.}~\bibnamefont {Chakrabarty}}, \bibinfo {author} {\bibfnamefont {K.~C.}\ \bibnamefont {Gendreau}}, \bibinfo {author} {\bibfnamefont {S.}~\bibnamefont {Guillot}}, \bibinfo {author} {\bibfnamefont {R.~M.}\ \bibnamefont {Ludlam}}, \bibinfo {author} {\bibfnamefont {M.}~\bibnamefont {Ng}}, \bibinfo
  {author} {\bibfnamefont {A.}~\bibnamefont {Sanna}}, \ and\ \bibinfo {author} {\bibfnamefont {J.}~\bibnamefont {Chenevez}},\ }\href {\doibase 10.3847/1538-4357/ad794e} {\bibfield  {journal} {\bibinfo  {journal} {Astrophys. J.}\ }\textbf {\bibinfo {volume} {975}},\ \bibinfo {pages} {67} (\bibinfo {year} {2024})}\BibitemShut {NoStop}%
\bibitem [{\citenamefont {Guver}\ \emph {et~al.}(2010)\citenamefont {Guver}, \citenamefont {Wroblewski}, \citenamefont {Camarota},\ and\ \citenamefont {Ozel}}]{Guver:2010td}%
  \BibitemOpen
  \bibfield  {author} {\bibinfo {author} {\bibfnamefont {T.}~\bibnamefont {Guver}}, \bibinfo {author} {\bibfnamefont {P.}~\bibnamefont {Wroblewski}}, \bibinfo {author} {\bibfnamefont {L.}~\bibnamefont {Camarota}}, \ and\ \bibinfo {author} {\bibfnamefont {F.}~\bibnamefont {Ozel}},\ }\href {\doibase 10.1088/0004-637X/719/2/1807} {\bibfield  {journal} {\bibinfo  {journal} {Astrophys. J.}\ }\textbf {\bibinfo {volume} {719}},\ \bibinfo {pages} {1807} (\bibinfo {year} {2010})},\ \Eprint {http://arxiv.org/abs/1002.3825} {arXiv:1002.3825 [astro-ph.HE]} \BibitemShut {NoStop}%
\bibitem [{\citenamefont {Poghosyan}\ \emph {et~al.}(2001)\citenamefont {Poghosyan}, \citenamefont {Grigorian},\ and\ \citenamefont {Blaschke}}]{Poghosyan:2000mr}%
  \BibitemOpen
  \bibfield  {author} {\bibinfo {author} {\bibfnamefont {G.~S.}\ \bibnamefont {Poghosyan}}, \bibinfo {author} {\bibfnamefont {H.}~\bibnamefont {Grigorian}}, \ and\ \bibinfo {author} {\bibfnamefont {D.}~\bibnamefont {Blaschke}},\ }\href {\doibase 10.1086/319851} {\bibfield  {journal} {\bibinfo  {journal} {Astrophys. J. Lett.}\ }\textbf {\bibinfo {volume} {551}},\ \bibinfo {pages} {L73} (\bibinfo {year} {2001})},\ \Eprint {http://arxiv.org/abs/astro-ph/0101002} {arXiv:astro-ph/0101002} \BibitemShut {NoStop}%
\bibitem [{\citenamefont {Fonseca}\ \emph {et~al.}(2021)\citenamefont {Fonseca} \emph {et~al.}}]{Fonseca:2021wxt}%
  \BibitemOpen
  \bibfield  {author} {\bibinfo {author} {\bibfnamefont {E.}~\bibnamefont {Fonseca}} \emph {et~al.},\ }\href {\doibase 10.3847/2041-8213/ac03b8} {\bibfield  {journal} {\bibinfo  {journal} {Astrophys. J. Lett.}\ }\textbf {\bibinfo {volume} {915}},\ \bibinfo {pages} {L12} (\bibinfo {year} {2021})},\ \Eprint {http://arxiv.org/abs/2104.00880} {arXiv:2104.00880 [astro-ph.HE]} \BibitemShut {NoStop}%
\bibitem [{\citenamefont {Koliogiannis}\ and\ \citenamefont {Moustakidis}(2021)}]{Koliogiannis:2020nhh}%
  \BibitemOpen
  \bibfield  {author} {\bibinfo {author} {\bibfnamefont {P.~S.}\ \bibnamefont {Koliogiannis}}\ and\ \bibinfo {author} {\bibfnamefont {C.~C.}\ \bibnamefont {Moustakidis}},\ }\href {\doibase 10.3847/1538-4357/abe542} {\bibfield  {journal} {\bibinfo  {journal} {Astrophys. J.}\ }\textbf {\bibinfo {volume} {912}},\ \bibinfo {pages} {69} (\bibinfo {year} {2021})},\ \Eprint {http://arxiv.org/abs/2007.10424} {arXiv:2007.10424 [astro-ph.HE]} \BibitemShut {NoStop}%
\bibitem [{\citenamefont {Managan}(1985)}]{1985ApJ...294..463M}%
  \BibitemOpen
  \bibfield  {author} {\bibinfo {author} {\bibfnamefont {R.~A.}\ \bibnamefont {Managan}},\ }\href {\doibase 10.1086/163312} {\bibfield  {journal} {\bibinfo  {journal} {Astrophys. J.}\ }\textbf {\bibinfo {volume} {294}},\ \bibinfo {pages} {463} (\bibinfo {year} {1985})}\BibitemShut {NoStop}%
\bibitem [{\citenamefont {Friedman}\ \emph {et~al.}(1986{\natexlab{b}})\citenamefont {Friedman}, \citenamefont {Ipser},\ and\ \citenamefont {Parker}}]{1986ApJ...304..115F}%
  \BibitemOpen
  \bibfield  {author} {\bibinfo {author} {\bibfnamefont {J.~L.}\ \bibnamefont {Friedman}}, \bibinfo {author} {\bibfnamefont {J.~R.}\ \bibnamefont {Ipser}}, \ and\ \bibinfo {author} {\bibfnamefont {L.}~\bibnamefont {Parker}},\ }\href {\doibase 10.1086/164149} {\bibfield  {journal} {\bibinfo  {journal} {Astrophys. J.}\ }\textbf {\bibinfo {volume} {304}},\ \bibinfo {pages} {115} (\bibinfo {year} {1986}{\natexlab{b}})}\BibitemShut {NoStop}%
\bibitem [{\citenamefont {Morsink}\ \emph {et~al.}(1999)\citenamefont {Morsink}, \citenamefont {Stergioulas},\ and\ \citenamefont {Blattnig}}]{Morsink:1998db}%
  \BibitemOpen
  \bibfield  {author} {\bibinfo {author} {\bibfnamefont {S.~M.}\ \bibnamefont {Morsink}}, \bibinfo {author} {\bibfnamefont {N.}~\bibnamefont {Stergioulas}}, \ and\ \bibinfo {author} {\bibfnamefont {S.~R.}\ \bibnamefont {Blattnig}},\ }\href {\doibase 10.1086/306630} {\bibfield  {journal} {\bibinfo  {journal} {Astrophys. J.}\ }\textbf {\bibinfo {volume} {510}},\ \bibinfo {pages} {854} (\bibinfo {year} {1999})},\ \Eprint {http://arxiv.org/abs/gr-qc/9806008} {arXiv:gr-qc/9806008} \BibitemShut {NoStop}%
\bibitem [{\citenamefont {Di~Clemente}\ \emph {et~al.}(2020)\citenamefont {Di~Clemente}, \citenamefont {Mannarelli},\ and\ \citenamefont {Tonelli}}]{DiClemente:2020szl}%
  \BibitemOpen
  \bibfield  {author} {\bibinfo {author} {\bibfnamefont {F.}~\bibnamefont {Di~Clemente}}, \bibinfo {author} {\bibfnamefont {M.}~\bibnamefont {Mannarelli}}, \ and\ \bibinfo {author} {\bibfnamefont {F.}~\bibnamefont {Tonelli}},\ }\href {\doibase 10.1103/PhysRevD.101.103003} {\bibfield  {journal} {\bibinfo  {journal} {Phys. Rev. D}\ }\textbf {\bibinfo {volume} {101}},\ \bibinfo {pages} {103003} (\bibinfo {year} {2020})},\ \Eprint {http://arxiv.org/abs/2002.09483} {arXiv:2002.09483 [gr-qc]} \BibitemShut {NoStop}%
\bibitem [{\citenamefont {Suwa}\ \emph {et~al.}(2018)\citenamefont {Suwa}, \citenamefont {Yoshida}, \citenamefont {Shibata}, \citenamefont {Umeda},\ and\ \citenamefont {Takahashi}}]{Suwa:2018uni}%
  \BibitemOpen
  \bibfield  {author} {\bibinfo {author} {\bibfnamefont {Y.}~\bibnamefont {Suwa}}, \bibinfo {author} {\bibfnamefont {T.}~\bibnamefont {Yoshida}}, \bibinfo {author} {\bibfnamefont {M.}~\bibnamefont {Shibata}}, \bibinfo {author} {\bibfnamefont {H.}~\bibnamefont {Umeda}}, \ and\ \bibinfo {author} {\bibfnamefont {K.}~\bibnamefont {Takahashi}},\ }\href {\doibase 10.1093/mnras/sty2460} {\bibfield  {journal} {\bibinfo  {journal} {Mon. Not. Roy. Astron. Soc.}\ }\textbf {\bibinfo {volume} {481}},\ \bibinfo {pages} {3305} (\bibinfo {year} {2018})},\ \Eprint {http://arxiv.org/abs/1808.02328} {arXiv:1808.02328 [astro-ph.HE]} \BibitemShut {NoStop}%
\bibitem [{\citenamefont {Sagun}\ \emph {et~al.}(2023)\citenamefont {Sagun}, \citenamefont {Giangrandi}, \citenamefont {Dietrich}, \citenamefont {Ivanytskyi}, \citenamefont {Negreiros},\ and\ \citenamefont {Provid\^encia}}]{Sagun:2023rzp}%
  \BibitemOpen
  \bibfield  {author} {\bibinfo {author} {\bibfnamefont {V.}~\bibnamefont {Sagun}}, \bibinfo {author} {\bibfnamefont {E.}~\bibnamefont {Giangrandi}}, \bibinfo {author} {\bibfnamefont {T.}~\bibnamefont {Dietrich}}, \bibinfo {author} {\bibfnamefont {O.}~\bibnamefont {Ivanytskyi}}, \bibinfo {author} {\bibfnamefont {R.}~\bibnamefont {Negreiros}}, \ and\ \bibinfo {author} {\bibfnamefont {C.}~\bibnamefont {Provid\^encia}},\ }\href {\doibase 10.3847/1538-4357/acfc9e} {\bibfield  {journal} {\bibinfo  {journal} {Astrophys. J.}\ }\textbf {\bibinfo {volume} {958}},\ \bibinfo {pages} {49} (\bibinfo {year} {2023})},\ \Eprint {http://arxiv.org/abs/2306.12326} {arXiv:2306.12326 [astro-ph.HE]} \BibitemShut {NoStop}%
\bibitem [{\citenamefont {Nan}\ \emph {et~al.}(2011)\citenamefont {Nan}, \citenamefont {Li}, \citenamefont {Jin}, \citenamefont {Wang}, \citenamefont {Zhu}, \citenamefont {Zhu}, \citenamefont {Zhang}, \citenamefont {Yue},\ and\ \citenamefont {Qian}}]{Nan:2011um}%
  \BibitemOpen
  \bibfield  {author} {\bibinfo {author} {\bibfnamefont {R.}~\bibnamefont {Nan}}, \bibinfo {author} {\bibfnamefont {D.}~\bibnamefont {Li}}, \bibinfo {author} {\bibfnamefont {C.}~\bibnamefont {Jin}}, \bibinfo {author} {\bibfnamefont {Q.}~\bibnamefont {Wang}}, \bibinfo {author} {\bibfnamefont {L.}~\bibnamefont {Zhu}}, \bibinfo {author} {\bibfnamefont {W.}~\bibnamefont {Zhu}}, \bibinfo {author} {\bibfnamefont {H.}~\bibnamefont {Zhang}}, \bibinfo {author} {\bibfnamefont {Y.}~\bibnamefont {Yue}}, \ and\ \bibinfo {author} {\bibfnamefont {L.}~\bibnamefont {Qian}},\ }\href {\doibase 10.1142/S0218271811019335} {\bibfield  {journal} {\bibinfo  {journal} {Int. J. Mod. Phys. D}\ }\textbf {\bibinfo {volume} {20}},\ \bibinfo {pages} {989} (\bibinfo {year} {2011})},\ \Eprint {http://arxiv.org/abs/1105.3794} {arXiv:1105.3794 [astro-ph.IM]} \BibitemShut {NoStop}%
\bibitem [{\citenamefont {Padmanabh}\ \emph {et~al.}(2023)\citenamefont {Padmanabh} \emph {et~al.}}]{Padmanabh:2023vma}%
  \BibitemOpen
  \bibfield  {author} {\bibinfo {author} {\bibfnamefont {P.~V.}\ \bibnamefont {Padmanabh}} \emph {et~al.},\ }\href {\doibase 10.1093/mnras/stad1900} {\  (\bibinfo {year} {2023}),\ 10.1093/mnras/stad1900},\ \Eprint {http://arxiv.org/abs/2303.09231} {arXiv:2303.09231 [astro-ph.HE]} \BibitemShut {NoStop}%
\bibitem [{\citenamefont {Watts}\ \emph {et~al.}(2015)\citenamefont {Watts} \emph {et~al.}}]{Watts:2014tja}%
  \BibitemOpen
  \bibfield  {author} {\bibinfo {author} {\bibfnamefont {A.}~\bibnamefont {Watts}} \emph {et~al.},\ }\href {\doibase 10.22323/1.215.0043} {\bibfield  {journal} {\bibinfo  {journal} {PoS}\ }\textbf {\bibinfo {volume} {AASKA14}},\ \bibinfo {pages} {043} (\bibinfo {year} {2015})},\ \Eprint {http://arxiv.org/abs/1501.00042} {arXiv:1501.00042 [astro-ph.SR]} \BibitemShut {NoStop}%
\bibitem [{\citenamefont {Barcons}\ \emph {et~al.}(2012)\citenamefont {Barcons} \emph {et~al.}}]{Barcons:2012zb}%
  \BibitemOpen
  \bibfield  {author} {\bibinfo {author} {\bibfnamefont {X.}~\bibnamefont {Barcons}} \emph {et~al.},\ }\href@noop {} {\  (\bibinfo {year} {2012})},\ \Eprint {http://arxiv.org/abs/1207.2745} {arXiv:1207.2745 [astro-ph.HE]} \BibitemShut {NoStop}%
\bibitem [{\citenamefont {Ray}\ \emph {et~al.}(2019)\citenamefont {Ray} \emph {et~al.}}]{STROBE-XScienceWorkingGroup:2019cyd}%
  \BibitemOpen
  \bibfield  {author} {\bibinfo {author} {\bibfnamefont {P.~S.}\ \bibnamefont {Ray}} \emph {et~al.} (\bibinfo {collaboration} {STROBE-X Science Working Group}),\ }\href@noop {} {\  (\bibinfo {year} {2019})},\ \Eprint {http://arxiv.org/abs/1903.03035} {arXiv:1903.03035 [astro-ph.IM]} \BibitemShut {NoStop}%
\bibitem [{\citenamefont {Zhang}\ \emph {et~al.}(2019)\citenamefont {Zhang} \emph {et~al.}}]{eXTP:2018anb}%
  \BibitemOpen
  \bibfield  {author} {\bibinfo {author} {\bibfnamefont {S.-N.}\ \bibnamefont {Zhang}} \emph {et~al.} (\bibinfo {collaboration} {eXTP}),\ }\href {\doibase 10.1007/s11433-018-9309-2} {\bibfield  {journal} {\bibinfo  {journal} {Sci. China Phys. Mech. Astron.}\ }\textbf {\bibinfo {volume} {62}},\ \bibinfo {pages} {29502} (\bibinfo {year} {2019})},\ \Eprint {http://arxiv.org/abs/1812.04020} {arXiv:1812.04020 [astro-ph.IM]} \BibitemShut {NoStop}%
\bibitem [{\citenamefont {Clark}\ \emph {et~al.}(2023)\citenamefont {Clark} \emph {et~al.}}]{Clark:2023owb}%
  \BibitemOpen
  \bibfield  {author} {\bibinfo {author} {\bibfnamefont {C.~J.}\ \bibnamefont {Clark}} \emph {et~al.},\ }\href {\doibase 10.1038/s41550-022-01874-x} {\bibfield  {journal} {\bibinfo  {journal} {Nature Astron.}\ }\textbf {\bibinfo {volume} {7}},\ \bibinfo {pages} {451} (\bibinfo {year} {2023})},\ \Eprint {http://arxiv.org/abs/2301.10995} {arXiv:2301.10995 [astro-ph.HE]} \BibitemShut {NoStop}%
\bibitem [{\citenamefont {Strader}\ \emph {et~al.}(2019)\citenamefont {Strader} \emph {et~al.}}]{Strader:2018qbi}%
  \BibitemOpen
  \bibfield  {author} {\bibinfo {author} {\bibfnamefont {J.}~\bibnamefont {Strader}} \emph {et~al.},\ }\href {\doibase 10.3847/1538-4357/aafbaa} {\bibfield  {journal} {\bibinfo  {journal} {Astrophys. J.}\ }\textbf {\bibinfo {volume} {872}},\ \bibinfo {pages} {42} (\bibinfo {year} {2019})},\ \Eprint {http://arxiv.org/abs/1812.04626} {arXiv:1812.04626 [astro-ph.HE]} \BibitemShut {NoStop}%
\bibitem [{\citenamefont {Serylak}\ \emph {et~al.}(2022)\citenamefont {Serylak} \emph {et~al.}}]{Serylak:2022kna}%
  \BibitemOpen
  \bibfield  {author} {\bibinfo {author} {\bibfnamefont {M.}~\bibnamefont {Serylak}} \emph {et~al.},\ }\href {\doibase 10.1051/0004-6361/202142670} {\bibfield  {journal} {\bibinfo  {journal} {Astron. Astrophys.}\ }\textbf {\bibinfo {volume} {665}},\ \bibinfo {pages} {A53} (\bibinfo {year} {2022})},\ \Eprint {http://arxiv.org/abs/2203.00607} {arXiv:2203.00607 [astro-ph.HE]} \BibitemShut {NoStop}%
\bibitem [{\citenamefont {Freire}\ \emph {et~al.}(2011)\citenamefont {Freire} \emph {et~al.}}]{Freire:2010tf}%
  \BibitemOpen
  \bibfield  {author} {\bibinfo {author} {\bibfnamefont {P.~C.~C.}\ \bibnamefont {Freire}} \emph {et~al.},\ }\href {\doibase 10.1111/j.1365-2966.2010.18109.x} {\bibfield  {journal} {\bibinfo  {journal} {Mon. Not. Roy. Astron. Soc.}\ }\textbf {\bibinfo {volume} {412}},\ \bibinfo {pages} {2763} (\bibinfo {year} {2011})},\ \Eprint {http://arxiv.org/abs/1011.5809} {arXiv:1011.5809 [astro-ph.GA]} \BibitemShut {NoStop}%
\bibitem [{\citenamefont {Padmanabh}\ \emph {et~al.}(2024)\citenamefont {Padmanabh} \emph {et~al.}}]{Padmanabh:2024bsz}%
  \BibitemOpen
  \bibfield  {author} {\bibinfo {author} {\bibfnamefont {P.~V.}\ \bibnamefont {Padmanabh}} \emph {et~al.},\ }\href {\doibase 10.1051/0004-6361/202449303} {\bibfield  {journal} {\bibinfo  {journal} {Astron. Astrophys.}\ }\textbf {\bibinfo {volume} {686}},\ \bibinfo {pages} {A166} (\bibinfo {year} {2024})},\ \Eprint {http://arxiv.org/abs/2403.17799} {arXiv:2403.17799 [astro-ph.HE]} \BibitemShut {NoStop}%
\bibitem [{\citenamefont {Arzoumanian}\ \emph {et~al.}(2018)\citenamefont {Arzoumanian} \emph {et~al.}}]{NANOGrav:2017wvv}%
  \BibitemOpen
  \bibfield  {author} {\bibinfo {author} {\bibfnamefont {Z.}~\bibnamefont {Arzoumanian}} \emph {et~al.} (\bibinfo {collaboration} {NANOGrav}),\ }\href {\doibase 10.3847/1538-4365/aab5b0} {\bibfield  {journal} {\bibinfo  {journal} {Astrophys. J. Suppl.}\ }\textbf {\bibinfo {volume} {235}},\ \bibinfo {pages} {37} (\bibinfo {year} {2018})},\ \Eprint {http://arxiv.org/abs/1801.01837} {arXiv:1801.01837 [astro-ph.HE]} \BibitemShut {NoStop}%
\bibitem [{\citenamefont {Romani}\ \emph {et~al.}(2015)\citenamefont {Romani}, \citenamefont {Filippenko},\ and\ \citenamefont {Cenko}}]{Romani:2015gaa}%
  \BibitemOpen
  \bibfield  {author} {\bibinfo {author} {\bibfnamefont {R.~W.}\ \bibnamefont {Romani}}, \bibinfo {author} {\bibfnamefont {A.~V.}\ \bibnamefont {Filippenko}}, \ and\ \bibinfo {author} {\bibfnamefont {S.~B.}\ \bibnamefont {Cenko}},\ }\href {\doibase 10.1088/0004-637X/804/2/115} {\bibfield  {journal} {\bibinfo  {journal} {Astrophys. J.}\ }\textbf {\bibinfo {volume} {804}},\ \bibinfo {pages} {115} (\bibinfo {year} {2015})},\ \Eprint {http://arxiv.org/abs/1503.05247} {arXiv:1503.05247 [astro-ph.HE]} \BibitemShut {NoStop}%
\bibitem [{\citenamefont {Shamohammadi}\ \emph {et~al.}(2023)\citenamefont {Shamohammadi} \emph {et~al.}}]{Shamohammadi:2022ttx}%
  \BibitemOpen
  \bibfield  {author} {\bibinfo {author} {\bibfnamefont {M.}~\bibnamefont {Shamohammadi}} \emph {et~al.},\ }\href {\doibase 10.1093/mnras/stac3719} {\bibfield  {journal} {\bibinfo  {journal} {Mon. Not. Roy. Astron. Soc.}\ }\textbf {\bibinfo {volume} {520}},\ \bibinfo {pages} {1789} (\bibinfo {year} {2023})},\ \Eprint {http://arxiv.org/abs/2212.04051} {arXiv:2212.04051 [astro-ph.HE]} \BibitemShut {NoStop}%
\bibitem [{\citenamefont {Ransom}\ \emph {et~al.}(2014)\citenamefont {Ransom} \emph {et~al.}}]{Ransom:2014xla}%
  \BibitemOpen
  \bibfield  {author} {\bibinfo {author} {\bibfnamefont {S.~M.}\ \bibnamefont {Ransom}} \emph {et~al.},\ }\href {\doibase 10.1038/nature12917} {\bibfield  {journal} {\bibinfo  {journal} {Nature}\ }\textbf {\bibinfo {volume} {505}},\ \bibinfo {pages} {520} (\bibinfo {year} {2014})},\ \Eprint {http://arxiv.org/abs/1401.0535} {arXiv:1401.0535 [astro-ph.SR]} \BibitemShut {NoStop}%
\bibitem [{\citenamefont {Andersen}\ and\ \citenamefont {Ransom}(2018)}]{Andersen:2018nsx}%
  \BibitemOpen
  \bibfield  {author} {\bibinfo {author} {\bibfnamefont {B.~C.}\ \bibnamefont {Andersen}}\ and\ \bibinfo {author} {\bibfnamefont {S.~M.}\ \bibnamefont {Ransom}},\ }\href {\doibase 10.3847/2041-8213/aad59f} {\bibfield  {journal} {\bibinfo  {journal} {Astrophys. J. Lett.}\ }\textbf {\bibinfo {volume} {863}},\ \bibinfo {pages} {L13} (\bibinfo {year} {2018})},\ \Eprint {http://arxiv.org/abs/1807.07900} {arXiv:1807.07900 [astro-ph.HE]} \BibitemShut {NoStop}%
\bibitem [{\citenamefont {Liu}\ \emph {et~al.}(2020)\citenamefont {Liu} \emph {et~al.}}]{Liu:2020hkx}%
  \BibitemOpen
  \bibfield  {author} {\bibinfo {author} {\bibfnamefont {K.}~\bibnamefont {Liu}} \emph {et~al.},\ }\href {\doibase 10.1093/mnras/staa2993} {\bibfield  {journal} {\bibinfo  {journal} {Mon. Not. Roy. Astron. Soc.}\ }\textbf {\bibinfo {volume} {499}},\ \bibinfo {pages} {2276} (\bibinfo {year} {2020})},\ \Eprint {http://arxiv.org/abs/2009.12544} {arXiv:2009.12544 [astro-ph.HE]} \BibitemShut {NoStop}%
\bibitem [{\citenamefont {Gautam}\ \emph {et~al.}(2024)\citenamefont {Gautam}, \citenamefont {Freire}, \citenamefont {Wu}, \citenamefont {Venkatraman~Krishnan}, \citenamefont {Kramer}, \citenamefont {Barr}, \citenamefont {Bailes},\ and\ \citenamefont {Cameron}}]{Gautam:2023ref}%
  \BibitemOpen
  \bibfield  {author} {\bibinfo {author} {\bibfnamefont {T.}~\bibnamefont {Gautam}}, \bibinfo {author} {\bibfnamefont {P.~C.~C.}\ \bibnamefont {Freire}}, \bibinfo {author} {\bibfnamefont {J.}~\bibnamefont {Wu}}, \bibinfo {author} {\bibfnamefont {V.}~\bibnamefont {Venkatraman~Krishnan}}, \bibinfo {author} {\bibfnamefont {M.}~\bibnamefont {Kramer}}, \bibinfo {author} {\bibfnamefont {E.~D.}\ \bibnamefont {Barr}}, \bibinfo {author} {\bibfnamefont {M.}~\bibnamefont {Bailes}}, \ and\ \bibinfo {author} {\bibfnamefont {A.~D.}\ \bibnamefont {Cameron}},\ }\href {\doibase 10.1051/0004-6361/202347836} {\bibfield  {journal} {\bibinfo  {journal} {Astron. Astrophys.}\ }\textbf {\bibinfo {volume} {682}},\ \bibinfo {pages} {A103} (\bibinfo {year} {2024})},\ \Eprint {http://arxiv.org/abs/2311.13563} {arXiv:2311.13563 [astro-ph.HE]} \BibitemShut {NoStop}%
\bibitem [{\citenamefont {Barr}\ \emph {et~al.}(2017)\citenamefont {Barr}, \citenamefont {Freire}, \citenamefont {Kramer}, \citenamefont {Champion}, \citenamefont {Berezina}, \citenamefont {Bassa}, \citenamefont {Lyne},\ and\ \citenamefont {Stappers}}]{Barr:2016vxv}%
  \BibitemOpen
  \bibfield  {author} {\bibinfo {author} {\bibfnamefont {E.~D.}\ \bibnamefont {Barr}}, \bibinfo {author} {\bibfnamefont {P.~C.~C.}\ \bibnamefont {Freire}}, \bibinfo {author} {\bibfnamefont {M.}~\bibnamefont {Kramer}}, \bibinfo {author} {\bibfnamefont {D.~J.}\ \bibnamefont {Champion}}, \bibinfo {author} {\bibfnamefont {M.}~\bibnamefont {Berezina}}, \bibinfo {author} {\bibfnamefont {C.~G.}\ \bibnamefont {Bassa}}, \bibinfo {author} {\bibfnamefont {A.~G.}\ \bibnamefont {Lyne}}, \ and\ \bibinfo {author} {\bibfnamefont {B.~W.}\ \bibnamefont {Stappers}},\ }\href {\doibase 10.1093/mnras/stw2947} {\bibfield  {journal} {\bibinfo  {journal} {Mon. Not. Roy. Astron. Soc.}\ }\textbf {\bibinfo {volume} {465}},\ \bibinfo {pages} {1711} (\bibinfo {year} {2017})},\ \Eprint {http://arxiv.org/abs/1611.03658} {arXiv:1611.03658 [astro-ph.HE]} \BibitemShut {NoStop}%
\bibitem [{\citenamefont {Freire}\ \emph {et~al.}(2017)\citenamefont {Freire} \emph {et~al.}}]{Freire:2017mgu}%
  \BibitemOpen
  \bibfield  {author} {\bibinfo {author} {\bibfnamefont {P.~C.~C.}\ \bibnamefont {Freire}} \emph {et~al.},\ }\href {\doibase 10.1093/mnras/stx1533} {\bibfield  {journal} {\bibinfo  {journal} {Mon. Not. Roy. Astron. Soc.}\ }\textbf {\bibinfo {volume} {471}},\ \bibinfo {pages} {857} (\bibinfo {year} {2017})},\ \Eprint {http://arxiv.org/abs/1706.04908} {arXiv:1706.04908 [astro-ph.HE]} \BibitemShut {NoStop}%
\bibitem [{\citenamefont {Corongiu}\ \emph {et~al.}(2023)\citenamefont {Corongiu} \emph {et~al.}}]{Corongiu:2023gft}%
  \BibitemOpen
  \bibfield  {author} {\bibinfo {author} {\bibfnamefont {A.}~\bibnamefont {Corongiu}} \emph {et~al.},\ }\href {\doibase 10.1051/0004-6361/202244418} {\bibfield  {journal} {\bibinfo  {journal} {Astron. Astrophys.}\ }\textbf {\bibinfo {volume} {671}},\ \bibinfo {pages} {A72} (\bibinfo {year} {2023})},\ \Eprint {http://arxiv.org/abs/2301.04055} {arXiv:2301.04055 [astro-ph.HE]} \BibitemShut {NoStop}%
\bibitem [{\citenamefont {Desvignes}\ \emph {et~al.}(2016)\citenamefont {Desvignes} \emph {et~al.}}]{EPTA:2016ndq}%
  \BibitemOpen
  \bibfield  {author} {\bibinfo {author} {\bibfnamefont {G.}~\bibnamefont {Desvignes}} \emph {et~al.} (\bibinfo {collaboration} {EPTA}),\ }\href {\doibase 10.1093/mnras/stw483} {\bibfield  {journal} {\bibinfo  {journal} {Mon. Not. Roy. Astron. Soc.}\ }\textbf {\bibinfo {volume} {458}},\ \bibinfo {pages} {3341} (\bibinfo {year} {2016})},\ \Eprint {http://arxiv.org/abs/1602.08511} {arXiv:1602.08511 [astro-ph.HE]} \BibitemShut {NoStop}%
\bibitem [{\citenamefont {Geyer}\ \emph {et~al.}(2023)\citenamefont {Geyer} \emph {et~al.}}]{Geyer:2023vng}%
  \BibitemOpen
  \bibfield  {author} {\bibinfo {author} {\bibfnamefont {M.}~\bibnamefont {Geyer}} \emph {et~al.},\ }\href {\doibase 10.1051/0004-6361/202244654} {\bibfield  {journal} {\bibinfo  {journal} {Astron. Astrophys.}\ }\textbf {\bibinfo {volume} {674}},\ \bibinfo {pages} {A169} (\bibinfo {year} {2023})},\ \Eprint {http://arxiv.org/abs/2304.09060} {arXiv:2304.09060 [astro-ph.HE]} \BibitemShut {NoStop}%
\bibitem [{\citenamefont {Stovall}\ \emph {et~al.}(2019)\citenamefont {Stovall} \emph {et~al.}}]{Stovall:2018rvy}%
  \BibitemOpen
  \bibfield  {author} {\bibinfo {author} {\bibfnamefont {K.}~\bibnamefont {Stovall}} \emph {et~al.},\ }\href {\doibase 10.3847/1538-4357/aaf37d} {\bibfield  {journal} {\bibinfo  {journal} {Astrophys. J.}\ }\textbf {\bibinfo {volume} {870}},\ \bibinfo {pages} {74} (\bibinfo {year} {2019})},\ \Eprint {http://arxiv.org/abs/1809.05064} {arXiv:1809.05064 [astro-ph.HE]} \BibitemShut {NoStop}%
\bibitem [{\citenamefont {Begin}(2006)}]{Begin_2006}%
  \BibitemOpen
  \bibfield  {author} {\bibinfo {author} {\bibfnamefont {S.}~\bibnamefont {Begin}},\ }\emph {\bibinfo {title} {A search for fast pulsars in globular clusters}},\ \href {\doibase http://dx.doi.org/10.14288/1.0066153} {Ph.D. thesis},\ \bibinfo  {school} {University of British Columbia} (\bibinfo {year} {2006})\BibitemShut {NoStop}%
\bibitem [{\citenamefont {Lynch}\ \emph {et~al.}(2012)\citenamefont {Lynch}, \citenamefont {Freire}, \citenamefont {Ransom},\ and\ \citenamefont {Jacoby}}]{Lynch:2011aa}%
  \BibitemOpen
  \bibfield  {author} {\bibinfo {author} {\bibfnamefont {R.~S.}\ \bibnamefont {Lynch}}, \bibinfo {author} {\bibfnamefont {P.~C.~C.}\ \bibnamefont {Freire}}, \bibinfo {author} {\bibfnamefont {S.~M.}\ \bibnamefont {Ransom}}, \ and\ \bibinfo {author} {\bibfnamefont {B.~A.}\ \bibnamefont {Jacoby}},\ }\href {\doibase 10.1088/0004-637X/745/2/109} {\bibfield  {journal} {\bibinfo  {journal} {Astrophys. J.}\ }\textbf {\bibinfo {volume} {745}},\ \bibinfo {pages} {109} (\bibinfo {year} {2012})},\ \Eprint {http://arxiv.org/abs/1112.2612} {arXiv:1112.2612 [astro-ph.HE]} \BibitemShut {NoStop}%
\bibitem [{\citenamefont {Zhu}\ \emph {et~al.}(2019)\citenamefont {Zhu} \emph {et~al.}}]{Zhu:2019oax}%
  \BibitemOpen
  \bibfield  {author} {\bibinfo {author} {\bibfnamefont {W.~W.}\ \bibnamefont {Zhu}} \emph {et~al.},\ }\href {\doibase 10.3847/1538-4357/ab2bef} {\bibfield  {journal} {\bibinfo  {journal} {Astrophys. J.}\ }\textbf {\bibinfo {volume} {881}},\ \bibinfo {pages} {165} (\bibinfo {year} {2019})},\ \Eprint {http://arxiv.org/abs/1907.05046} {arXiv:1907.05046 [astro-ph.SR]} \BibitemShut {NoStop}%
\bibitem [{\citenamefont {Ridolfi}\ \emph {et~al.}(2019)\citenamefont {Ridolfi}, \citenamefont {Freire}, \citenamefont {Gupta},\ and\ \citenamefont {Ransom}}]{Ridolfi:2019wgs}%
  \BibitemOpen
  \bibfield  {author} {\bibinfo {author} {\bibfnamefont {A.}~\bibnamefont {Ridolfi}}, \bibinfo {author} {\bibfnamefont {P.~C.~C.}\ \bibnamefont {Freire}}, \bibinfo {author} {\bibfnamefont {Y.}~\bibnamefont {Gupta}}, \ and\ \bibinfo {author} {\bibfnamefont {S.~M.}\ \bibnamefont {Ransom}},\ }\href {\doibase 10.1093/mnras/stz2645} {\bibfield  {journal} {\bibinfo  {journal} {Mon. Not. Roy. Astron. Soc.}\ }\textbf {\bibinfo {volume} {490}},\ \bibinfo {pages} {3860} (\bibinfo {year} {2019})},\ \Eprint {http://arxiv.org/abs/1909.06163} {arXiv:1909.06163 [astro-ph.HE]} \BibitemShut {NoStop}%
\end{thebibliography}%

\end{document}